\documentclass{article}

\usepackage{lipsum} 
\usepackage{amsmath} 
\usepackage{subcaption} 

\usepackage[utf8]{inputenc} 
\usepackage[T1]{fontenc}    
\usepackage{hyperref}       
\usepackage{url}            
\usepackage{booktabs}       
\usepackage{amsfonts}       
\usepackage{nicefrac}       
\usepackage{microtype}      
\usepackage{xcolor}    

\usepackage{cleveref}

\usepackage{graphicx}
\usepackage{xr}
\usepackage{xcite}
\usepackage{float}
\makeatletter
\newcommand*{\addFileDependency}[1]{
  \typeout{(#1)}
  \@addtofilelist{#1}
  \IfFileExists{#1}{}{\typeout{No file #1.}}
}
\makeatother

\usepackage{hyperref}
\usepackage[margin=1in]{geometry}
\crefname{appfig}{Supplementary Figure}{Supplementary Figures}
\usepackage{rotating}

\usepackage[numbers,sort&compress]{natbib}

\title{Large-scale spatial variable gene atlas for spatial transcriptomics}
\author{Jiawen Chen$^{1,2*}$, Jinwei Zhang$^{3*}$, Dongshen Peng$^{4,5*}$, Yutong Song$^{3*}$,\\
Aitong Ruan$^3$, Yun Li$^{3,6}$, Didong Li$^{3,5,7}$\\
Gladstone Institutes$^1$\\
Department of Biomedical Data Science, Stanford University$^2$\\
Department of Biostatistics$^3$, Computer Science$^4$, Statistics and Operations Research$^5$,\\ Genetics$^6$, Lineberger Comprehensive Cancer Center$^{7}$,\\ University of North Carolina at Chapel Hill\\
}
\date{}

\begin{document}

\maketitle
\begingroup
\renewcommand\thefootnote{}\footnotetext{* These authors contributed equally to this work.}
\endgroup

\setcounter{footnote}{0} 

\begin{abstract}

Spatial variable genes (SVGs) reveal critical information about tissue architecture, cellular interactions, and disease microenvironments. As spatial transcriptomics (ST) technologies proliferate, accurately identifying SVGs across diverse platforms, tissue types, and disease contexts has become both a major opportunity and a significant computational challenge. Here, we present a comprehensive benchmarking study of 20 state-of-the-art SVG detection methods using human slides from STimage-1K4M, a large-scale resource of ST data comprising 662 slides from more than 18 tissue types. We evaluate each method across a range of biologically and technically meaningful criteria, including recovery of pathologist-annotated domain-specific markers, cross-slide reproducibility, scalability to high-resolution data, and robustness to technical variation. Our results reveal marked differences in performance depending on tissue type, spatial resolution, and study design. Beyond benchmarking, we construct the first cross-tissue atlas of SVGs, enabling comparative analysis of spatial gene programs across cancer and normal tissues. We observe similarities between pairs of tissues that reflect developmental and functional relationships, such as high overlap between thymus and lymph node, and uncover spatial gene programs associated with metastasis, immune infiltration, and tissue-of-origin identity in cancer. Together, our work defines a framework for evaluating and interpreting spatial gene expression and establishes a reference resource for the ST community.

\end{abstract}

\section{Introduction}

Spatial transcriptomics (ST) technologies have revolutionized our ability to study gene expression within the spatial context of intact tissues. Unlike traditional single-cell RNA sequencing, which dissociates cells and loses positional information, ST allows researchers to investigate how gene expression patterns are organized in space, revealing how cell types, states, and microenvironments are arranged within complex tissues~\citep{staahl2016visualization,chen2022comprehensive}. This spatially resolved view is especially critical for understanding tissue organization~\citep{Maynard2021-wa}, developmental biology~\citep{Asp2019-ga}, and disease processes such as tumor heterogeneity~\citep{chen2023cell}, immune infiltration~\citep{oliveira2025high}, and tissue remodeling~\citep{vannan2025spatial}.

One of the foundational analytical tasks in ST is the identification of spatial variable genes (SVGs), genes whose expression varies in a spatially structured manner across the tissue~\citep{yan2025categorization}. SVGs serve as powerful markers for spatial domains, enabling unsupervised discovery of biologically meaningful regions such as cortical layers in the brain or tumor margins in cancer. Identifying SVGs is often the first step in downstream analyses including spatial clustering~\citep{shang2022spatially}, trajectory inference~\citep{shang2025statistical}, spatial deconvolution~\citep{cai2023spanve}, and multi-modal integration~\citep{liang2024multi}. They also play a critical role in guiding tissue-level interpretation and have been used for dimensionality reduction to facilitate computational scalability in large-scale ST datasets~\citep{shang2022spatially}. Given their foundational roles, the accuracy and robustness of SVG detection directly impact the biological interpretations drawn from ST data. 

To meet this demand, a growing number of computational methods have been proposed to identify SVGs~\citep{moran1950notes, svensson2018spatialde,govek2019clustering,sun2020statistical,vandenbon2020clustering, dries2021giotto,andersson2021sepal,zhu2021spark,hao2021somde,hu2021spagcn,kats2021spatialde2,wu2022highly,liu2022scalable,zhang2022identification,jiang2023sinfonia,weber2023nnsvg,wang2023dimension,yuan2024heartsvg,chang2024graph,cai2023spanve,arnol2019modeling,cable2022cell,edsgard2018identification,bae2021discovery,detomaso2021hotspot,miller2021characterizing,li2021bayesian,bintayyash2021non,moehlin2021inferring,yu2022identification,boost-mi,hong2023spatiotemporal,seal2023smash,liang2024prost,yang2024bayesian}, drawing on tools from spatial statistics, graph-based modeling, and kernel-based modeling. While each method brings different strengths and assumptions, their performance can vary widely depending on spatial scale, tissue complexity, and technological platform. Over the years, several benchmarking and summary efforts have emerged to address these challenges. For example, recent comparisons by \cite{chen2024evaluating,chen2025benchmarking} and categorization efforts by \cite{yan2025categorization}, but these evaluations have been limited in scope, often focusing on small datasets, single tissue types, or simulated settings. A systematic, large-scale evaluation of SVG methods across real-world biological and technological diversity is still lacking.

In this study, we present, to the best of our knowledge, the most comprehensive benchmarking of SVG detection methods to date, utilizing STimage-1K4M, a large-scale ST database comprising 662 spatially resolved slides from 18 human tissue types. This dataset includes samples generated with both 10X Visium and Spatial Transcriptomics (ST) platforms~\citep{staahl2016visualization}, and contains pathologist annotations for spatial domains in both cancerous and non-cancerous tissues. Drawing on this unprecedented scale and enrichness, we evaluate 20 representative SVG detection methods~\citep{moran1950notes, svensson2018spatialde,govek2019clustering,sun2020statistical,vandenbon2020clustering, dries2021giotto,andersson2021sepal,zhu2021spark,hao2021somde,hu2021spagcn,kats2021spatialde2,wu2022highly,liu2022scalable,zhang2022identification,jiang2023sinfonia,weber2023nnsvg,wang2023dimension,yuan2024heartsvg,chang2024graph,cai2023spanve} across a broad set of criteria, including biological relevance, robustness regarding tissue type, technology platform, rotation, and computational efficiency. 

Our results reveal method-specific patterns of strength and weakness, and highlight key challenges in the current landscape of SVG detection. Beyond per-method comparisons, the large scale of STimage-1K4M analysis enables meta-level insights into SVG behavior across biological and technical axes, often overlooked by existing relatively small-scale benchmark studies. For instance, we identify conditions under which most methods consistently fail, such as when spatial domains are highly imbalanced or poorly defined, and explore similarities and differences in method outputs across tissues and platforms. Notably, we also uncover meaningful cross-tissue relationships in SVG sets, offering a new lens into shared and distinct spatial gene programs across organs. These findings have important implications for multi-organ studies, spatial disease modeling, and the design of pan-tissue spatial atlases.

To guide readers through the structure of the study, we organize the manuscript into the following sections. In \Cref{sec:SVG_overview}, we provide an overview of SVG detection methods, categorizing them, and summarizing our evaluation criteria, with an emphasis on usability. \Cref{sec:STimage_overview} introduces the STimage-1k4M dataset used for benchmarking, and compares the computational costs across methods. In \Cref{sec:DEgene}, we evaluate each method’s ability to recover domain-specific markers using pathologist-annotated slides from STimage-1K4M. \Cref{sec:tissue_robustness} investigates the robustness of SVG detection within tissue types, both within-study and across-study. We then present a tissue-to-tissue similarity atlas in \Cref{sec:tissue_sim}, revealing insights into the concordance of SVGs across cancer and non-cancer tissue types. In \Cref{sec:rotation}, we assess robustness to spatial coordinate rotation. \Cref{sec:number_of_svgs} analyzes the number of SVGs identified by each method, while \Cref{sec:method_sim} explores the pairwise similarity between methods based on shared SVGs. Finally, \Cref{sec:visiumHD} evaluates computational scalability on high-resolution slides from VisiumHD, offering practical insights into runtime and memory efficiency, followed by a discussion in \Cref{sec:discussion}. Additional experimental details are in \Cref{sec:online} and the Supplement. 

\section{SVG methods overview} \label{sec:SVG_overview}

\begin{figure}[!h]
    \centering
    \includegraphics[width=\textwidth]{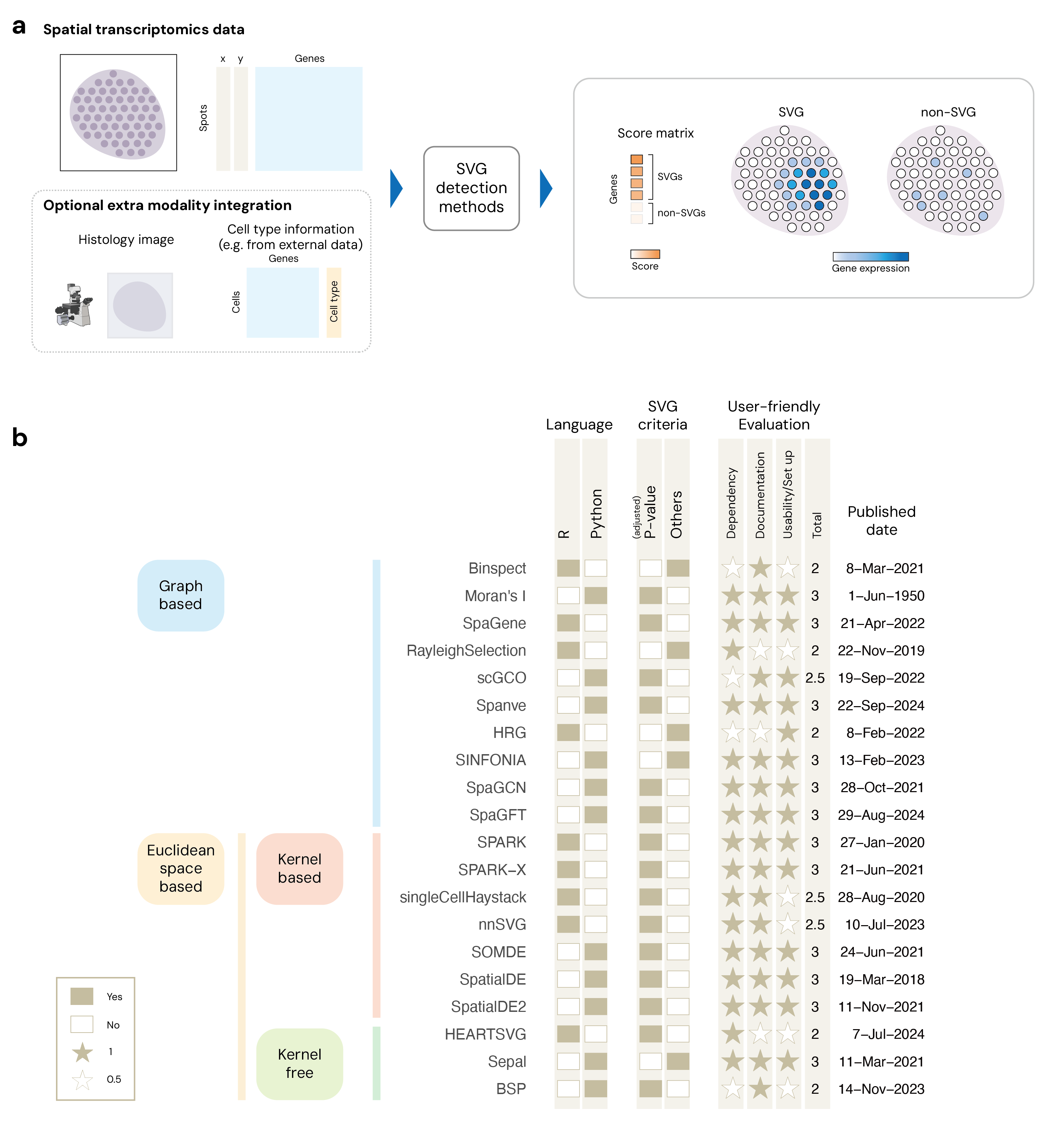}
    \caption{SVG method overview. (a) Conceptual overview of SVG detection methods. Most methods take as input a ST dataset containing gene expression and spatial coordinates, optionally incorporating additional modalities such as histology images or cell-type labels. The core output of each method is a score matrix or ranking, used to classify genes as SVGs or not. (b) Comparative summary of 20 SVG detection methods evaluated in this study.}
    \label{fig:svg_overview}
\end{figure}

A wide range of computational methods have been developed to identify SVGs, which are essential for downstream tasks, as they serve to highlight genes that delineate distinct cellular neighborhoods or microenvironments. Given their central role, understanding the assumptions, modeling strategies, and implementation features of SVG detection methods is essential for both methodological advancement and practical application. SVG detection methods typically take ST data as input, consisting of gene expression values mapped to spot or pixel coordinates within the tissue (\Cref{fig:svg_overview}a). Some methods also allow the incorporation of auxiliary data modalities, such as histology images utilized in~\citep{hu2021spagcn} or cell-type annotations utilized in~\citep{cable2022cell,yu2022identification} derived from cell type deconvolution~\citep{cable2022robust}, to improve sensitivity or interpretability. The core output is a ranked list of genes scored by a metric of spatial variability, indicating whether the genes are spatial variable or not, from which the top SVGs are selected for downstream analysis (\Cref{fig:svg_overview}a).

To capture the methodological diversity of existing approaches, we categorized 20 representative SVG detection methods into the following three broad families based on their core modeling strategies \cite{yan2025categorization}: graph-based, Euclidean space-based, and kernel-free approaches (\Cref{fig:svg_overview}b). Graph-based methods construct spatial graphs over the tissue, where spots are treated as nodes and edges reflect spatial distance or shared tissue features (e.g., gene expression). Examples include classical spatial statistics such as Moran’s I~\citep{moran1950notes} and RayleighSelection~\citep{govek2019clustering}, as well as more recent techniques such as SpaGCN~\citep{hu2021spagcn}, HRG~\citep{wu2022highly}, and scGCO~\citep{zhang2022identification}, which integrate graph convolution, hierarchical modeling, or graphical models to detect spatial structure. Euclidean space-based methods model gene expression as a function of spatial coordinates, often leveraging Gaussian Process regression or other kernel-based techniques to assess spatial smoothness and autocorrelation. Notable examples include SPARK~\citep{sun2020statistical}, SPARK-X~\citep{zhu2021spark}, and nnSVG~\citep{weber2023nnsvg}. Kernel-free methods bypass explicit spatial modeling by employing specially designed statistical or geometric metrics to assess spatial structure. For example, Sepal~\citep{andersson2021sepal} computes a diffusion-based score that quantifies the ``effort'' required to transform a spatially random expression pattern into the observed structured pattern, while BSP~\citep{wang2023dimension} evaluates spatial variability by analyzing the variance of gene expression across multiple spatial resolutions, enabling the detection of both fine-grained and broad spatial structures without relying on explicit spatial kernels. Among the 20 benchmarked methods, 9 are implemented in R and 11 in Python. 15 methods report formal statistical significance via p-values (or adjusted p-value, q-value, adjusted q-value), facilitating downstream interpretation and thresholding, whereas 5 methods rely on custom scoring systems to quantify spatial variability. For instance, RayleighSelection~\citep{govek2019clustering} computes 0- and 1-dimensional combinatorial Laplacian scores derived from topological features of the gene expression graph. Meanwhile, Sepal~\citep{andersson2021sepal} and HRG~\citep{wu2022highly} each introduce their own distinct scoring metrics that do not correspond to conventional p-values but are instead designed to capture specific forms of spatial structure. These scoring schemes are often interpreted empirically, with performance judged by ranking or manual inspection of top-scoring genes, rather than strict statistical significance. This lack of formal thresholding may limit their interpretability and complicate downstream comparative analyses, especially in large-scale or automated workflows.

In addition to methodological differences, SVG detection methods also differ substantially in software implementation quality and user accessibility. Especially in the context of ST, which is a field attracting increasingly more experimental biologists and clinicians, computational tools must be accessible to users with diverse technical backgrounds, including those without formal training in programming. To systematically assess usability, we evaluated all 20 methods along three key criteria (see \Cref{online_methods:user_friendly} for details): 1. documentation of the dependency list, 2. documentation quality of the software, and 3. ease of installation and set up. Each method was scored on a 0–1 scale (0 = lacking, 0.5 = partial, 1 = fully satisfactory) in each category, yielding a maximum composite score of 3. There are 12 methods achieving full score, including Moran's I, SpaGene, Spanve, SINFONIA, SpaGCN, SpaGFT, SPARK, SPARK-X, SOMDE, SpatialDE, SpatialDE2, and Sepal.

\section{Overview: Evaluate SVG methods with STimage-1K4M} \label{sec:STimage_overview}

\begin{figure}[H]
    \centering
    \includegraphics[width=\textwidth]{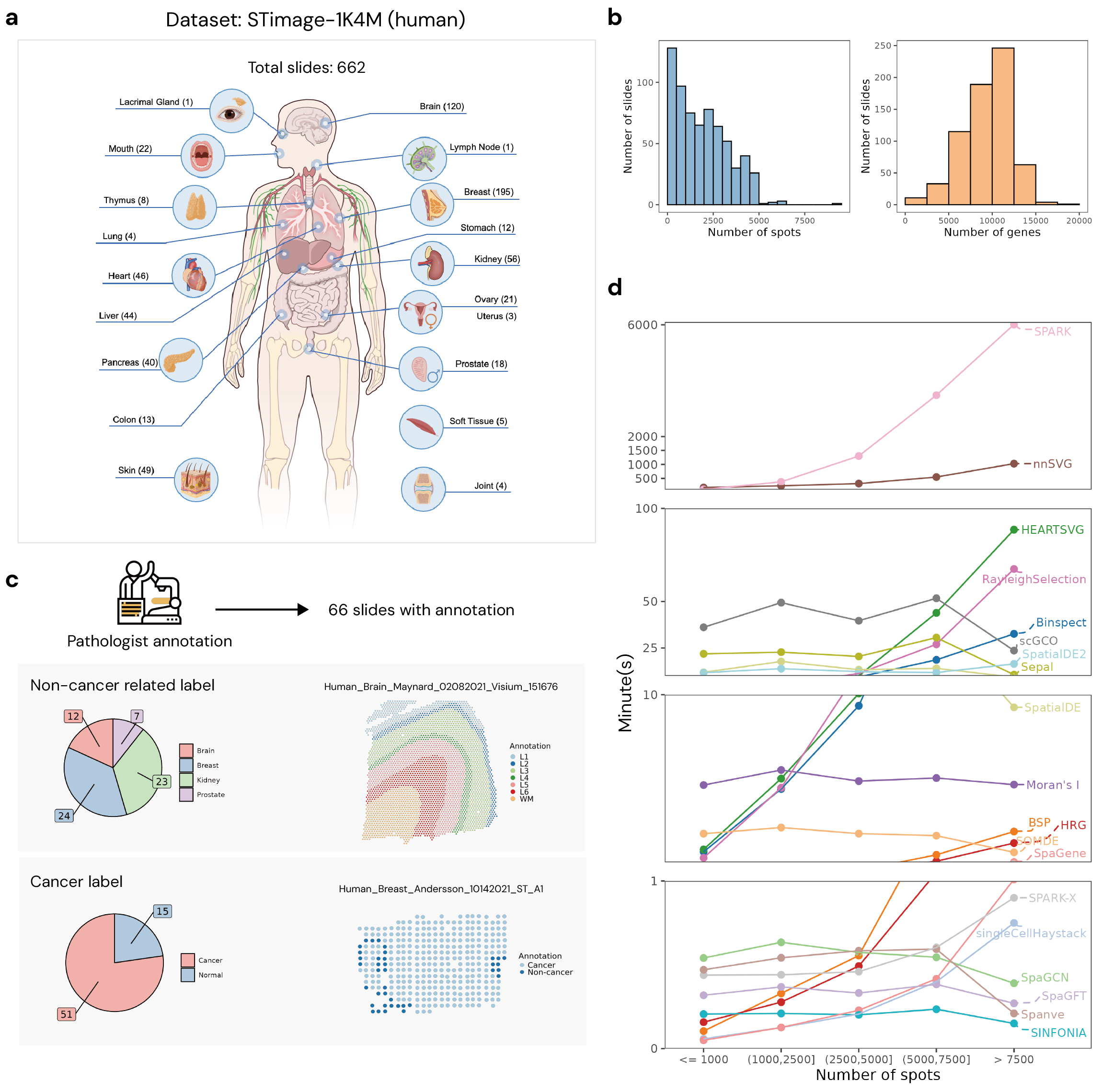}
    \caption{Overview of the STimage-1K4M dataset and computational time evaluation. (a) Summary of the human portion of the STimage-1K4M dataset. (b) Distribution of number of spots (left) and number of genes (right) per slide. (c) Subset of 66 slides were annotated by pathologists to provide ground-truth labels. (d) Benchmarking of computational cost for 20 SVG detection methods.}
    \label{fig:stimage_time}
\end{figure}

We evaluated 20 SVG detection methods using the extensive STimage-1K4M dataset, comprising 662 ST slides from human tissues. To our knowledge, this represents the largest systematic evaluation of SVG methods to date. Our evaluation spans 18 distinct tissue types, providing a uniquely comprehensive landscape to examine methodological performance across varied biological contexts. 
The two most represented tissues are breast ($n=196$) and brain ($n=120$). In addition to tissue-type annotations, we manually classified each slide into cancer and non-cancer categories, resulting in 399 cancer slides and 263 non-cancer slides~(\Cref{fig:stimage_time}a).

A particularly valuable subset of the STimage-1K4M dataset includes 66 slides with pathologist annotations. These include both non-cancer-specific labels - such as the seven-layer cortical structure (L1–L6) and white matter (WM) in human brain slides from \cite{Maynard2021-wa} - and cancer-related region annotations (e.g., from \cite{Andersson2021-nr}). Of these 66 slides, 51 are from cancer tissues. We further curated these annotations to derive a binary cancer vs. non-cancer label at the spot level for benchmarking analysis~(\Cref{fig:stimage_time}c).

The dataset includes slides generated from two primary ST technologies: 135 slides from the original Spatial Transcriptomics platform (grid layout) and 527 slides from the 10x Genomics Visium platform (hexagonal layout)~\cite{staahl2016visualization}. The vast majority of slides (655, or 98.9\%) contain fewer than 5,000 spots after preprocessing. The average number of genes per slide after preprocessing is 9,440~(\Cref{fig:stimage_time}b).

We first assessed computational efficiency by evaluating runtime across all slides with gene number greater than 10,000 and smaller than 20,000 for easier comparison (full computational cost table in Online Methods data availability section). Most methods exhibit non-linear increases in computation time relative to the number of spots~(\Cref{fig:stimage_time}d). Specifically, 9 methods completed SVG detection computations for slides containing $\leq$1,000 spots within 1 minute, while 14 methods accomplished the task within 10 minutes. However, for larger slides with approximately 10,000 spots (specifically 9,080 spots), most methods experienced significantly increased computational demands. Nevertheless, 6 methods - SINFONIA, Spanve, SpaGFT, SpaGCN, singlecellHaystack, and SPARK-X - maintained exceptional computational efficiency with computation time controlled under 1 minute even with  ~10,000 spots. We note that methods like nnSVG, which support multi-core parallelization, could see substantially reduced runtimes under multi-threaded settings, though this was not evaluated in our single-core experiments for consistency across methods. Additionally, during our implementation phase, we initially planned to include 35 methods in this benchmark. However, we ultimately excluded several due to practical limitations. Some methods failed to complete on large slides due to excessive memory (we filtered out certain methods using a small slide GSE144239\_GSM4565824 contained 648 spots and 10,923 genes) or runtime requirements, or raised errors during execution. Others were skipped due to incompatible input requirements. For example, SPADE requires raw data from the Space Ranger pipeline, which was not always available in ST datasets. Similarly, C-SIDE and CTSV require cell type information, which were unavailable for many slides and would introduce unfair comparisons to other unsupervised methods. See Supplementary Table 1 for details. 

\section{Evaluate SVG detection methods through DE gene analysis} \label{sec:DEgene}

\begin{figure}[!h]
    \centering
    \includegraphics[width=\textwidth]{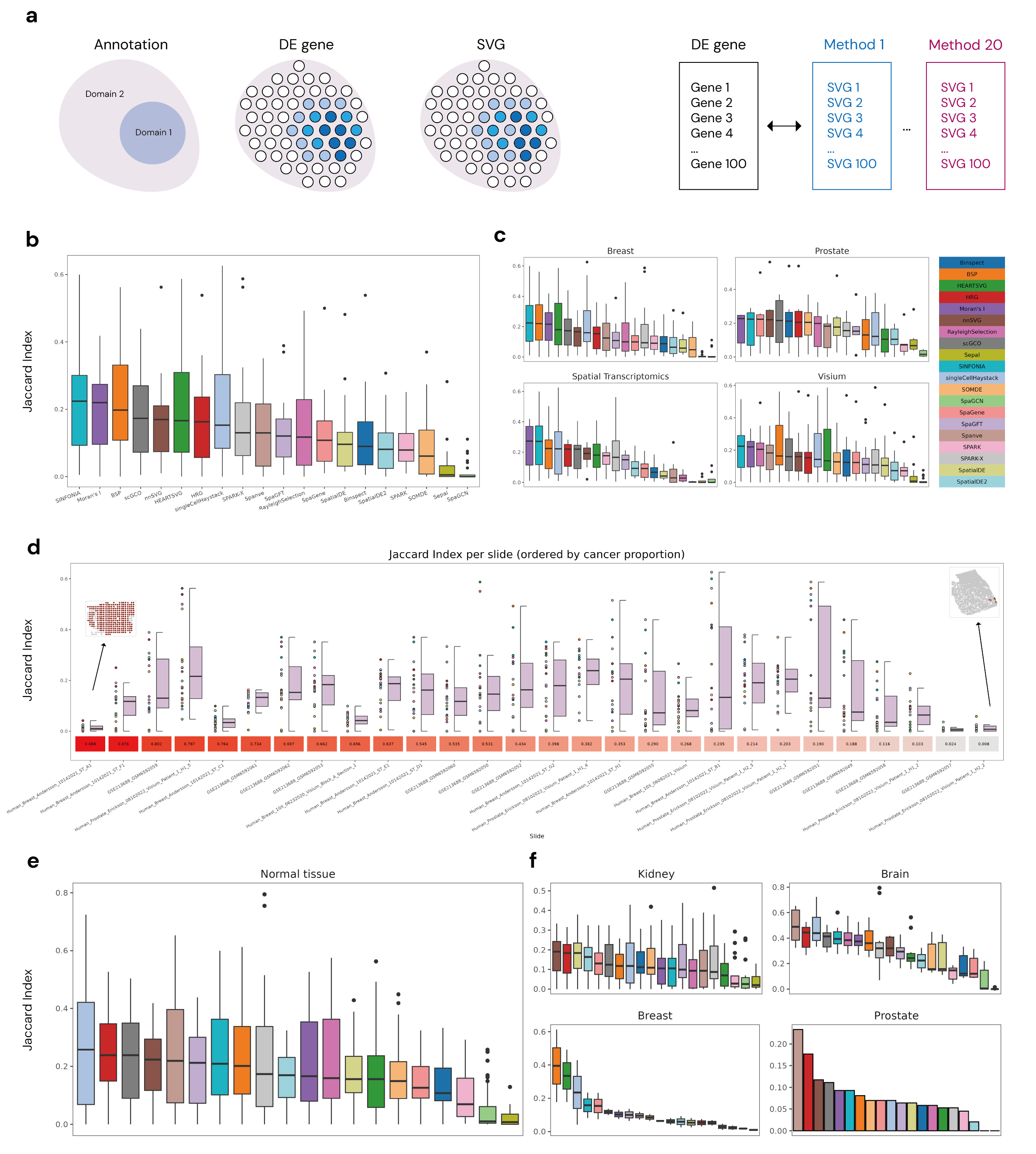}
    \caption{Benchmarking SVG detection methods using domain-specific DE genes. (a) Illustration of the evaluation framework. Cancer slides: (b) Overall performance of 20 methods across 28 cancer-annotated slides, sorted by median Jaccard Index. (c) Method performance stratified by tissue type (breast and prostate) and technological platform (Spatial Transcriptomics and Visium). (d) Slide-level performance sorted by cancer proportion (bottom). Non-cancer slides: (e) Method performance across 32 non-cancer-annotated tissues. (f) Method performance stratified by tissue type (kidney, brain, breast, and prostate).}
    \label{fig:degene}
\end{figure}

One of the key goals of ST is to identify genes that delineate functionally distinct spatial domains, such as cortical layers in the brain or tumor versus normal regions in cancer. However, such domain structures may not always be known a priori. In this case, SVG methods are particularly valuable, as they enable unsupervised identification of genes with biologically meaningful spatial patterns. By extracting SVGs, these methods can reveal underlying spatial domains and guide downstream interpretation in tissue organization and disease pathology. To assess the biological relevance and practical utility of SVG methods, we evaluated their ability to detect domain-specific genes that are differentially expressed across anatomically or pathologically defined tissue regions (\Cref{fig:degene}a). 

We utilized 60 slides with pathologist-provided annotations to evaluate the performance of SVG detection methods in identifying domain-specific genes (slides with only one type of region were excluded). These slides include both cancer and non-cancer tissues, enabling us to design two complementary evaluations: (i) detection of cancer-associated genes based on annotated tumor boundaries, and (ii) identification of genes corresponding to anatomical tissue organization in healthy samples.

For the cancer-focused analysis, we manually curated the annotation labels to derive a binary classification of cancer versus non-cancer regions. Using these labels, we performed differential expression (DE) analysis to identify genes significantly enriched in each domain. The top 100 DE genes ranked by p-values in ascending order were used as ``ground-truth'' domain-specific reference genes. For each SVG detection method, we compared its top 100 identified SVGs against these reference genes and computed the Jaccard Index~\citep{jaccard1901etude} as the overlap rate, providing a measure of how well each method recovers biologically relevant cancer-associated markers. Among all methods (\Cref{fig:degene}b), SINFONIA demonstrated the highest performance, with highest median Jaccard Index over 0.2, indicating strong concordance with domain-specific DE patterns. Moran's I and BSP ranked second and third, respectively, both showing robust performance across multiple tissue contexts. A second tier of methods, including scGCO, nnSVG, HEARTSVG, HRG, and singleCellHaystack, achieved moderate overlap rates, consistently recovering approximately 30\% of the DE-derived cancer markers. 
It is important to note that the Jaccard Index, while a useful measure of overlap, is sensitive not only to gene set agreement but also to the number of non-overlapping elements. Some methods return test statistics with ties will have more than 100 SVGs in the top 100 SVGs, which can penalize their Jaccard scores even if the methods recover many DE genes. We discuss this behavior in more detail in \Cref{sec:number_of_svgs}.

The overall performance across all methods on cancer slides remained largely consistent across technological platforms (Spatial Transcriptomics vs. Visium), as shown in the stratified boxplots (\Cref{fig:degene}c), with SINFONIA and Moran's I as the top 2 methods. Nonetheless, we observed performance variability across different tissue types. Among the 28 cancer-annotated slides, there are 22 breast and 6 prostate slides. We note here that although 51 slides in our dataset originated from cancer patients, only those with both cancer and non-cancer annotations were included in this evaluation. Slides with other type of annotations, such as those lacking a cancer region or labeled only with tumor-associated tertiary lymphoid structures (TLS)-related labels, were excluded. All prostate slides were generated using the 10X Visium platform, whereas the breast slides were drawn from both platforms (8 Spatial Transcriptomics and 14 Visium). Notably, most methods demonstrated similar performance across both platforms. For instance, Moran's I (1$^{st}$ on Spatial Transciptomics, 2$^{nd}$ on Visium), SINFONIA (2$^{nd}$ on Spatial Transciptomics, 1$^{st}$ on Visium), and BSP (3$^{rd}$ on Spatial Transciptomics, 5$^{th}$ on Visium) have similar high performance on both slides. However, some methods exhibited clear platform preferences. For example, RayleighSelection performed markedly better on Visium (17$^{th}$ on Spatial Transciptomics, 3$^{rd}$ on Visium), and singleCellHaystack has the opposite trend, with better performance on Spatial Transcriptomics (4$^{th}$ on Spatial Transciptomics, 9$^{th}$ on Visium). This divergence highlights that technological platform is an important factor influencing SVG method performance for certain methods, and underscores the need to consider platform-specific behaviors when benchmarking and applying ST tools.

During our evaluation of cancer-related gene recovery, we identified several slides on which most SVG detection methods consistently failed to recover domain-specific markers (\Cref{fig:degene}d, \Cref{supp_fig:DE_cancer_median}). These poorly performing slides are primarily found at the extremes of cancer proportion (left and right side of the figure), where the cancer proportion is either extremely high or extremely low (indicated by the red gradient). For instance, in slide A1 from \cite{Andersson2021-nr} (the leftmost slide), where the cancer proportion is 88.8\%, most methods achieved Jaccard Index below 0.03. Similarly, slide H2 from patient 1 in \cite{Erickson2022-ia} (the rightmost slide), with only 0.8\% cancer coverage, also resulted in Jaccard Index below 0.03 across most methods. This trend suggests that extreme imbalance in cancer vs. non-cancer spatial domains impairs the ability of unsupervised SVG methods to detect meaningful markers. The reason behind this is that, SVG methods operate in an unsupervised manner, relying on spatial variation to infer gene relevance. When one spatial domain is disproportionately small or large, key signals may be diluted or localized to only a few spots, making them harder to distinguish from noise. For example, genes over-expressed in a small tumor region may appear as weak or noisy spatial signals, while highly homogeneous slides lack sufficient contrast to drive strong SVG detection. These findings highlight a fundamental limitation of current SVG approaches: their sensitivity to domain imbalance. Consequently, spatial domain proportion needs to be carefully considered when inferring SVG results, especially in datasets with heterogeneous or skewed tissue composition.


We also performed the DE-based evaluation on non-cancer tissue slides, which were annotated by pathologists but lacked cancer-related labels (\Cref{fig:degene}e,f). This subset included 32 slides from four tissue types: 12 brain, 17 kidney, 2 breast, and 1 prostate slide. As with the cancer tissue analysis, we compared each method’s top 100 SVGs to the top 100 domain-specific DE genes derived from the annotated tissue labels. The performance of SVG methods varied notably across tissues. Overall, singleCellHaystack achieved the highest median Jaccard Index, followed by HRG and scGCO. Method performance also differed across tissue types. For instance, nnSVG performed well on kidney and prostate but underperformed on brain and breast tissues. In contrast, HRG performed consistently well except on the breast slides, while singleCellHaystack excelled on brain and breast but underperformed on kidney and prostate.

Across tissue types, brain slides from \cite{Maynard2021-wa} yielded the highest Jaccard Index values, while kidney slides had the lowest. In addition, for same tissue type, the Jaccard Index is lower on the non-cancer slides compared to the cancer slides (\Cref{supp_fig:DE_comparison_cancer_noncancer}). This discrepancy may stem from differences in domain complexity. In cancer slides, DE analysis was based on a binary classification (cancer vs. non-cancer), whereas normal tissue slides often had more granular annotations. For example, the breast slide (GSE213688\_GSM6592054) includes six region types: adipose tissue, fibrosis, lymphocytes, normal epithelium, peripheral nerve, and vascular. The increased number of spatial domains likely introduces greater complexity and makes SVG recovery more challenging. Moreover, tissues with regular spatial architecture, such as the layered structure of the brain cortex, tended to support higher detection performance, highlighting the influence of anatomical organization on SVG method performance.

\section{SVG methods' robustness within tissue type} \label{sec:tissue_robustness}

\begin{figure}[!h]
    \centering
    \includegraphics[width=\textwidth]{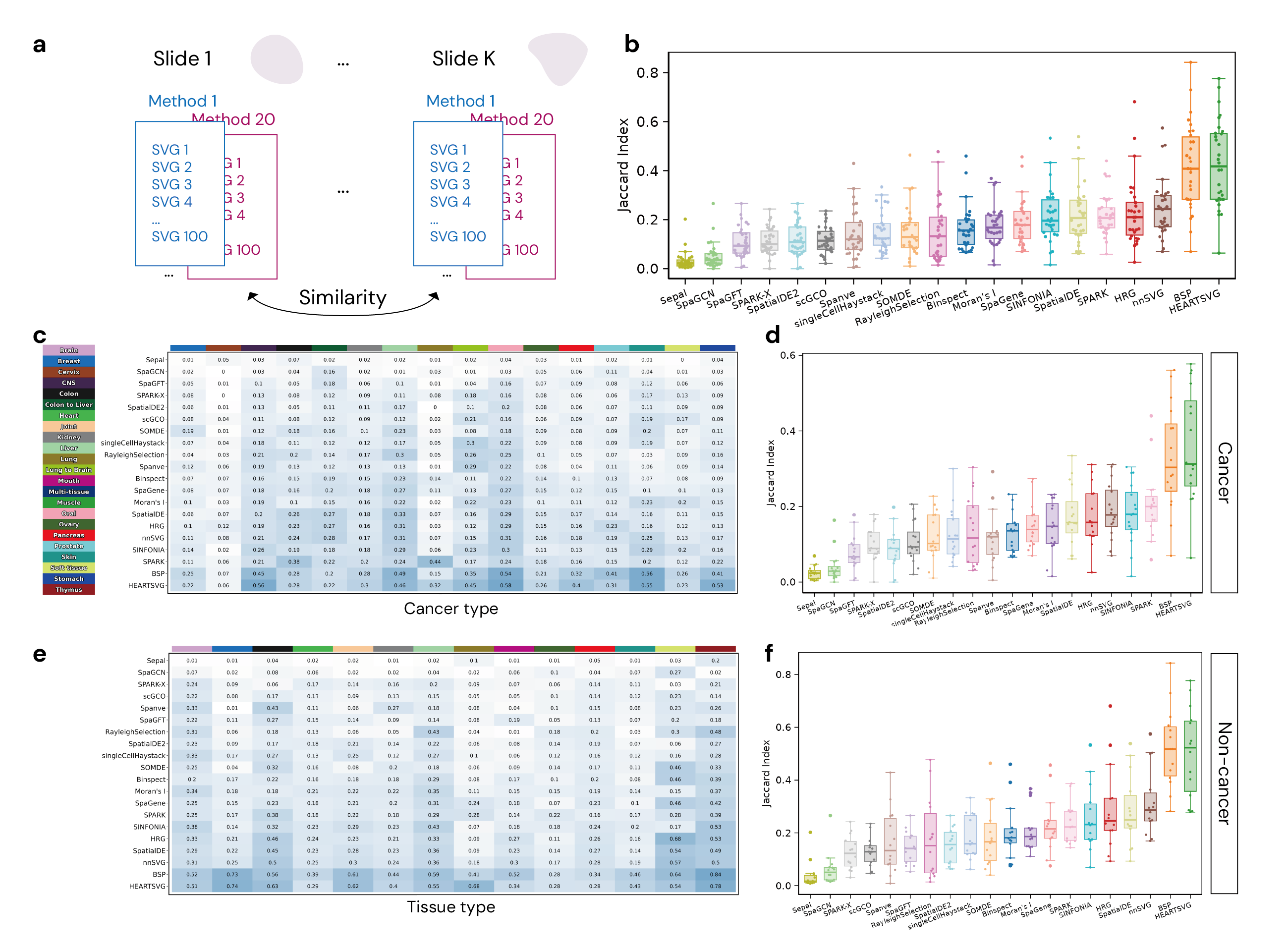}
    \caption{SVG detection robustness across tissues. (a) Illustration of the comparison of SVGs across multiple slides and methods. (b) Jaccard Index of within-tissue robustness for each method, ranked by median Jaccard Index. Each dot represents a tissue type. (c) Average pairwise Jaccard index between slides of the same cancer tissue type for each method. (d) Jaccard Index of within-cancer-tissue robustness for each method, ranked by median Jaccard Index. Each dot represents a cancer type. (e) Average pairwise Jaccard index between slides of the same non-cancer tissue type. (f) Jaccard Index of within-normal-tissue robustness for each method, ranked by median Jaccard Index. Each dot represents a normal tissue type.}
    \label{fig:tissue_robustness}
\end{figure}

In this section, we evaluate the robustness of SVG methods by examining the consistency of their identified SVG sets across different slides within the same tissue type. Assessing cross-slide consistency is critical for understanding whether a method can reliably capture underlying biological patterns that are reproducible across individuals or experimental replicates. Importantly, this type of robustness evaluation is only feasible with large-scale datasets like STimage-1K4M, which includes a wide range of tissues and a substantial number of slides per tissue type. To perform this analysis, we used the full collection of 662 slides and stratified our evaluation by cancer status to account for their distinct biological and spatial characteristics. To quantify cross-slide consistency, we computed the pairwise Jaccard Index between the top 100 SVGs identified by each method across slides within the same tissue type (\Cref{fig:tissue_robustness}a). We further compared three conditions: (1) slides from the same tissue type across all studies, (2) within-study: slides within the same study series, and (3) across-study: slides from different study series but the same tissue type. Comparing across-study performance with within study performance allowed us to distinguish method robustness in the presence of biological variability from robustness under potential batch or study-specific effects.

Across all methods, we observed substantial heterogeneity in robustness (\Cref{fig:tissue_robustness}b). Methods such as HEARTSVG, BSP, and nnSVG consistently achieved high reproducibility, while others like Sepal, SpaGCN, and SpaGFT showed markedly lower cross-slide consistency. Notably, these differences were not only method-specific but also tissue- and context-dependent. 

Focusing on cancer samples (\Cref{fig:tissue_robustness}c,d), we found that the relative performance ranking of methods largely persisted. HEARTSVG and BSP again excelled in consistency across most cancer types, with SPARK and SINFONIA following closely behind. Importantly, within-study comparisons generally exhibited higher consistency than comparisons across different studies of the same cancer type, highlighting the impact of batch effects and study-specific factors on SVG detection. Interestingly, methods that performed well overall tended to show even better performance in the within study setting. This was especially evident for HEARTSVG and BSP, suggesting that these methods are highly effective at capturing spatial structure when the data quality or structure is favorable, i.e., when samples are generated under consistent protocols~(\Cref{supp_fig:robustness_cancer_study_heatmap,supp_fig:robustness_cancer_by_tissue_method}). 

In non-cancer tissues (\Cref{fig:tissue_robustness}e,f), the same set of top-performing methods, HEARTSVG, BSP, nnSVG, SpatialDE, HRG, SINFONIA, and SPARK, remained consistently robust. Moreover, consistent with cancer tissues, we again observed higher robustness within study series than across studies (\Cref{supp_fig:robustness_noncancer_by_tissue_method,supp_fig:robustness_noncancer_study_heatmap}). Among these high-performing methods, the performance advantage was particularly pronounced in the within-study setting, significantly outperforming other methods (\Cref{supp_fig:robustness_noncancer_by_tissue_method}a). However, in the more challenging cross-study setting, this performance gap narrowed, reinforcing the observation that even the most robust tools remain sensitive to study-specific variations such as batch effects, tissue processing, and experimental protocols.

In addition to evaluating method-level performance, we also aimed to assess tissue-level robustness and heterogeneity. From this perspective, we identified distinct patterns across tissue types. For cancer slides, tissues such as liver, oral, and central nervous system (CNS) exhibited the highest cross-slide consistency across methods, whereas breast, cervix, and soft tissue showed notably lower reproducibility (\Cref{supp_fig:robustness_cancer_by_tissue_method}b). Among non-cancer slides (\Cref{supp_fig:robustness_noncancer_by_tissue_method}b), tissues like thymus, muscle, and soft tissue demonstrated high robustness, whereas heart, skin, and pancreas had relatively poor consistency. Notably, some tissues exhibited divergent robustness patterns depending on disease status. For example, soft tissue showed high cross-slide robustness in cancer samples but low robustness in normal samples. This suggests that certain biological contexts or pathological transformations may accentuate or obscure spatial signals, even within the same organ system (\Cref{supp_fig:robustness_cancer_noncancer_similarity_by_tissue}).

Finally, when comparing overall robustness across cancer and non-cancer tissue types, we found that non-cancer tissues generally exhibited higher cross-slide consistency across all methods (\Cref{supp_fig:robustness_cancer_noncancer_similarity}). This may reflect the fact that spatial structure in healthy tissues, particularly those with well-defined histological architecture, is often more reproducible and less noisy than in heterogeneous tumor environments. Together, these results emphasize the need to carefully consider tissue type, disease context, and study design when interpreting SVG results, and highlight the advantages of benchmarking methods in large and diverse reference datasets.

\section{Tissue-tissue similarity atlas} \label{sec:tissue_sim}



\begin{figure}[!h]
    \centering
    \includegraphics[width=\textwidth]{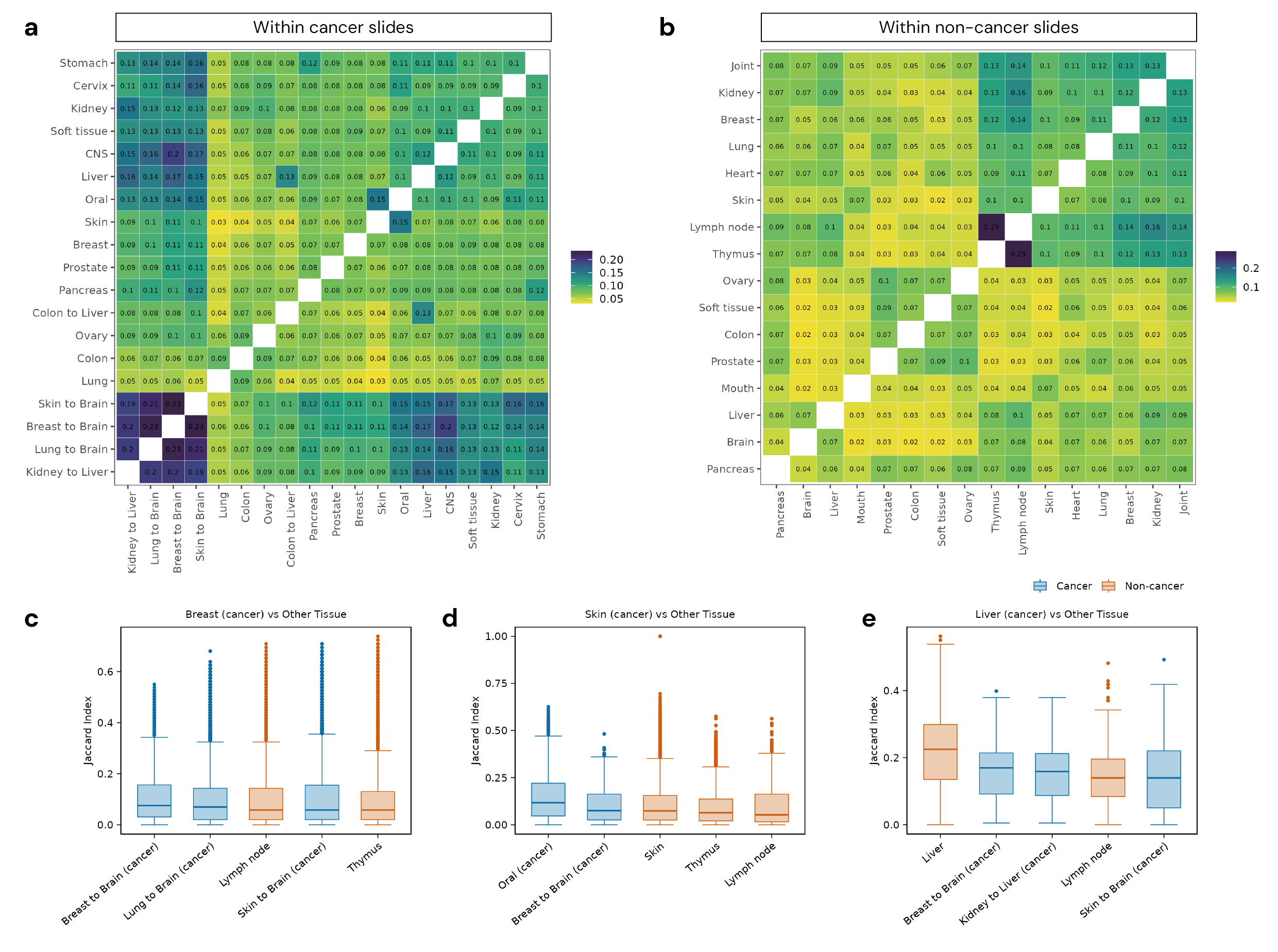}
    \caption{Cross-tissue similarity of SVG sets across (a) cancer slides and (b) non-cancer slides. Boxplots of pairwise Jaccard Index between (c) breast cancer, (d) skin cancer, (d) liver cancer and other tissues.}
    \label{fig:tissue_tissue_sim}
\end{figure}

With this massive benchmarking effort, we are uniquely positioned to construct a large-scale atlas of SVG sets for diverse tissue types. These tissue-specific SVG profiles offer valuable biological insights into both normal tissue architecture and disease-specific spatial organization. To systematically assess the similarity of spatial expression programs across tissues, we computed pairwise Jaccard indices between the top 100 SVGs for each tissue type across different methods (\Cref{fig:tissue_tissue_sim}).

Within cancer tissues (\Cref{fig:tissue_tissue_sim}a), we observed notable SVG similarity between tissue pairs involved in known metastatic pathways. For instance, breast-to-brain metastasis slides exhibited strong overlap with other tissue-to-brain metastases, including lung-to-brain and skin-to-brain cases. Likewise, kidney-to-liver metastasis slides showed high similarity to both liver cancer and several other metastases, such as lung-to-brain and skin-to-brain. These findings suggest that the SVG profiles of metastatic cancers capture both tissue-of-origin signals and microenvironmental adaptation patterns, where the SVG sets in these cases likely reflect a mixture of metastatic driver genes and target-site-specific spatial programs. In contrast, non-cancer tissues demonstrated lower and more homogeneous SVG similarity overall (\Cref{fig:tissue_tissue_sim}b). The strongest overlap was observed between thymus and lymph node, two immune-related tissues that share similar functional roles. This strong correlation underscores the ability of SVG sets to capture conserved immunological architecture.

We further examined the relationship between cancer tissues and their corresponding normal counterparts to assess whether cancers retain tissue-of-origin identity or adopt cancer-specific programs (\Cref{supp_fig:tissue_sim_tissue_tumor}). Notably, there was no universal pattern. Some cancer types were more similar to their normal tissue counterparts, while others showed higher similarity to other cancer types. For example, breast cancer showed the highest SVG similarity to breast-to-brain metastasis, rather than to normal breast tissue (\Cref{fig:tissue_tissue_sim}c, \Cref{supp_fig:tissue_sim_Breast__cancer__vsother}). Skin cancer exhibited high similarity to oral cancer, with SVG sets from normal skin slides ranking second, indicating a partial reservation of tissue identity (\Cref{fig:tissue_tissue_sim}d, \Cref{supp_fig:tissue_sim_Skin__cancer__vsother}). Liver cancer, on the other hand, shared the highest similarity with normal liver, followed by breast-to-brain cancer, suggesting a stronger tissue-specific signature (\Cref{fig:tissue_tissue_sim}e, \Cref{supp_fig:tissue_sim_Liver__cancer__vsother}). Across multiple cancer types, we also observed high similarity with thymus and lymph node SVGs, potentially reflecting shared patterns of immune infiltration in tumors.

\section{Robustness to slide rotation} \label{sec:rotation}

\begin{figure}[H]
    \centering
    \includegraphics[width=\textwidth]{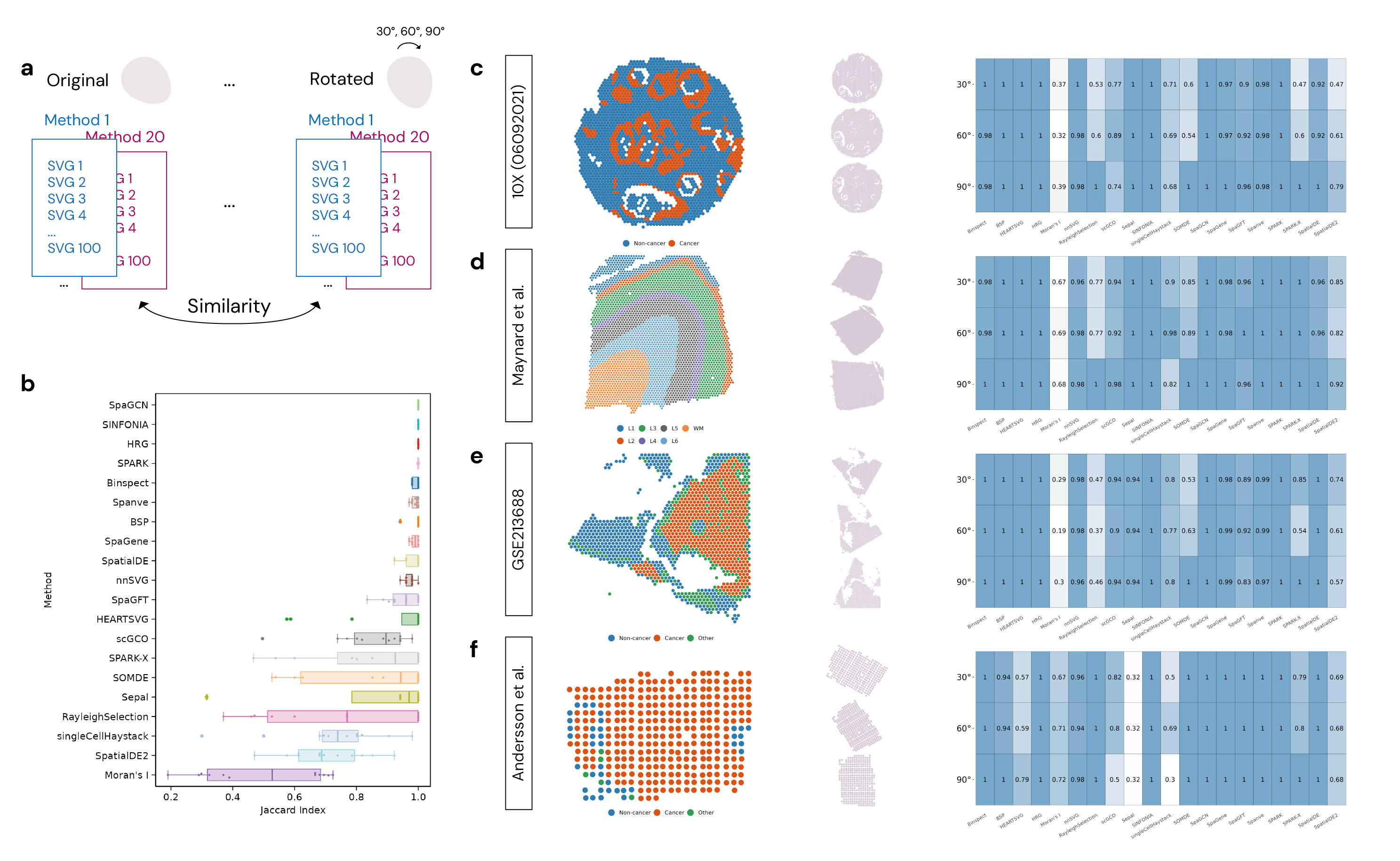}
    \caption{Rotation robustness analysis of SVG detection methods. (a) Rotation robustness evaluation framework. (b) Distribution of Jaccard Index scores across all methods, ranked by mean Jaccard Index. Jaccard Index of each method for each rotation on (c) Human\_Breast\_10X\_06092021\_Visium, (d) Human\_Brain\_Maynard\_02082021\_Visium\_151676, (e) GSE213688\_GSM6592060, (f) Human\_Breast\_Andersson\_10142021\_ST\_A1.}
    \label{fig:rotation}
\end{figure}

To evaluate the robustness of SVG detection methods against meaningless changes in spatial coordinates, we conducted a rotation-based consistency analysis. Specifically, we applied geometric transformations (rotations of 30$^\circ$, 60$^\circ$, and 90$^\circ$) to the spatial coordinates of four representative annotated slides. Each method was then re-applied to the rotated coordinates, and the resulting top 100 SVGs were compared to those identified from the original (unrotated) slide. We used the Jaccard Index to quantify consistency between original and rotated outputs for each method and rotation angle (\Cref{fig:rotation}a).

Our results revealed that several methods demonstrated high robustness across all rotation angles. SpaGCN, SINFONIS, HRG, and SPARK exhibited highest mean Jaccard Index equal/close to 1.0 (\Cref{fig:rotation}b), indicating that their SVG outputs remained stable under spatial transformation. These methods appear invariant to coordinate rotation, a desirable property in real-world applications where spatial orientation may vary across experiments. Other methods such as Binspect, Spanve, BSP, and SpaGene also performed relatively well, though with slightly increased variability. In contrast, methods including RayleighSelection, singleCellHaystack, SpatialDE2, and Moran's I showed low robustness, with median Jaccard Index below 0.8, suggesting that their SVG results are sensitive to spatial orientation. In addition, the degree of rotation also influenced method performance. Rotation of 90$^\circ$ generally yielded the highest robustness across methods (mean Jaccard Index = 0.911), while rotations of 60$^\circ$ and 30$^\circ$ produced slightly worse performance (mean Jaccard Index = 0.879 for 60$^\circ$ and 0.877 for 30$^\circ$). This is likely because a 90$^\circ$ rotation is a simple axis swap between x and y axes, which does not significantly alter pairwise distances or neighborhood structures, thereby preserving spatial relationships more effectively than arbitrary-angle rotations. However, methods including Moran's I, HEARTSVG, Sepal, singleCellHaystack, SpaGFT, RayleighSelection, and SpatialDE2 could not produce robust SVG detection under 90$^\circ$ rotation on certain slides.

Regarding different spatial patterns and tissue layouts, in this analysis, we selected 4 structurally distinct slides (\Cref{fig:rotation}c–f), including three breast cancer tissues and one brain tissue. Slide 1, a round breast cancer section from 10X Genomics (\Cref{fig:rotation}c), contains dispersed cancer regions across the tissue. Slide 2 is a brain tissue with clearly layered spatial domains, which remains visually coherent upon rotation (\Cref{fig:rotation}d). Slide 3 is a sparsely populated breast cancer tissue with large empty regions (\Cref{fig:rotation}e), while Slide 4 is a dense breast cancer sample with nearly the entire slide labeled as tumor, which serves as a particularly difficult challenge for SVG detection (\Cref{fig:rotation}f).

Performance varied across these tissue types and landscapes (\Cref{fig:rotation}c-f, \Cref{supp_fig:rotation_by_tissue,supp_fig:rotation_by_method}). For example, SpatialDE2 showed relatively better robustness on Slide 2 compared to other slides. On Slide 2, due to the clear layered structure, most methods achieved their highest consistency. In contrast, Slide 4 posed significant challenges: the overwhelming presence of cancer regions reduced spatial contrast, which makes both detection (as stated in DE gene comparison section) and robustness tasks harder for all methods. Methods such as singelCellHaystack and Sepal were particularly sensitive, showing large drops in Jaccard similarity across all angles on this slide.


\section{Number of SVGs} \label{sec:number_of_svgs}

In addition to evaluating biological relevance and robustness, we examined the number of SVGs identified by each method to better understand their thresholding behavior and practical usability. While the “ground truth” number of spatial variable genes is unknown in real datasets, understanding the relative conservativeness or liberality of different methods provides important context for users applying these tools in practice.

\begin{figure}[H]
    \centering
    \includegraphics[width=\textwidth]{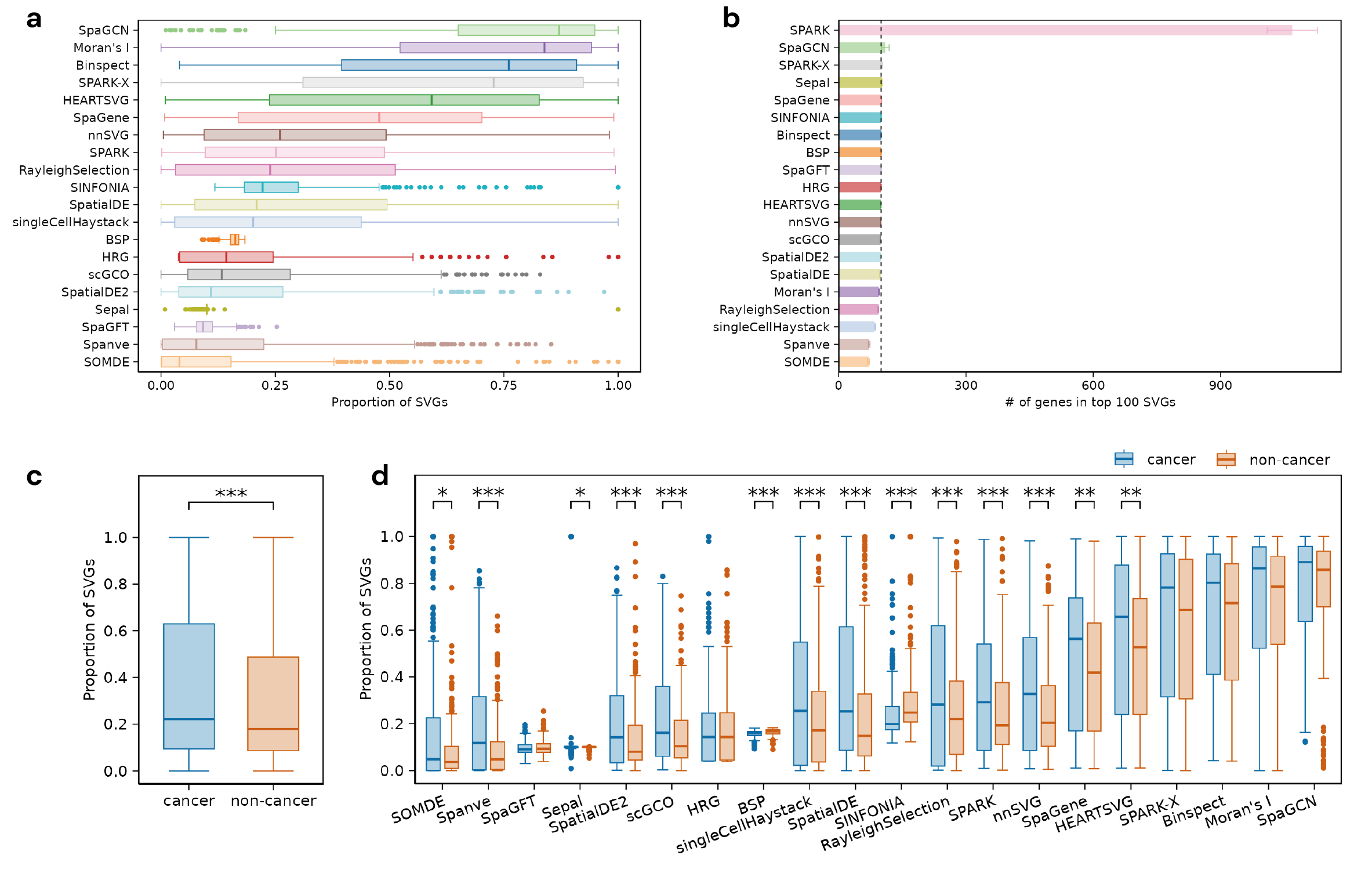}
    \caption{SVG number across slides. (a) Proportion of SVGs detected by each method across all slides. (b) Number of genes in the top 100 SVGs. (c) Comparison of overall SVG proportions between cancer and non-cancer slides. (d) SVG proportions stratified by method and cancer status.}
    \label{fig:svg_number}
\end{figure}

Across the 662 STimage-1K4M slides, we observed substantial variation in the proportion of SVGs reported by each method (\Cref{fig:svg_number}a). Some methods, such as SpaGCN, Moran’s I, Binspect, SPARK-X, and HEARTSVG, consistently identified a large fraction of genes as spatially variable, often exceeding 50\% of the number of genes on a given slide. These methods are relatively liberal in their SVG calling, which may be beneficial in exploratory analyses but could also introduce false positives. In contrast, methods such as SOMDE, Spanve, SpaGFT, and Sepal were markedly more conservative, with median SVG proportions below 10\%. It is worth noting that some methods provide built-in mechanisms for further SVG refinement. For instance, SpaGCN includes a secondary domain-specific thresholding pipeline designed to identify SVGs that are spatially enriched within a user-defined target domain. However, we did not apply this refinement step in our benchmarking, as our study design did not specify a target domain.

When stratifying results by cancer status, we found that most methods tended to call more SVGs in cancer slides than in non-cancer slides (\Cref{fig:svg_number}c). This likely reflects the increased spatial heterogeneity and domain boundaries present in tumor tissues. However, the difference in SVG quantity was not uniform across methods: several methods including SpaGFT, Sepal, and HRG showed only modest differences, while others, such as Spanve and SpatialDE2, reported significantly more SVGs in cancer slides (\Cref{fig:svg_number}d). These results emphasize that SVG quantity is not only method-specific but also tissue- and context-dependent.

We also examined the issue of tied scores among top-ranked genes, which has important implications for practical downstream usage, particularly in scenarios where biologists select a fixed number of top SVGs (e.g., the top 100 genes) for visualization, functional annotation, or experimental validation. To do so, we counted the number of genes that were tied (i.e., assigned identical p-values or scores) within the top 100 SVGs returned by each method. Notably, SPARK exhibited an large tie effect, returning over 900 genes across many slides, far exceeding the expected top-100 cutoff (\Cref{fig:svg_number}b). This behavior arises from the internal file I/O mechanism of SPARK, which converts extremely small p-values (e.g., <1e-16) are truncated to a constant 5.55e-17, resulting in many genes sharing the same p-value. Similar tie-related issues were observed in SpaGCN, which reports the adjusted p-values and sometimes returns a large number of genes with identical adjusted p-values at the significance threshold. One potential way to mitigate this issue is to apply target-domain-based filtering as originally suggested in the paper. However, we did not adopt this setting in this benchmark, as we did not have specify domain of interest in our setting.

\section{Similarity of SVG methods based on their SVGs identified } \label{sec:method_sim}

\begin{figure}[!h]
    \centering
    \includegraphics[width=\textwidth]{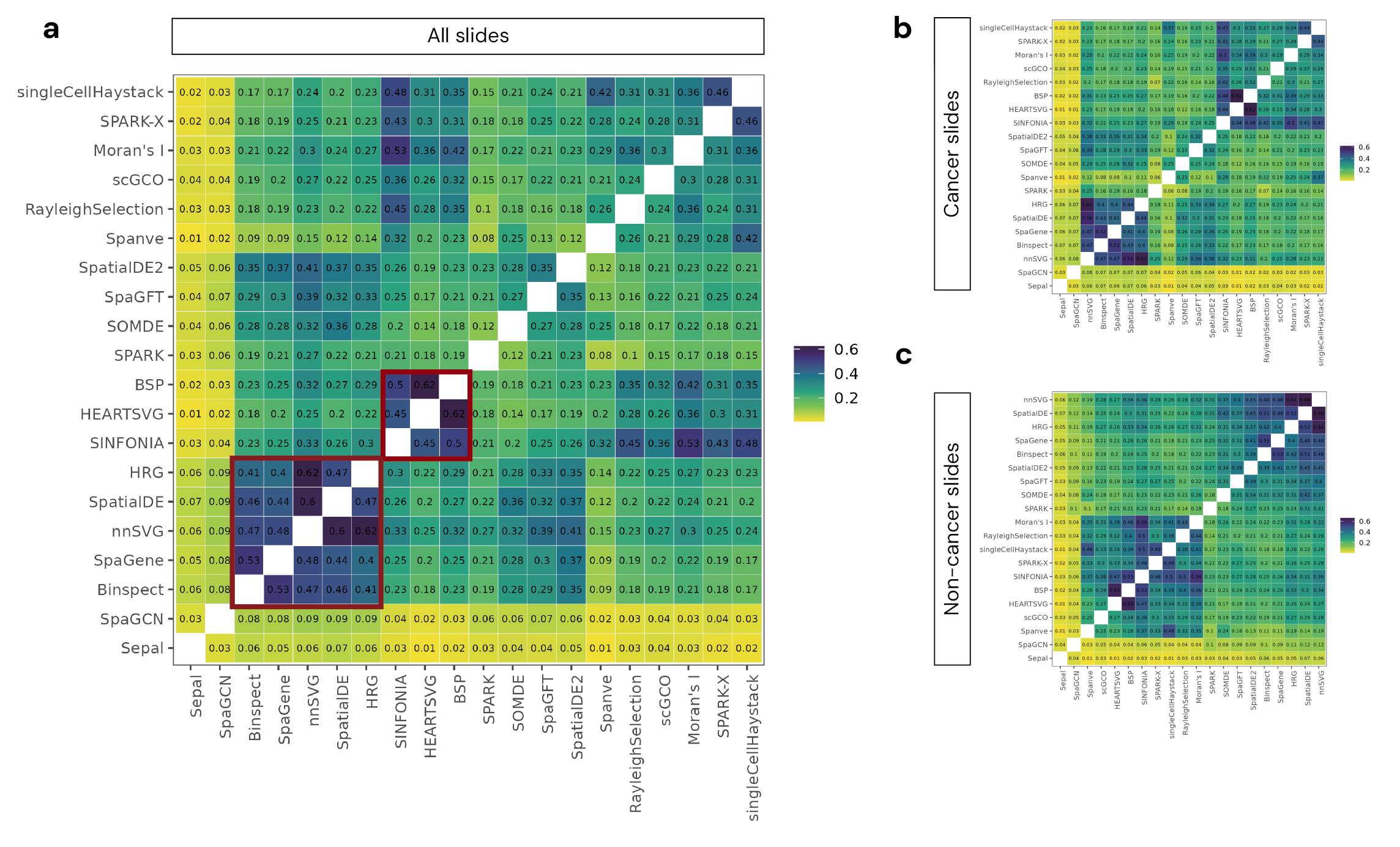}
    \caption{Method–method similarity analysis. Jaccard similarity between methods across (a) all 662 STimage-1K4M slides, (b) cancer slides, (c) non-cancer slides.}
    \label{fig:method_similarity}
\end{figure}

To better understand the relationships among SVG detection methods, we examined the similarity between the sets of top-ranked SVGs identified by each method across all slides. Specifically, we computed pairwise Jaccard Index using the top 100 SVGs per slide and averaged the results across slides to construct a method–method similarity matrix (\Cref{fig:method_similarity}a).

Across all slides, we observed two prominent clusters of methods (\Cref{fig:method_similarity}a). The first cluster includes HRG, SpatialDE, nnSVG, SpaGene, and Binspect, which tend to rely on generalized kernel-based and graph-based modeling and statistical testing to identify spatial structure. These methods often emphasize variance modeling or discrete pattern enrichment. The second cluster is composed of HEARTSVG, BSP, and SINFONIA, which incorporate spatial autocorrelation, local variance, or hierarchical spatial resolution frameworks. When restricting to cancer slides (\Cref{fig:method_similarity}b), we again identified the same two clusters. The overall structure of the clusters remained stable, suggesting that these methods capture consistent biological signals across cancer types. In contrast, in non-cancer slides (\Cref{fig:method_similarity}c), which featured more homogeneous and well-organized tissues, the similarity structure became larger and more diffuse. The two clusters combined together and formed a larger cluster (\Cref{fig:method_similarity}c). This reflects greater consensus among methods when applied to regular tissue architectures with clear spatial domains (e.g., layered cortex).

Among all method pairs, three pairs exhibit notably high similarity in their output SVGs, reflecting shared conceptual underpinnings despite different implementations (\Cref{fig:method_similarity}). First, HRG and nnSVG both aim to identify genes that maximize spatial signal strength. nnSVG fits a Gaussian Process model and performs a likelihood ratio test to determine whether the inclusion of spatial variance parameter significantly improves model fit compared to a non-spatial baseline. HRG shares a similar philosophy: it constructs a graph and iteratively refines it using genes that maximize a regional distribution score, thereby enhancing spatial specificity. This alignment in their focus on spatial signal optimization likely accounts for their strong agreement. Next, we observe striking similarity between HEARTSVG and BSP, two methods that emphasize local spatial variance via pooling strategies. BSP performs multi-resolution pooling by aggregating gene expression over nested grids of varying scales, capturing spatial structure across different resolutions. In contrast, HEARTSVG applies a form of directional marginal pooling, aggregating expression across axes and testing for spatial patterning along those marginal distributions. Despite differences in implementation, both methods enhance detection power by reducing noise through local averaging, enabling more stable assessment of spatial variation. Finally, the close similarity between SINFONIA and Moran’s I arises from direct methodological overlap. SINFONIA employs both Moran’s I and Geary’s C as part of an ensemble framework for SVG ranking, explicitly incorporating the output of Moran’s I.


\section{Scalability for high resolution slides} \label{sec:visiumHD}

\begin{figure}[!h]
    \centering
    \includegraphics[width=\textwidth]{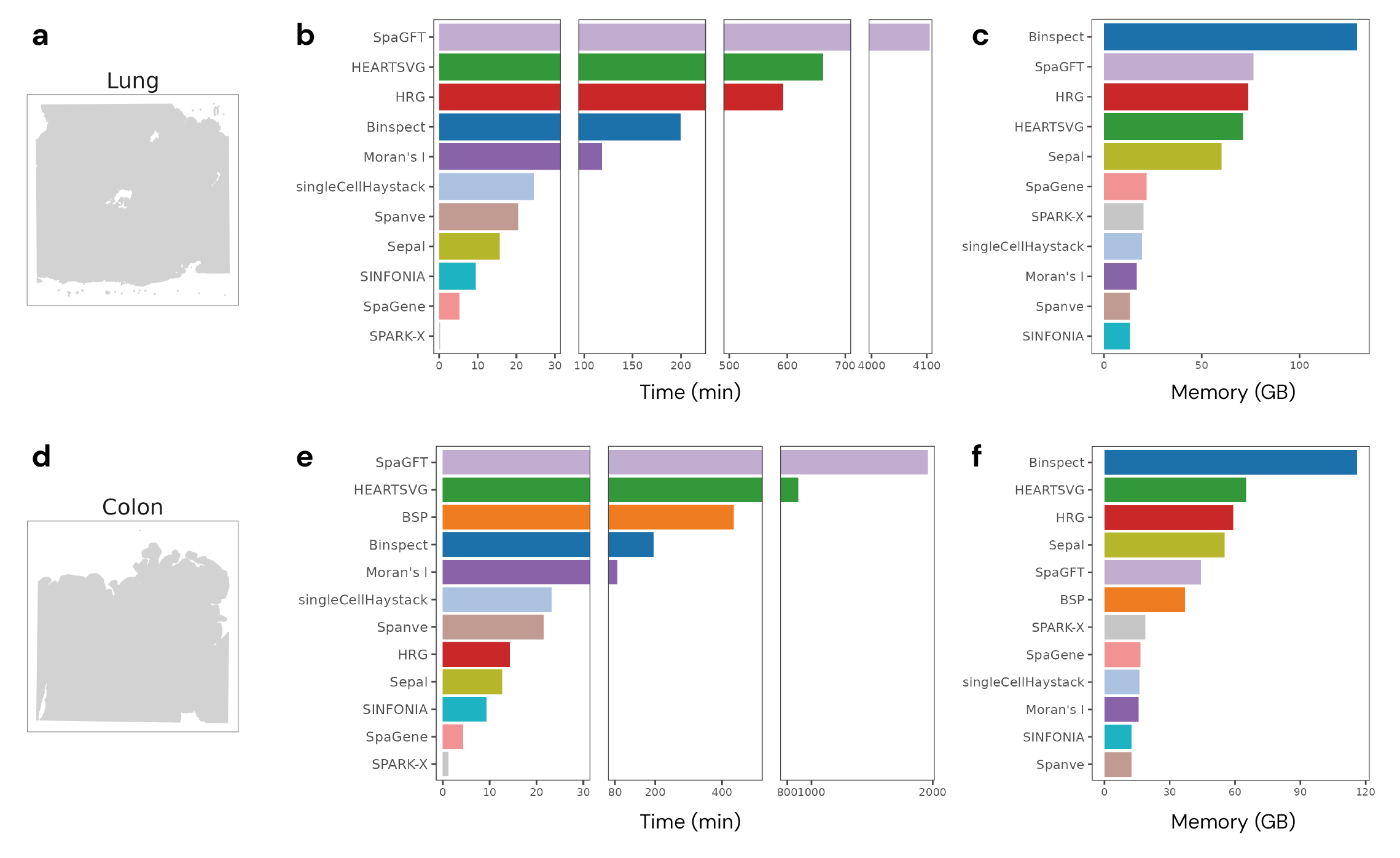}
    \caption{Evaluation of computational efficiency across Visium HD slides. Human\_Lung\_10X\_03292024\_VisiumHD: (a) tissue landscape, (b) computation time, (c) memory used. Human\_Colon\_10X\_03252024\_VisiumHD: (d) tissue landscape, (e) computation time, (f) memory used.}
    \label{fig:visiumHD}
\end{figure}

As ST technologies continue to evolve, platforms such as 10x Genomics’ Visium HD now enable transcriptome-wide profiling at subcellular resolution, generating datasets with hundreds of thousands of spatial barcodes per tissue section. These large-scale data pose a considerable computational challenge for existing SVG detection methods. To assess scalability and practical usability on such massive datasets, we evaluated all 20 SVG methods using two Visium HD samples: one from lung tissue (516,356 spots, 1623 genes) and another from colon tissue (348,783 spots, 2256 genes) (\Cref{fig:visiumHD}a,d).

Despite allocating generous computational resources (200 GB of memory limit), only 11 out of 20 methods successfully completed the analysis on both samples. Among the remaining 9 methods, 8 failed due to out-of-memory errors, and SOMDE did not finish within its 5-day internal time limit. Among the 11 runnable methods, SPARK-X and SpaGene were the most computationally efficient, completing both lung and colon sample analyses in under 10 minutes and with relatively modest memory usage (around 10 GB) (\Cref{fig:visiumHD}b,e). Several other methods, including SINFONIA, Sepal, and Spanve, also demonstrated feasible runtimes (under 1 hour) and controlled memory usage (less than 20 GB), making them practical options for high-throughput data analysis.

\section{Discussion}\label{sec:discussion}

In this comprehensive benchmarking study, we systematically evaluated 20 SVG detection methods across 662 ST slides from human tissue samples in STimage-1K4M, spanning a wide spectrum of tissue types, experimental platforms, and biological conditions. Our results provide a unified landscape of method performance in terms of spatial signal detection, computational efficiency, tissue coverage, and robustness. This work represents one of the most extensive benchmarking efforts to date for ST analysis and offers crucial guidance for method selection and future tool development.

Through a comprehensive benchmarking of 60 pathologist-annotated ST slides, we systematically assessed the ability of the SVG detection methods to recover domain-specific marker genes. Our analysis revealed that several methods, including SINFONIA, and Moran's I, consistently exhibited strong concordance with ``ground-truth'' markers derived from DE analyses, particularly in cancer slides. 
Notably, method rankings were largely consistent across technological platforms, although platform-specific biases were evident in certain methods. We also observed that method performance varied across tissue types, with higher accuracy observed in tissues with well-defined spatial structures, such as the cortical layers of the brain. A key limitation we identified is the sensitivity of current SVG methods to spatial domain imbalance: slides with extremely high or low proportions of cancerous regions frequently led to poor marker recovery, likely due to diluted or localized spatial signals that challenge unsupervised detection frameworks. In non-cancerous tissues, overall performance was more variable and generally lower, likely reflecting the increased anatomical heterogeneity and subtler expression gradients typical of normal physiological organization. These findings underscore the importance of considering both tissue context and technical platform when applying and interpreting the results of SVG detection algorithms.

Leveraging the scale and diversity of the STimage-1K4M dataset, our benchmarking framework enables us to evaluate the robustness of SVG detection methods across diverse biological and technical conditions. We systematically assessed the reproducibility of SVG sets across individuals and studies. Methods such as HEARTSVG, BSP, and nnSVG consistently produced highly reproducible SVGs across slides of the same tissue type, indicating strong robustness to biological heterogeneity. Comparisons within the same study series yielded higher consistency than across different studies. Tissue-level analysis revealed that reproducibility also depends on biological context. Well-organized tissues like thymus or brain exhibited high robustness, whereas structurally complex or variable tissues like cervix and lung showed lower consistency. Furthermore, non-cancer tissues generally supported more reproducible SVG identification than cancer tissues, likely due to clearer anatomical organization and reduced spatial noise.

We also evaluated the methods on other technical aspects. We evaluated methods on the robustness to spatial coordinate rotation, which should not affect the performance. SpaGCN, SINFONIS, HRG, and SPARK remain robust under rotation while others are not robust on certain slides. Tissue complexity also affects rotation robustness. In addition, we systematically evaluated the computational time and memory cost for 662 slides from Spatial Transcriptomics and Visium. SINFONIA, SPanve, SpaGFT, singleCellHaystack and SPARK-X remained the most efficient methods that for slides with over 7500 spots, they still could manage the time under 1 minute. We also evaluate the computational cost on 2 Visium HD slides with only 11 methods runnable among 20 methods. On these two slides with over 300,000 spots, SPARK-X, SPaGene and SINFONIA remain most efficient regarding computational time, where Spanve, SINFONIA and Moran's I use the least memory.

This study not only serves as a benchmark, but also builds a large-scale resource cataloging SVG sets across diverse tissue types and disease contexts, enabling systematic biological analyses beyond benchmarking. By comparing the top 100 SVGs identified in each tissue, we constructed a cross-tissue similarity atlas that reveals biologically meaningful relationships. Within cancer tissues, we found high SVG similarity between samples involved in known metastatic pathways. For example, breast-to-brain metastases closely resemble lung-to-brain and skin-to-brain metastases, while kidney-to-liver metastases overlap significantly with both liver cancer and other metastases. These results indicate that SVG sets in metastatic tissues capture both tissue-of-origin and target-site-specific spatial programs. In contrast, non-cancer tissues exhibit lower and more homogeneous similarity overall, with the strongest concordance between thymus and lymph node, reflecting their shared immunological functions. Interestingly, comparisons between cancer types and their matched normal tissues revealed no consistent preference: some cancers retained high similarity with their normal counterparts (e.g., liver cancer with normal liver), while others aligned more closely with unrelated cancers (e.g., breast cancer with breast-to-brain metastasis, skin cancer with oral cancer). Across multiple cancers, SVG sets also showed strong similarity with lymphoid tissues, likely reflecting common immune-related spatial signatures. These findings demonstrate how the resulting SVG database can be leveraged to explore questions of tissue identity, tumor evolution, and immune microenvironment in a spatial context.

Our large-scale benchmarking effort lays the groundwork for both method development and biological discovery in ST, but also opens several directions for future research. First, the observed variability in SVG detection performance across tissue types, spatial contexts, and technological platforms underscores the need for more adaptive and context-aware algorithms. Current methods struggle in scenarios with skewed tissue composition, such as when a cancer region is spatially confined or occupies only a small fraction of the slide, where signal dilution hampers spatial gene detection. Future methods could address this by introducing parameters that explicitly account for tissue structure or spatial heterogeneity. Designing such parameters also remains an open challenge, particularly in unsupervised settings where domain labels are unavailable. Second, the comprehensive SVG atlas generated in this study represents a unique and valuable resource for the ST community. Beyond benchmarking, it offers great potential for biological research, facilitating cross-tissue comparisons, revealing conserved or divergent spatial programs, and enabling the study of spatial gene expression across disease conditions and anatomical systems. Furthermore, the atlas itself could serve as prior knowledge to guide new method development, including the construction of adaptive parameters or models tailored to specific tissue contexts. 

\subsection*{Acknowledgments}
JC was supported by the National Institutes of Health (NIH) R01MH136055, R01AG079291; YL was supported by R01AG079291, R01AG085581, R01AR083790, and P50HD103573; DL was supported by NIH R01 LM014407, P50 CA058223, P42 ES031007. DL acknowledges the support from the NVIDIA Academic Grant Program.

\section{Online methods}\label{sec:online}

\subsection{Data availability}

The STimage-1K4M dataset can be accessed at \url{https://huggingface.co/datasets/jiawennnn/STimage-1K4M}. All the results in our analysis including the SVG atlas are available at \url{https://huggingface.co/spaces/jiawennnn/STimage-benchmark}.

\subsection{Technical details for SVG methods}

For SINFONIA, we followed the \href{https://github.com/BioX-NKU/SINFONIA/blob/main/docs/source/10X_Brain_Coronal.ipynb}{tutorial} accessed in Nov 2024. 

\noindent For scGCO, we followed the \href{https://github.com/WangPeng-Lab/scGCO/blob/master/code/Tutorial/scGCO_tutorial.ipynb}{tutorial} accessed in Feb 2025. 

\noindent For Spanve, we followed the \href{https://github.com/gx-Cai/Spanve/blob/main/tutorial.ipynb}{tutorial} accessed in Nov 2024. 

\noindent For Moran's I, we followed the \href{https://scanpy.readthedocs.io/en/stable/generated/scanpy.metrics.morans_i.html}{Scanpy package tutorial} accessed in Nov 2024. 

\noindent For SOMDE, we followed the \href{https://github.com/WhirlFirst/somde/blob/master/slide_seq0819_11_SOM.ipynb}{tutorial} accessed in Nov 2024. 

\noindent For SpatialDE2, we followed the \href{https://github.com/PMBio/spatialde2-paper/blob/master/figure_2.ipynb}{tutorial}) accessed in Nov 2024. 

\noindent For SpatialDE, we followed the \href{https://github.com/Teichlab/SpatialDE/blob/master/README.ipynb}{tutorial} accessed in Nov 2024. 

\noindent For sepal, we followed the \href{https://squidpy.readthedocs.io/en/stable/notebooks/examples/graph/compute_sepal.html#compute-sepal-score}{Squidpy pacakge tutorial} accessed in Nov 2024. 

\noindent For SpaGCN, we followed the \href{https://github.com/jianhuupenn/SpaGCN/blob/master/tutorial/tutorial.ipynb}{tutorial} accessed in Nov 2024. In addition, SpaGCN suggest extra thresholding for SVG selection which we did not apply in our study, which suggests to apply the following filter conditions if the target domain exists: in-fraction $>0.8$, in/out fraction ratio $>1$, and fold change $>1.5$. 

\noindent For SpaGFT, we didn't fully followed the \href{https://spagft.readthedocs.io/en/latest/spatial/lymphnode_tutorial.html}{tutorial} accessed in Feb 2025 because slides with fewer than 500 spots consistently failed during the eigenvalue selection step. We found that the \texttt{kneed\_select\_values(eigvals\_l, increasing=False)} function returned \texttt{None} when eigenvalues were ordered increasingly rather than decreasingly. To address this, we set \texttt{increasing=True} in the function call, which enabled the successful processing of most small slides, while two slides (original sizes: $(57, 30479)$ and $(55, 30479)$; filtered sizes: $(57, 7323)$ and $(55, 7098)$) were excluded from downstream evaluation. 

\noindent For SpaGene, we followed the \href{https://htmlpreview.github.io/?https://github.com/liuqivandy/SpaGene/blob/master/Tutorial/mob.html}{documentation} accessed in Mar 2025. 

\noindent For BSP, we followed the \href{https://github.com/juexinwang/BSP/blob/main/BSP.py}{tutorial} accessed in Mar 2025. 

\noindent For HRG, we followed the \href{https://github.com/JulieBaker1/HighlyRegionalGenes/wiki}{tutorial} accessed in Feb 2025. 

\noindent For HEARTSVG, we followed the \href{https://github.com/cz0316/HEARTSVG}{tutorial} accessed in Feb 2025. 

\noindent For BinSpect, we followed the \href{https://rubd.github.io/Giotto_site/articles/mouse_visium_brain_201226.html}{tutorial} accessed in May 2025. 

\noindent For singleCellHaystack, we followed the \href{https://alexisvdb.github.io/singleCellHaystack/articles/examples/a03_example_spatial_visium.html}{tutorial} accessed in Mar 2025. 

\noindent For RayleighSelection, we followed the \href{https://github.com/CamaraLab/RayleighSelection/blob/master/examples/vr_cycle_example.md}{tutorial} accessed in Mar 2025.

\noindent For RayleighSelection, we set the variable radius to be the 0.1 percent quantile of distances. Large radius will filter out all spots and small radius can not filter out any spots, which makes the RAM explode. 

\noindent For nnSVG, we followed the \href{https://github.com/lmweber/nnSVG}{tutorial} accessed in Feb 2025. Here we set the number of threads as 1 (default), but we note here that nnSVG can use multi-threads. 

\noindent For SPARK, we followed the \href{https://xzhoulab.github.io/SPARK/}{tutorial} accessed in Feb 2025. We also set the number of cores used to be 1. We note here that SPARK can use multiple cores. 

\noindent For SPARKX, we followed the \href{https://xzhoulab.github.io/SPARK/}{tutorial} accessed in Feb 2025.We also set the number of cores used to be 1. We note here that SPARK-X can use multiple cores.

\subsection{User-friendly evaluation criteria}\label{online_methods:user_friendly}

We evalutated the SVG methods' user-friendly evaluation criteria for the following three categories. For each categories, we score the methods from \{0, 0.5, 1\}.

\noindent  \textbf{(1) Dependency list}: \textbf{1:} Includes a complete dependency list with suggestions and ensures the setup works. \textbf{0.5:} Partially defined dependencies (e.g. assumes the user already has common packages like numpy and pandas). \textbf{0:} No dependencies are provided.

\noindent \textbf{(2) Documentation quality}: \textbf{1:} Well-documented, including: a clear \texttt{README.md} with installation instructions, examples, and use cases. Additional documentation files (e.g. \texttt{docs /}) or links to external documentation, if applicable. \textbf{0.5:} Basic \texttt{README.md} exists but lacks important details (e.g., no example usage, unclear setup instructions). \textbf{0:} Poor or no documentation.

\noindent \textbf{(3) Usability and setup}: \textbf{1:} Easy to setup and works with our data. \textbf{0.5:} Requires debugging (exclusive to missing dependency) or adjustments during setup, but eventually works. \textbf{0:} Problematic setup and eventually doesn't work with our data.


\begin{thebibliography}{10}

\bibitem{staahl2016visualization}
Patrik~L St{\aa}hl, Fredrik Salm{\'e}n, Sanja Vickovic, Anna Lundmark, Jos{\'e}~Fern{\'a}ndez Navarro, Jens Magnusson, Stefania Giacomello, Michaela Asp, Jakub~O Westholm, Mikael Huss, et~al.
\newblock {Visualization and analysis of gene expression in tissue sections by spatial transcriptomics}.
\newblock {\em Science}, 353(6294):78--82, 2016.

\bibitem{chen2022comprehensive}
Jiawen Chen, Weifang Liu, Tianyou Luo, Zhentao Yu, Minzhi Jiang, Jia Wen, Gaorav~P Gupta, Paola Giusti, Hongtu Zhu, Yuchen Yang, et~al.
\newblock {A comprehensive comparison on cell-type composition inference for spatial transcriptomics data}.
\newblock {\em Briefings in Bioinformatics}, 23(4):bbac245, 2022.

\bibitem{Maynard2021-wa}
Kristen~R Maynard, Leonardo Collado-Torres, Lukas~M Weber, Cedric Uytingco, Brianna~K Barry, Stephen~R Williams, Joseph~L Catallini, 2nd, Matthew~N Tran, Zachary Besich, Madhavi Tippani, Jennifer Chew, Yifeng Yin, Joel~E Kleinman, Thomas~M Hyde, Nikhil Rao, Stephanie~C Hicks, Keri Martinowich, and Andrew~E Jaffe.
\newblock Transcriptome-scale spatial gene expression in the human dorsolateral prefrontal cortex.
\newblock {\em Nat. Neurosci.}, 24(3):425--436, March 2021.

\bibitem{Asp2019-ga}
Michaela Asp, Stefania Giacomello, Ludvig Larsson, Chenglin Wu, Daniel F{\"u}rth, Xiaoyan Qian, Eva W{\"a}rdell, Joaquin Custodio, Johan Reimeg{\aa}rd, Fredrik Salm{\'e}n, Cecilia {\"O}sterholm, Patrik~L St{\aa}hl, Erik Sundstr{\"o}m, Elisabet {\AA}kesson, Olaf Bergmann, Magda Bienko, Agneta M{\aa}nsson-Broberg, Mats Nilsson, Christer Sylv{\'e}n, and Joakim Lundeberg.
\newblock A spatiotemporal organ-wide gene expression and cell atlas of the developing human heart.
\newblock {\em Cell}, 179(7):1647--1660.e19, December 2019.

\bibitem{chen2023cell}
Jiawen Chen, Tianyou Luo, Minzhi Jiang, Jiandong Liu, Gaorav~P Gupta, and Yun Li.
\newblock {Cell composition inference and identification of layer-specific spatial transcriptional profiles with {POLARIS}}.
\newblock {\em Science Advances}, 9(9):eadd9818, 2023.

\bibitem{oliveira2025high}
Michelli Faria~de Oliveira, Juan~Pablo Romero, Meii Chung, Stephen~R Williams, Andrew~D Gottscho, Anushka Gupta, Susan~E Pilipauskas, Seayar Mohabbat, Nandhini Raman, David~J Sukovich, et~al.
\newblock {High-definition spatial transcriptomic profiling of immune cell populations in colorectal cancer}.
\newblock {\em Nature Genetics}, pages 1--12, 2025.

\bibitem{vannan2025spatial}
Annika Vannan, Ruqian Lyu, Arianna~L Williams, Nicholas~M Negretti, Evan~D Mee, Joseph Hirsh, Samuel Hirsh, Niran Hadad, David~S Nichols, Carla~L Calvi, et~al.
\newblock {Spatial transcriptomics identifies molecular niche dysregulation associated with distal lung remodeling in pulmonary fibrosis}.
\newblock {\em Nature genetics}, 57(3):647--658, 2025.

\bibitem{yan2025categorization}
Guanao Yan, Shuo~Harper Hua, and Jingyi~Jessica Li.
\newblock {Categorization of 34 computational methods to detect spatially variable genes from spatially resolved transcriptomics data}.
\newblock {\em Nature Communications}, 16(1):1141, 2025.

\bibitem{shang2022spatially}
Lulu Shang and Xiang Zhou.
\newblock {Spatially aware dimension reduction for spatial transcriptomics}.
\newblock {\em Nature communications}, 13(1):7203, 2022.

\bibitem{shang2025statistical}
Lulu Shang, Peijun Wu, and Xiang Zhou.
\newblock {Statistical identification of cell type-specific spatially variable genes in spatial transcriptomics}.
\newblock {\em Nature communications}, 16(1):1059, 2025.

\bibitem{cai2023spanve}
Guoxin Cai, Yichang Chen, Shuqing Chen, Xun Gu, and Zhan Zhou.
\newblock {Spanve: A Statistical Method for Detecting Downstream-Friendly Spatially Variable Genes in Large-Scale Spatial Transcriptomic Data}.
\newblock {\em bioRxiv}, pages 2023--02, 2023.

\bibitem{liang2024multi}
Xiao Liang, Pei Liu, Li~Xue, Baiyun Chen, Wei Liu, Wanwan Shi, Yongwang Wang, Xiangtao Chen, and Jiawei Luo.
\newblock {A multi-modality and multi-granularity collaborative learning framework for identifying spatial domains and spatially variable genes}.
\newblock {\em Bioinformatics}, 40(10):btae607, 2024.

\bibitem{moran1950notes}
P.~A.~P. Moran.
\newblock {Notes on continuous stochastic phenomena}.
\newblock {\em Biometrika}, 37(1-2):17--23, 1950.

\bibitem{svensson2018spatialde}
Valentine Svensson, Sarah~A Teichmann, and Oliver Stegle.
\newblock {SpatialDE: identification of spatially variable genes}.
\newblock {\em Nature methods}, 15(5):343--346, 2018.

\bibitem{govek2019clustering}
Kiya~W Govek, Venkata~S Yamajala, and Pablo~G Camara.
\newblock {Clustering-independent analysis of genomic data using spectral simplicial theory}.
\newblock {\em PLoS computational biology}, 15(11):e1007509, 2019.

\bibitem{sun2020statistical}
Shiquan Sun, Jiaqiang Zhu, and Xiang Zhou.
\newblock {Statistical analysis of spatial expression patterns for spatially resolved transcriptomic studies}.
\newblock {\em Nature methods}, 17(2):193--200, 2020.

\bibitem{vandenbon2020clustering}
Alexis Vandenbon and Diego Diez.
\newblock {A clustering-independent method for finding differentially expressed genes in single-cell transcriptome data}.
\newblock {\em Nature communications}, 11(1):4318, 2020.

\bibitem{dries2021giotto}
Ruben Dries, Qian Zhu, Rui Dong, Chee-Huat~Linus Eng, Huipeng Li, Kan Liu, Yuntian Fu, Tianxiao Zhao, Arpan Sarkar, Feng Bao, et~al.
\newblock {Giotto: a toolbox for integrative analysis and visualization of spatial expression data}.
\newblock {\em Genome biology}, 22(1):78, 2021.

\bibitem{andersson2021sepal}
Alma Andersson and Joakim Lundeberg.
\newblock {{sepal: Identifying transcript profiles with spatial patterns by diffusion-based modeling}}.
\newblock {\em Bioinformatics}, 37(17):2644--2650, 2021.

\bibitem{zhu2021spark}
Jiaqiang Zhu, Shiquan Sun, and Xiang Zhou.
\newblock {SPARK-X: non-parametric modeling enables scalable and robust detection of spatial expression patterns for large spatial transcriptomic studies}.
\newblock {\em Genome biology}, 22(1):184, 2021.

\bibitem{hao2021somde}
Minsheng Hao, Kui Hua, and Xuegong Zhang.
\newblock {SOMDE: a scalable method for identifying spatially variable genes with self-organizing map}.
\newblock {\em Bioinformatics}, 37(23):4392--4398, 2021.

\bibitem{hu2021spagcn}
Jian Hu, Xiangjie Li, Kyle Coleman, Amelia Schroeder, Nan Ma, David~J Irwin, Edward~B Lee, Russell~T Shinohara, and Mingyao Li.
\newblock {{SpaGCN: Integrating gene expression, spatial location and histology to identify spatial domains and spatially variable genes by graph convolutional network}}.
\newblock {\em Nature methods}, 18(11):1342--1351, 2021.

\bibitem{kats2021spatialde2}
Ilia Kats, Roser Vento-Tormo, and Oliver Stegle.
\newblock {SpatialDE2: fast and localized variance component analysis of spatial transcriptomics}.
\newblock {\em Biorxiv}, pages 2021--10, 2021.

\bibitem{wu2022highly}
Yanhong Wu, Qifan Hu, Shicheng Wang, Changyi Liu, Yiran Shan, Wenbo Guo, Rui Jiang, Xiaowo Wang, and Jin Gu.
\newblock {Highly Regional Genes: graph-based gene selection for single-cell RNA-seq data}.
\newblock {\em Journal of Genetics and Genomics}, 49(9):891--899, 2022.

\bibitem{liu2022scalable}
Qi~Liu, Chih-Yuan Hsu, and Yu~Shyr.
\newblock {Scalable and model-free detection of spatial patterns and colocalization}.
\newblock {\em Genome research}, 32(9):1736--1745, 2022.

\bibitem{zhang2022identification}
Ke~Zhang, Wanwan Feng, and Peng Wang.
\newblock {Identification of spatially variable genes with graph cuts}.
\newblock {\em Nature Communications}, 13(1):5488, 2022.

\bibitem{jiang2023sinfonia}
Rui Jiang, Zhen Li, Yuhang Jia, Siyu Li, and Shengquan Chen.
\newblock {SINFONIA: scalable identification of spatially variable genes for deciphering spatial domains}.
\newblock {\em Cells}, 12(4):604, 2023.

\bibitem{weber2023nnsvg}
Lukas~M Weber, Arkajyoti Saha, Abhirup Datta, Kasper~D Hansen, and Stephanie~C Hicks.
\newblock {nnSVG for the scalable identification of spatially variable genes using nearest-neighbor Gaussian processes}.
\newblock {\em Nature communications}, 14(1):4059, 2023.

\bibitem{wang2023dimension}
Juexin Wang, Jinpu Li, Skyler~T Kramer, Li~Su, Yuzhou Chang, Chunhui Xu, Michael~T Eadon, Krzysztof Kiryluk, Qin Ma, and Dong Xu.
\newblock {Dimension-agnostic and granularity-based spatially variable gene identification using BSP}.
\newblock {\em Nature communications}, 14(1):7367, 2023.

\bibitem{yuan2024heartsvg}
Xin Yuan, Yanran Ma, Ruitian Gao, Shuya Cui, Yifan Wang, Botao Fa, Shiyang Ma, Ting Wei, Shuangge Ma, and Zhangsheng Yu.
\newblock {HEARTSVG: a fast and accurate method for identifying spatially variable genes in large-scale spatial transcriptomics}.
\newblock {\em Nature Communications}, 15(1):5700, 2024.

\bibitem{chang2024graph}
Yuzhou Chang, Jixin Liu, Yi~Jiang, Anjun Ma, Yao~Yu Yeo, Qi~Guo, Megan McNutt, Jordan~E Krull, Scott~J Rodig, Dan~H Barouch, et~al.
\newblock {Graph Fourier transform for spatial omics representation and analyses of complex organs}.
\newblock {\em Nature Communications}, 15(1):7467, 2024.

\bibitem{arnol2019modeling}
Damien Arnol, Denis Schapiro, Bernd Bodenmiller, Julio Saez-Rodriguez, and Oliver Stegle.
\newblock {Modeling cell-cell interactions from spatial molecular data with spatial variance component analysis}.
\newblock {\em Cell reports}, 29(1):202--211, 2019.

\bibitem{cable2022cell}
Dylan~M Cable, Evan Murray, Vignesh Shanmugam, Simon Zhang, Luli~S Zou, Michael Diao, Haiqi Chen, Evan~Z Macosko, Rafael~A Irizarry, and Fei Chen.
\newblock {Cell type-specific inference of differential expression in spatial transcriptomics}.
\newblock {\em Nature methods}, 19(9):1076--1087, 2022.

\bibitem{edsgard2018identification}
Daniel Edsg{\"a}rd, Per Johnsson, and Rickard Sandberg.
\newblock {Identification of spatial expression trends in single-cell gene expression data}.
\newblock {\em Nature methods}, 15(5):339--342, 2018.

\bibitem{bae2021discovery}
Sungwoo Bae, Hongyoon Choi, and Dong~Soo Lee.
\newblock {Discovery of molecular features underlying the morphological landscape by integrating spatial transcriptomic data with deep features of tissue images}.
\newblock {\em Nucleic acids research}, 49(10):e55--e55, 2021.

\bibitem{detomaso2021hotspot}
David DeTomaso and Nir Yosef.
\newblock {Hotspot identifies informative gene modules across modalities of single-cell genomics}.
\newblock {\em Cell systems}, 12(5):446--456, 2021.

\bibitem{miller2021characterizing}
Brendan~F Miller, Dhananjay Bambah-Mukku, Catherine Dulac, Xiaowei Zhuang, and Jean Fan.
\newblock {Characterizing spatial gene expression heterogeneity in spatially resolved single-cell transcriptomic data with nonuniform cellular densities}.
\newblock {\em Genome research}, 31(10):1843--1855, 2021.

\bibitem{li2021bayesian}
Qiwei Li, Minzhe Zhang, Yang Xie, and Guanghua Xiao.
\newblock {Bayesian modeling of spatial molecular profiling data via Gaussian process}.
\newblock {\em Bioinformatics}, 37(22):4129--4136, 2021.

\bibitem{bintayyash2021non}
Nuha BinTayyash, Sokratia Georgaka, ST~John, Sumon Ahmed, Alexis Boukouvalas, James Hensman, and Magnus Rattray.
\newblock {Non-parametric modelling of temporal and spatial counts data from RNA-seq experiments}.
\newblock {\em Bioinformatics}, 37(21):3788--3795, 2021.

\bibitem{moehlin2021inferring}
Julien Moehlin, Bastien Mollet, Bruno~Maria Colombo, and Marco~Antonio Mendoza-Parra.
\newblock {Inferring biologically relevant molecular tissue substructures by agglomerative clustering of digitized spatial transcriptomes with multilayer}.
\newblock {\em Cell Systems}, 12(7):694--705, 2021.

\bibitem{yu2022identification}
Jinge Yu and Xiangyu Luo.
\newblock {Identification of cell-type-specific spatially variable genes accounting for excess zeros}.
\newblock {\em Bioinformatics}, 38(17):4135--4144, 2022.

\bibitem{boost-mi}
Xi~Jiang, Guanghua Xiao, and Qiwei Li.
\newblock A bayesian modified ising model for identifying spatially variable genes from spatial transcriptomics data.
\newblock {\em Statistics in Medicine}, 41(23):4647--4665, 2022.

\bibitem{hong2023spatiotemporal}
Yingzhou Hong, Kai Song, Zongbo Zhang, Yuxia Deng, Xue Zhang, Jinqian Zhao, Jun Jiang, Qing Zhang, Chunming Guo, and Cheng Peng.
\newblock {The spatiotemporal dynamics of spatially variable genes in developing mouse brain revealed by a novel computational scheme}.
\newblock {\em Cell Death Discovery}, 9(1):264, 2023.

\bibitem{seal2023smash}
Souvik Seal, Benjamin~G Bitler, and Debashis Ghosh.
\newblock {SMASH: Scalable Method for Analyzing Spatial Heterogeneity of genes in spatial transcriptomics data}.
\newblock {\em PLoS Genetics}, 19(10):e1010983, 2023.

\bibitem{liang2024prost}
Yuchen Liang, Guowei Shi, Runlin Cai, Yuchen Yuan, Ziying Xie, Long Yu, Yingjian Huang, Qian Shi, Lizhe Wang, Jun Li, et~al.
\newblock {PROST: quantitative identification of spatially variable genes and domain detection in spatial transcriptomics}.
\newblock {\em Nature Communications}, 15(1):600, 2024.

\bibitem{yang2024bayesian}
Jie Yang, Xi~Jiang, Kevin~Wang Jin, Sunyoung Shin, and Qiwei Li.
\newblock {Bayesian hidden mark interaction model for detecting spatially variable genes in imaging-based spatially resolved transcriptomics data}.
\newblock {\em Frontiers in Genetics}, 15:1356709, 2024.

\bibitem{chen2024evaluating}
Carissa Chen, Hani~Jieun Kim, and Pengyi Yang.
\newblock {Evaluating spatially variable gene detection methods for spatial transcriptomics data}.
\newblock {\em Genome Biology}, 25(1):18, 2024.

\bibitem{chen2025benchmarking}
Xuanwei Chen, Qinghua Ran, Junjie Tang, Zihao Chen, Siyuan Huang, Xingjie Shi, and Ruibin Xi.
\newblock {Benchmarking algorithms for spatially variable gene identification in spatial transcriptomics}.
\newblock {\em Bioinformatics}, 41(4):btaf131, 2025.

\bibitem{cable2022robust}
Dylan~M Cable, Evan Murray, Luli~S Zou, Aleksandrina Goeva, Evan~Z Macosko, Fei Chen, and Rafael~A Irizarry.
\newblock {Robust decomposition of cell type mixtures in spatial transcriptomics}.
\newblock {\em Nature biotechnology}, 40(4):517--526, 2022.

\bibitem{Andersson2021-nr}
Alma Andersson, Ludvig Larsson, Linnea Stenbeck, Fredrik Salm{\'e}n, Anna Ehinger, Sunny~Z Wu, Ghamdan Al-Eryani, Daniel Roden, Alex Swarbrick, {\AA}ke Borg, Jonas Fris{\'e}n, Camilla Engblom, and Joakim Lundeberg.
\newblock Spatial deconvolution of {HER2-positive} breast cancer delineates tumor-associated cell type interactions.
\newblock {\em Nat. Commun.}, 12(1):6012, October 2021.

\bibitem{jaccard1901etude}
Paul Jaccard.
\newblock {{\'E}tude comparative de la distribution florale dans une portion des Alpes et des Jura}.
\newblock {\em Bull Soc Vaudoise Sci Nat}, 37:547--579, 1901.

\bibitem{Erickson2022-ia}
Andrew Erickson, Mengxiao He, Emelie Berglund, Maja Marklund, Reza Mirzazadeh, Niklas Schultz, Linda Kvastad, Alma Andersson, Ludvig Bergenstr{\aa}hle, Joseph Bergenstr{\aa}hle, Ludvig Larsson, Leire Alonso~Galicia, Alia Shamikh, Elisa Basmaci, Teresita D{\'\i}az De~St{\aa}hl, Timothy Rajakumar, Dimitrios Doultsinos, Kim Thrane, Andrew~L Ji, Paul~A Khavari, Firaz Tarish, Anna Tanoglidi, Jonas Maaskola, Richard Colling, Tuomas Mirtti, Freddie~C Hamdy, Dan~J Woodcock, Thomas Helleday, Ian~G Mills, Alastair~D Lamb, and Joakim Lundeberg.
\newblock Spatially resolved clonal copy number alterations in benign and malignant tissue.
\newblock {\em Nature}, 608(7922):360--367, August 2022.

\end{thebibliography}

\newpage
\appendix
\newgeometry{top=1in, bottom=1in,left=1in,right=1in}
\renewcommand{\thefigure}{S\arabic{figure}}
\renewcommand\figurename{Supplementary Figure}
\setcounter{figure}{0}    
\crefalias{figure}{appfig}

\section{Supplementary Figures}


\begin{figure}[H]
    \centering
    \includegraphics[width=\textwidth]{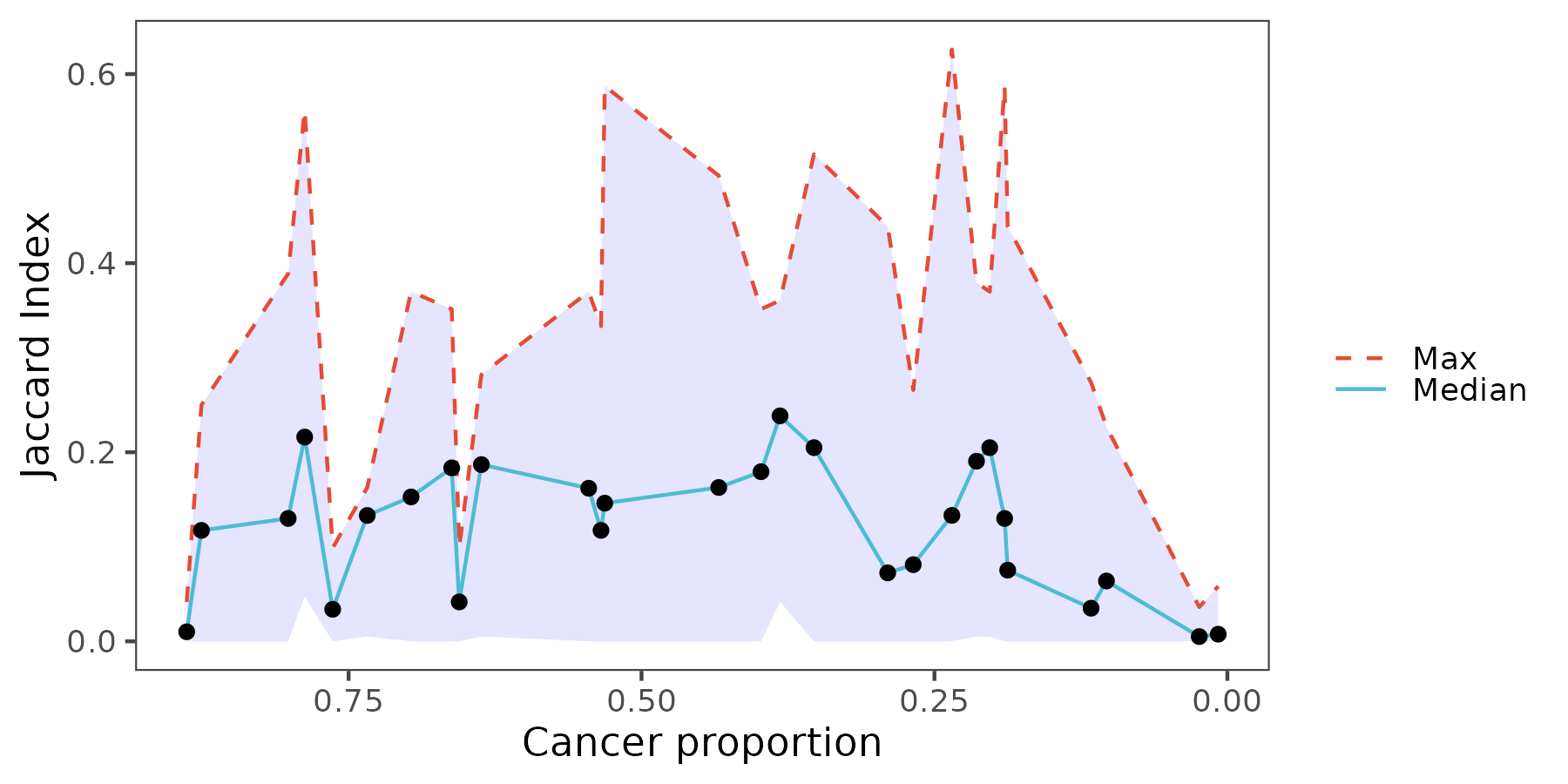}
    \caption{Slide-level performance sorted by cancer proportion. Shaded area indicates the minimal and maximal Jaccard Index within methods, and the black dots indicate the median Jaccard Index.}
    \label{supp_fig:DE_cancer_median}
\end{figure}

\begin{figure}[H]
    \centering
    \includegraphics[width=\textwidth]{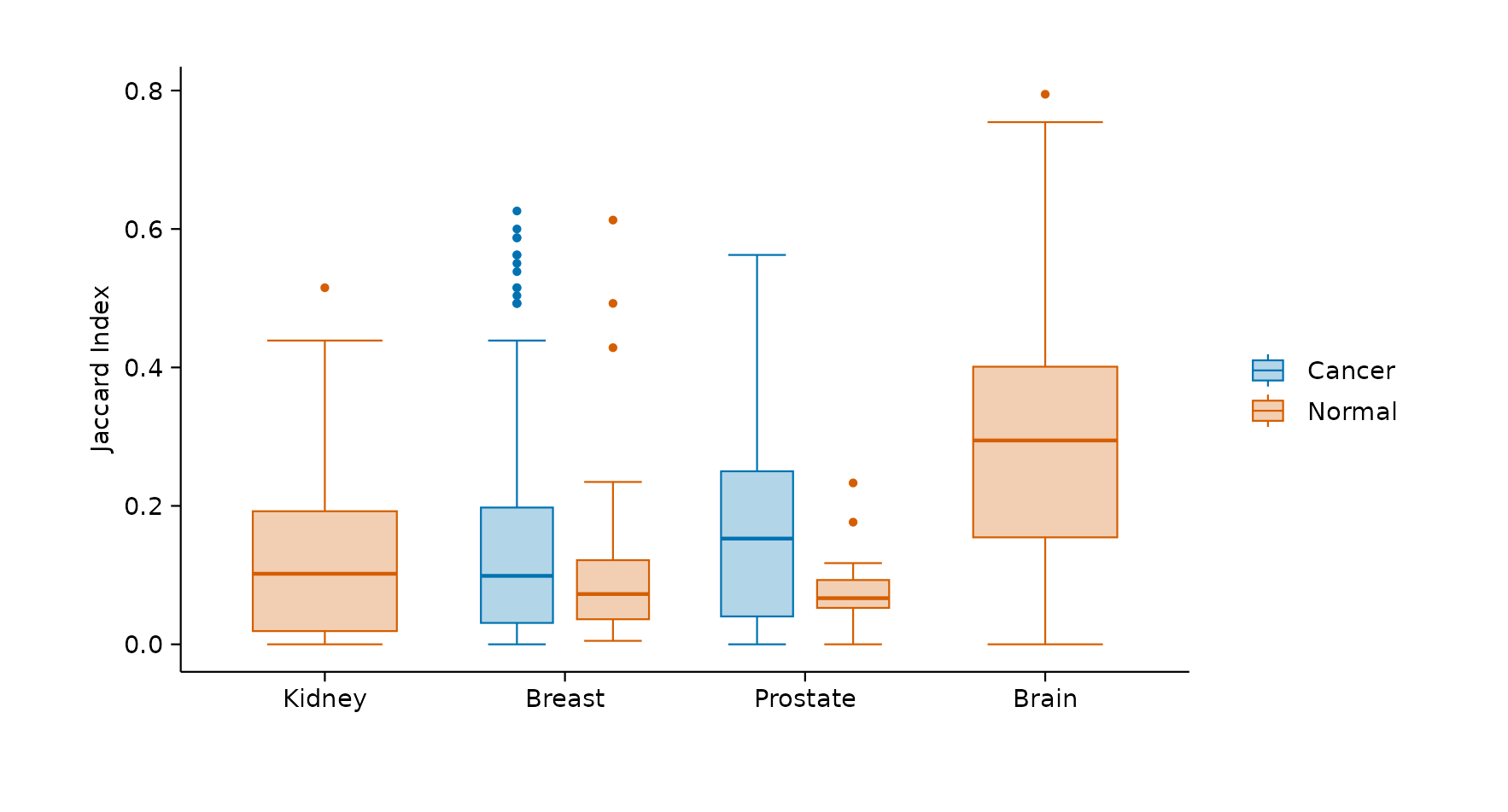}
    \caption{Jaccard Index between SVG sets and DE gene sets across all methods, stratified by tissue and disease status.}
    \label{supp_fig:DE_comparison_cancer_noncancer}
\end{figure}


\begin{figure}[H]
    \centering
    \includegraphics[width=\textwidth]{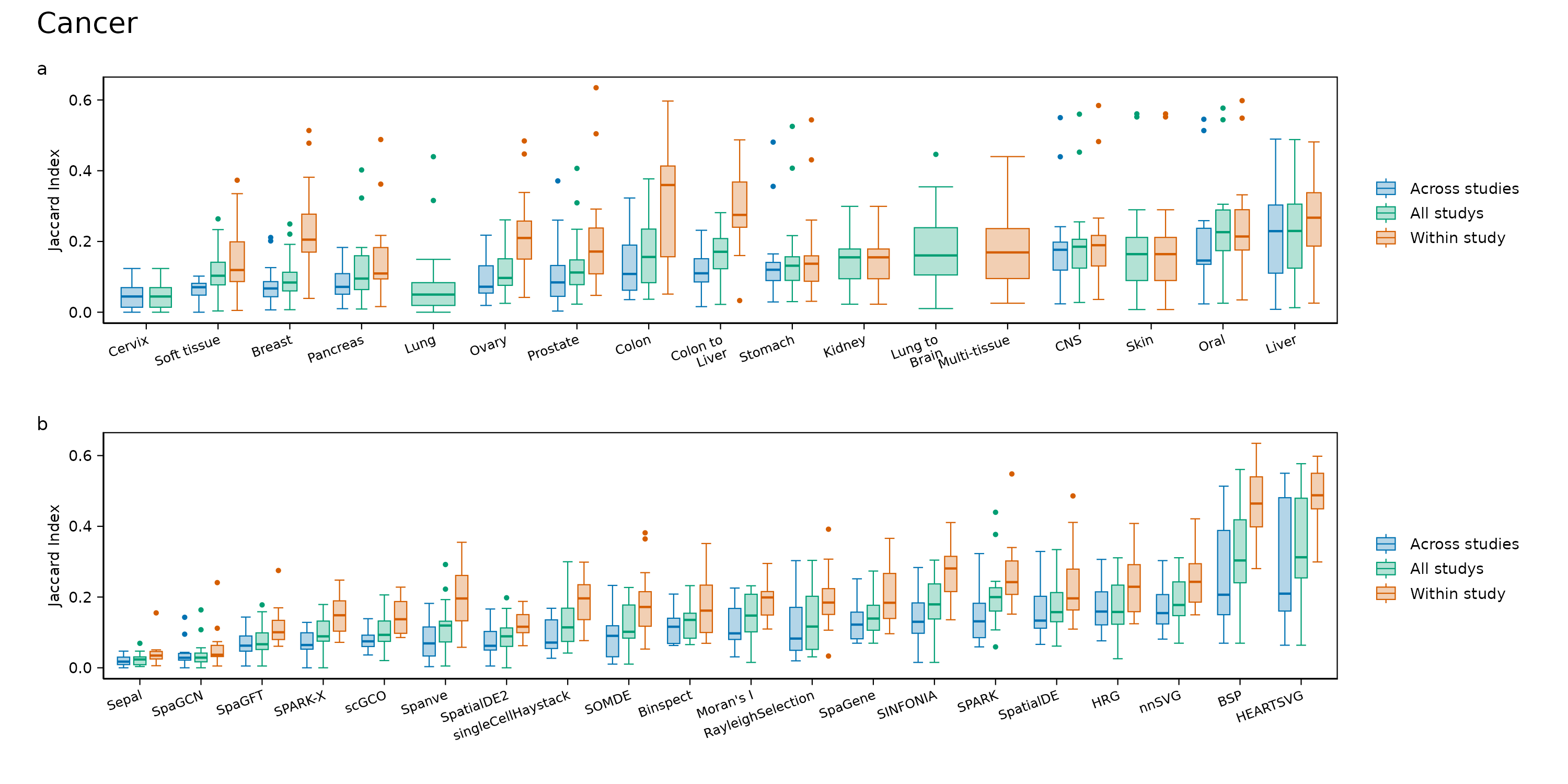}
    \caption{Jaccard Index between SVG sets and DE gene sets on cancer slides, stratified by (a) tissue type and study types and (b) methods and study types.}
    \label{supp_fig:robustness_cancer_by_tissue_method}
\end{figure}

\begin{figure}[H]
    \centering
    \includegraphics[width=\textwidth]{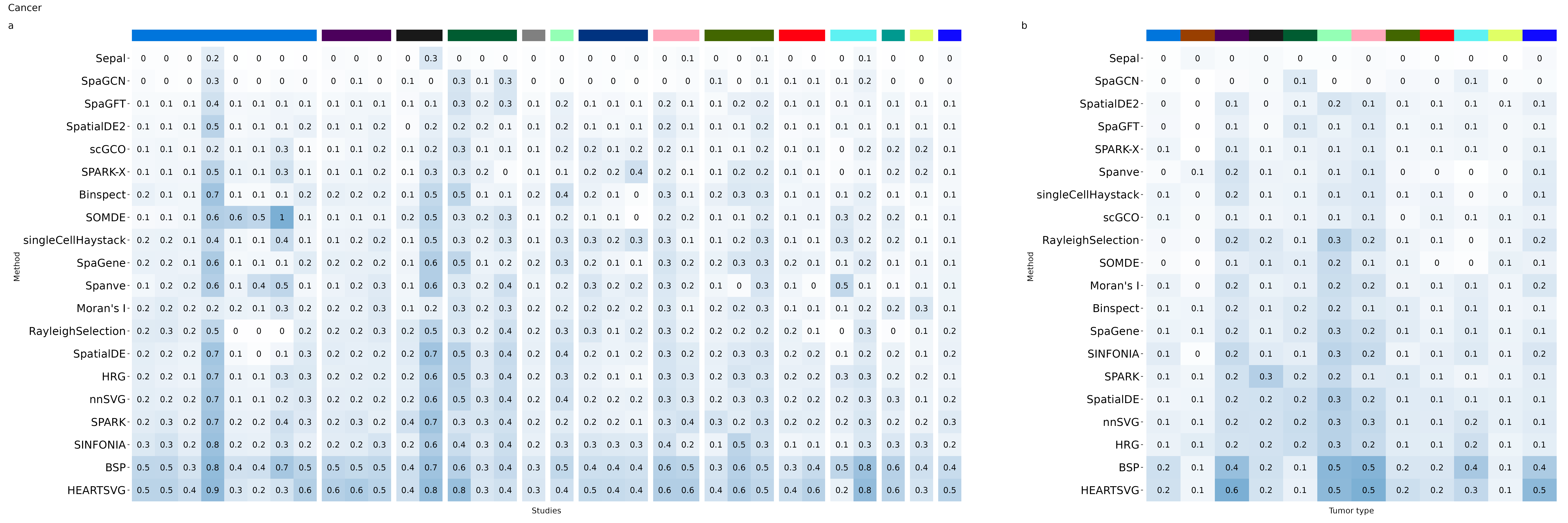}
    \caption{Average Jaccard Index between SVG sets and DE gene sets on cancer slides of each method for (a) with-in study evaluation (b) across study evaluation.}
    \label{supp_fig:robustness_cancer_study_heatmap}
\end{figure}

\begin{figure}[H]
    \centering
    \includegraphics[width=\textwidth]{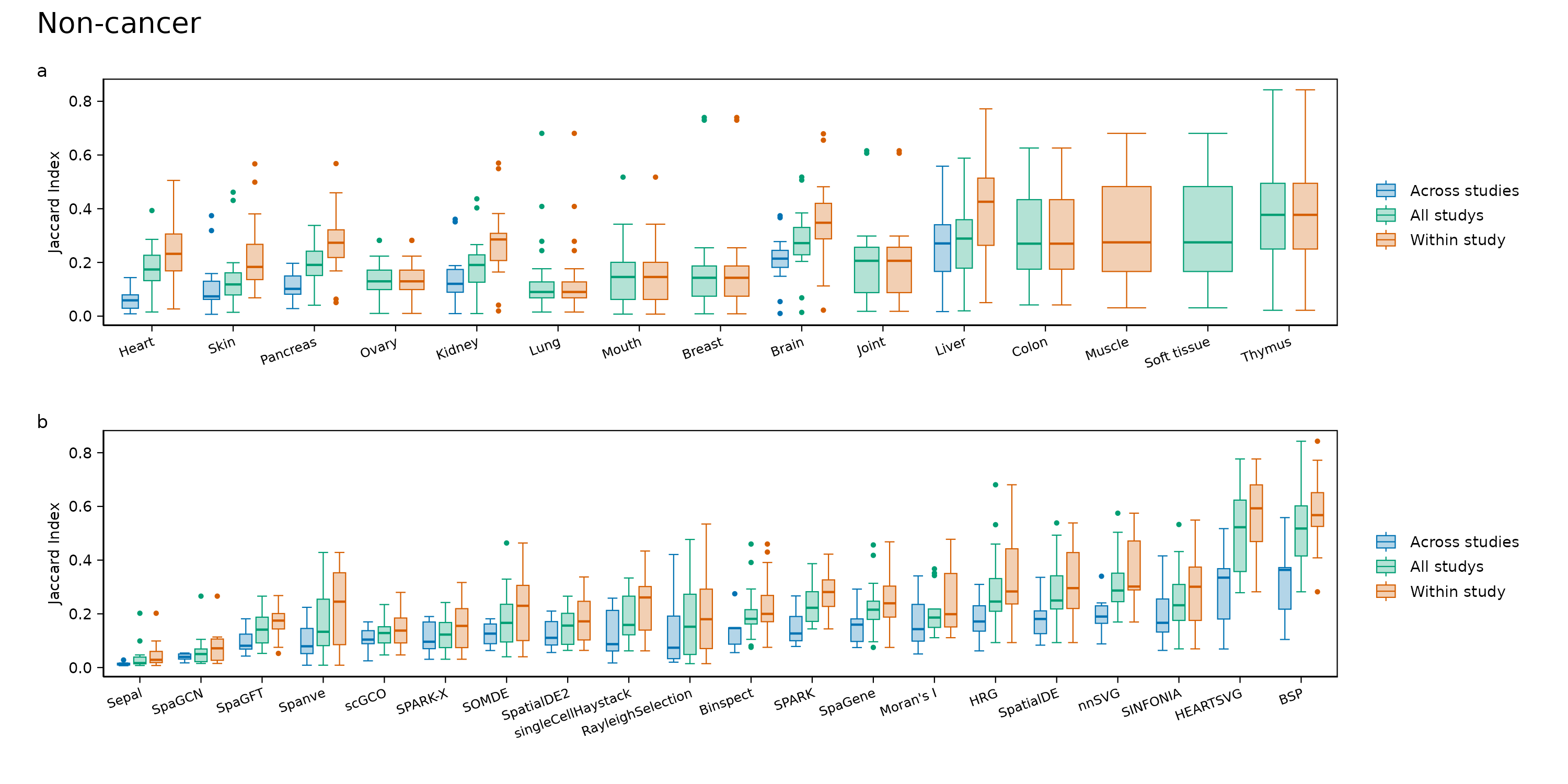}
    \caption{Jaccard Index between SVG sets and DE gene sets on non-cancer slides, stratified by (a) tissue type and study types and (b) methods and study types.}
    \label{supp_fig:robustness_noncancer_by_tissue_method}
\end{figure}

\begin{figure}[H]
    \centering
    \includegraphics[width=\textwidth]{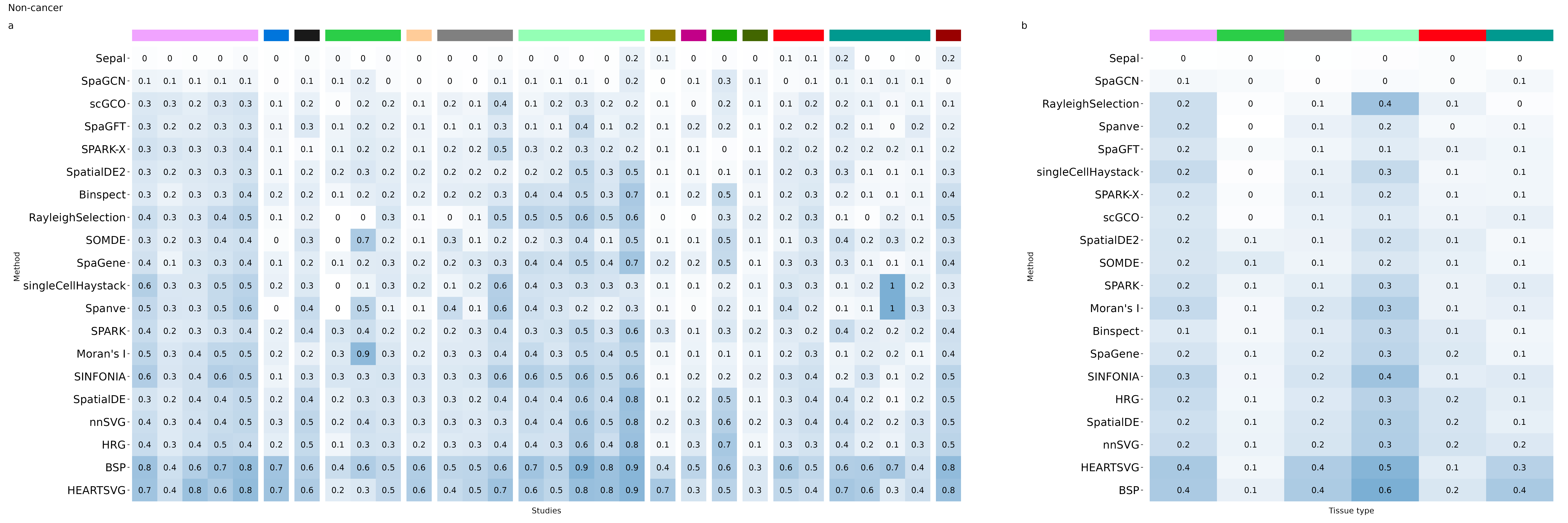}
    \caption{Average Jaccard Index between SVG sets and DE gene sets on non-cancer slides of each method for (a) with-in study evaluation (b) across study evaluation.}
    \label{supp_fig:robustness_noncancer_study_heatmap}
\end{figure}

\begin{figure}[H]
    \centering
    \includegraphics[]{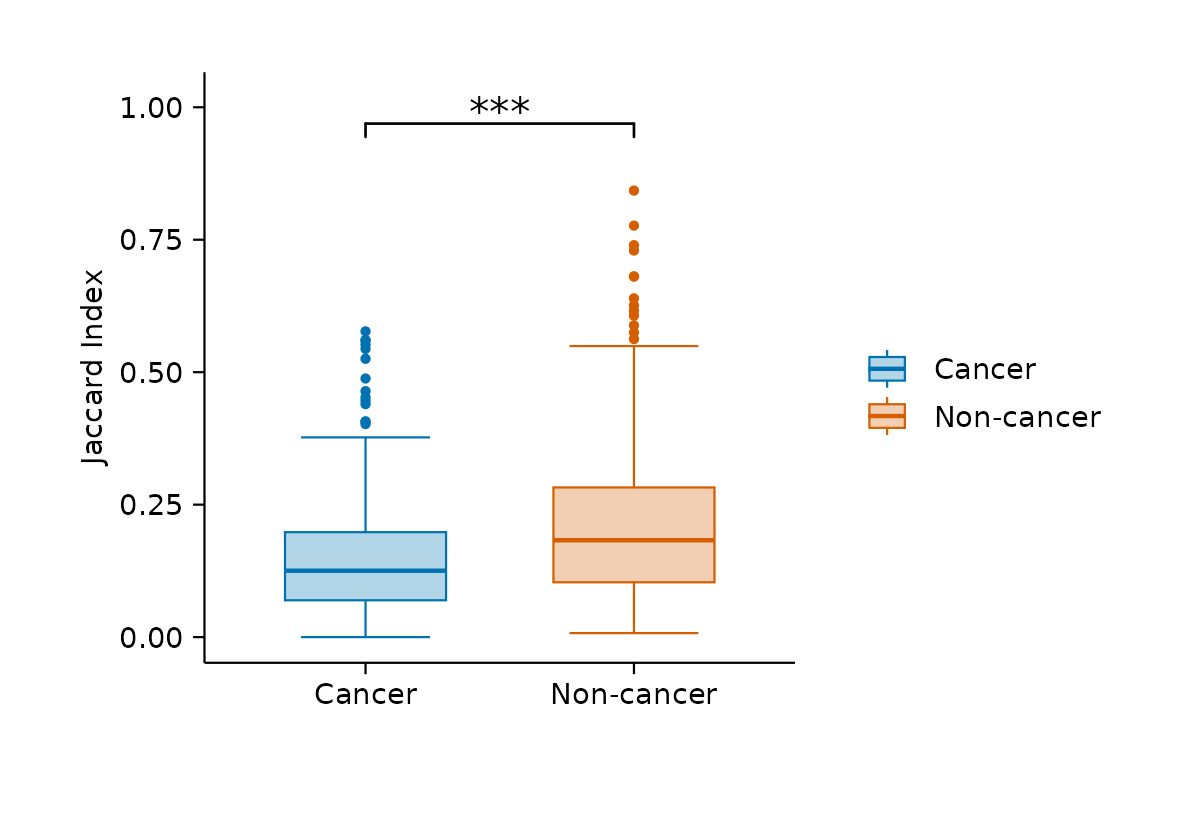}
    \caption{Jaccard Index between SVG sets and DE gene sets, stratified by disease status.}
    \label{supp_fig:robustness_cancer_noncancer_similarity}
\end{figure}

\begin{figure}[H]
    \centering
    \includegraphics[width=\textwidth]{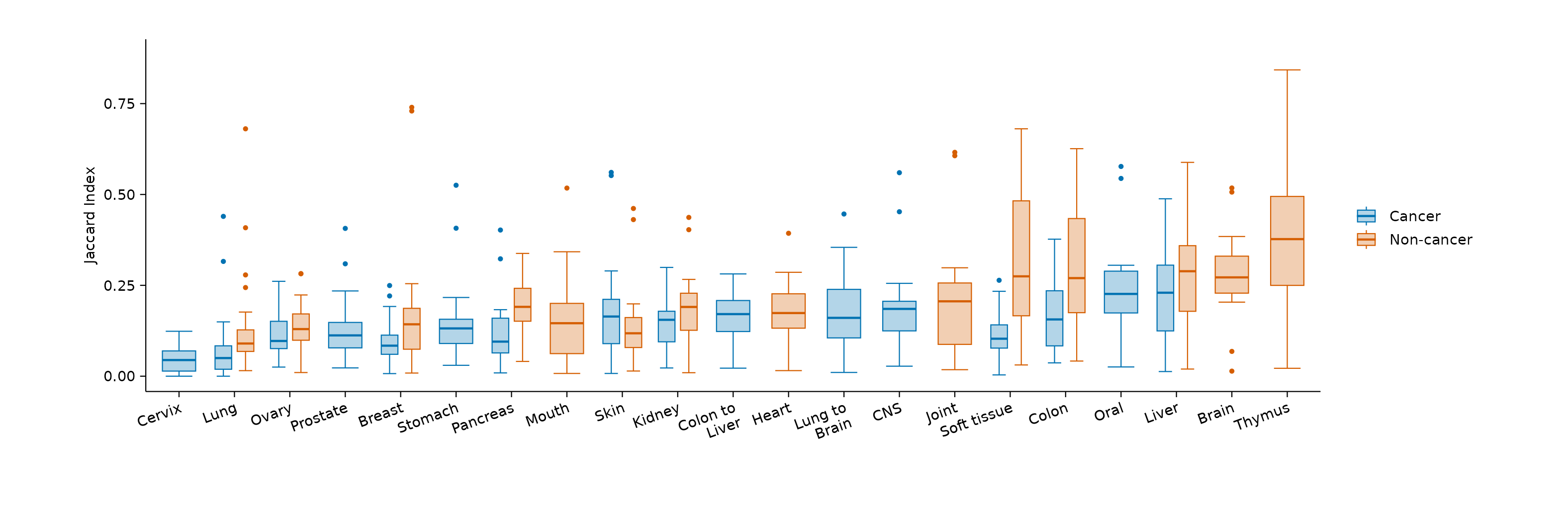}
    \caption{Jaccard Index between SVG sets and DE gene sets, stratified by disease status and tissue types.}
    \label{supp_fig:robustness_cancer_noncancer_similarity_by_tissue}
\end{figure}


\begin{figure}[H]
    \centering
    \includegraphics[width=0.8\textwidth]{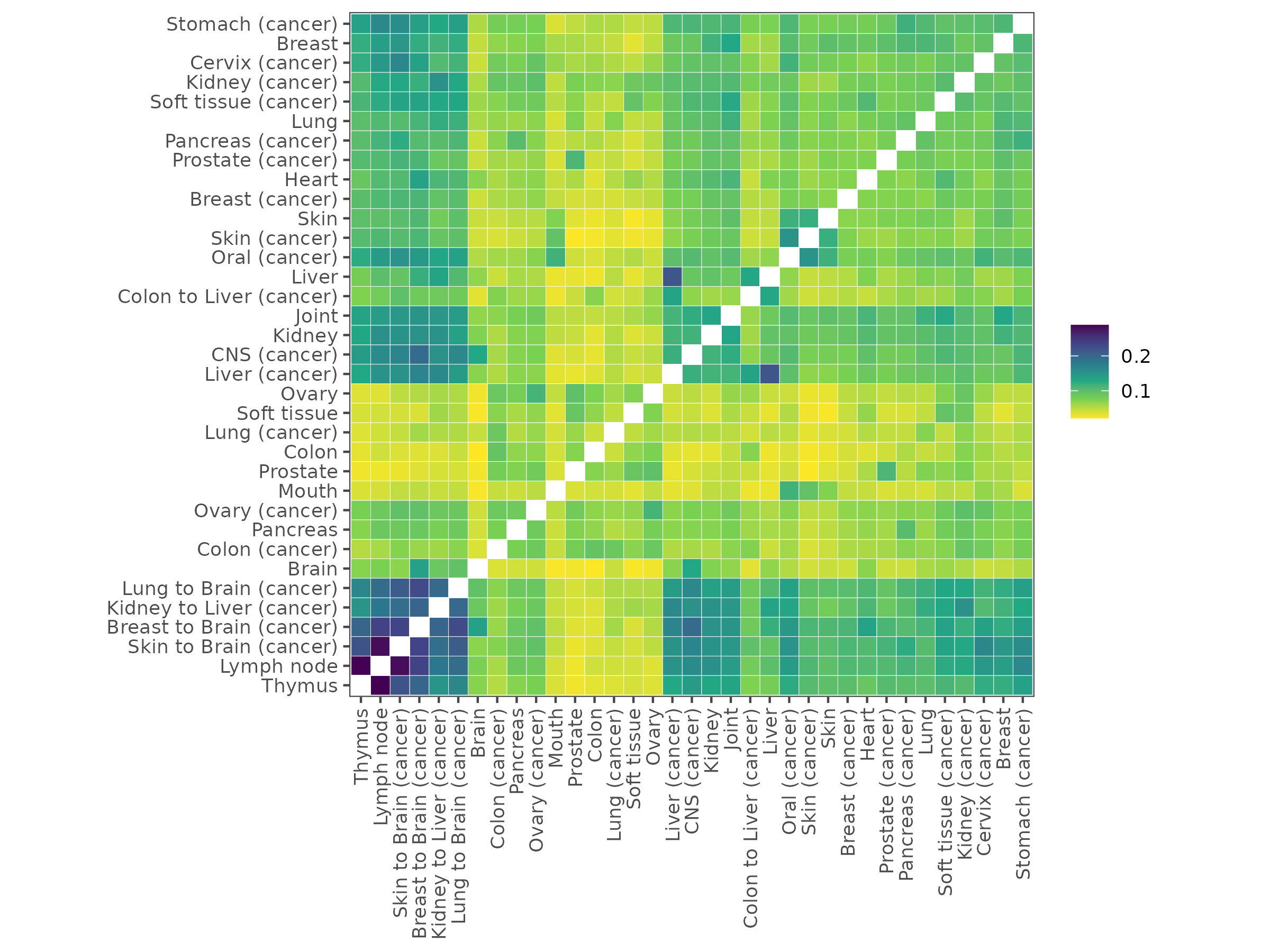}
    \caption{Jaccard Index similarity between all tissue types}
    \label{supp_fig:tissue_sim_tissue_tumor}
\end{figure}

\begin{figure}[H]
    \centering
    \includegraphics[width=0.8\textwidth]{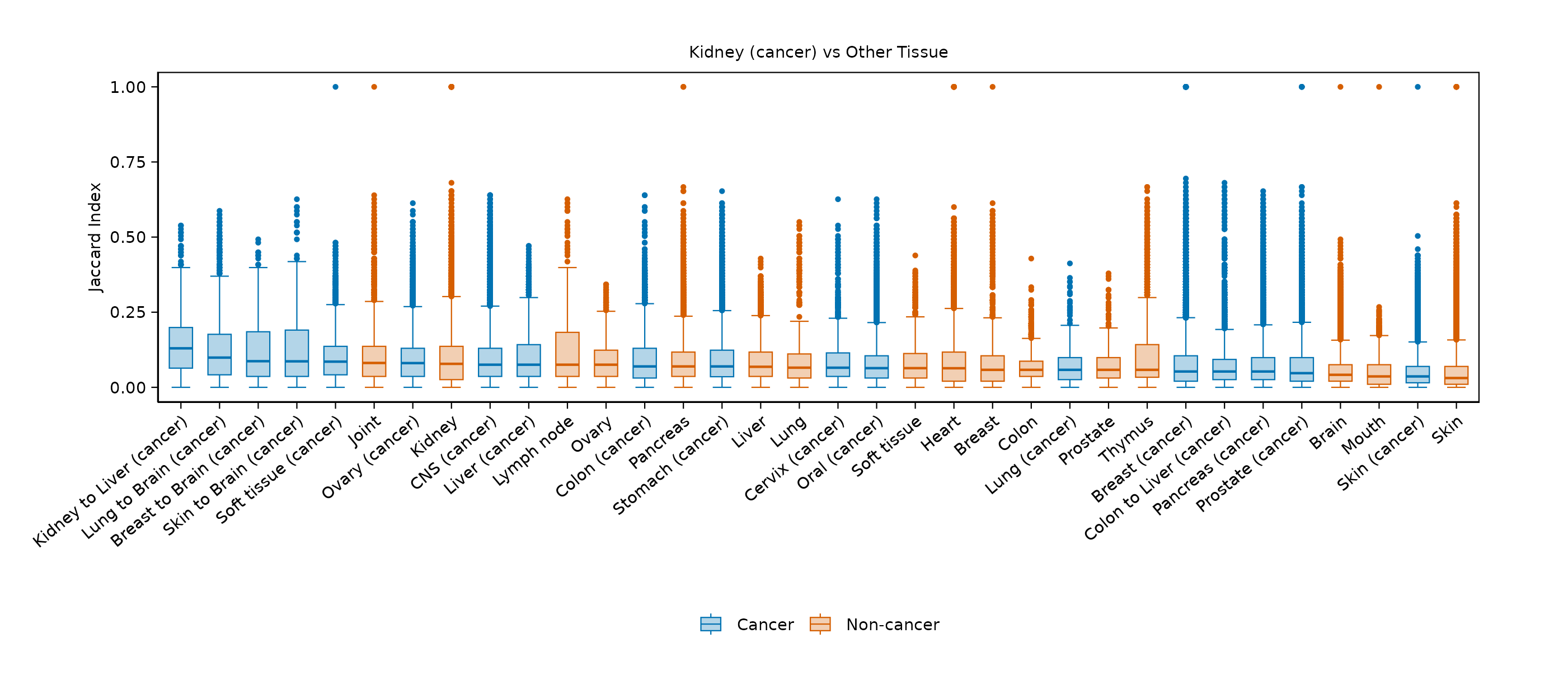}
    \caption{ Jaccard Index similarity between Kidney (cancer) vs Other Tissue}
    \label{supp_fig:tissue_sim_Kidney__cancer__vsother}
\end{figure}

\begin{figure}[H]
    \centering
    \includegraphics[width=0.8\textwidth]{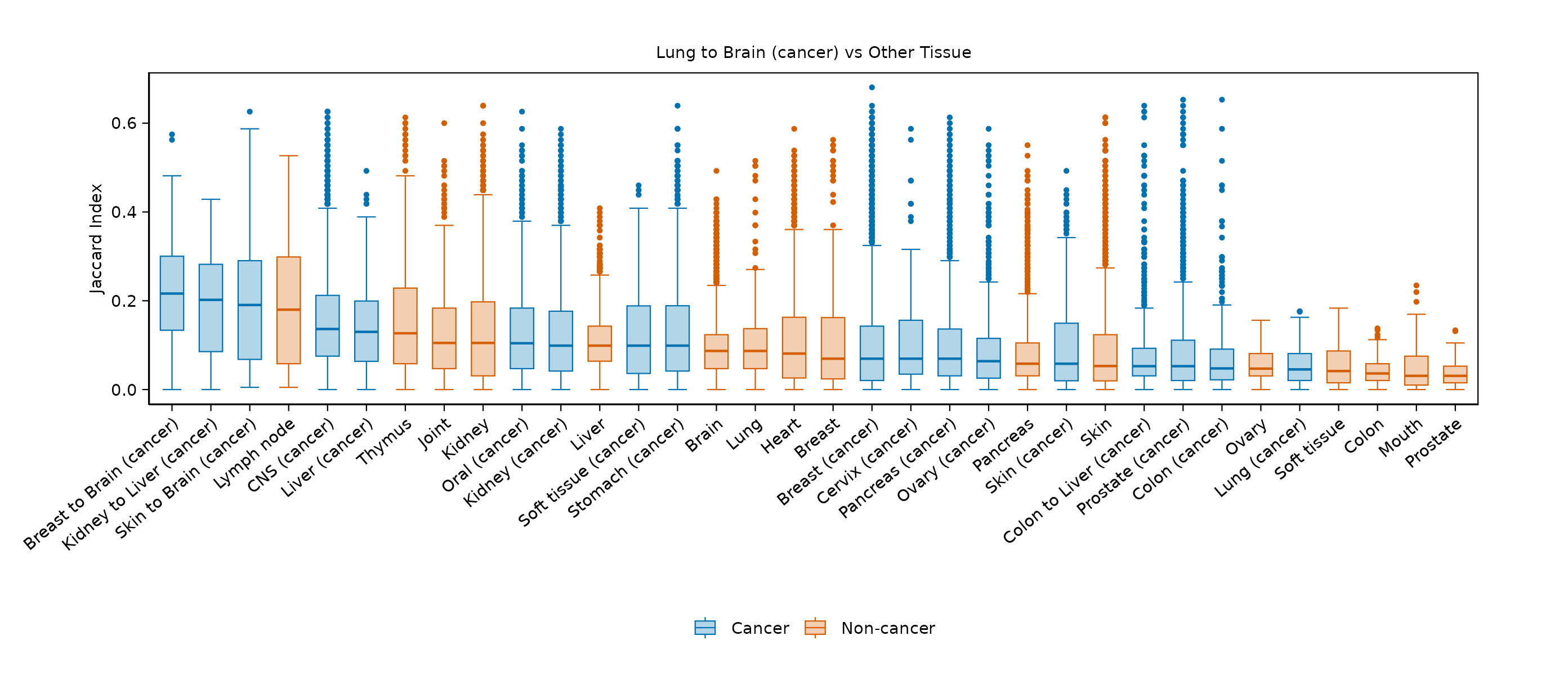}
    \caption{ Jaccard Index similarity between Lung to Brain (cancer) vs Other Tissue}
    \label{supp_fig:tissue_sim_Lung_to_Brain__cancer__vsother}
\end{figure}

\begin{figure}[H]
    \centering
    \includegraphics[width=0.8\textwidth]{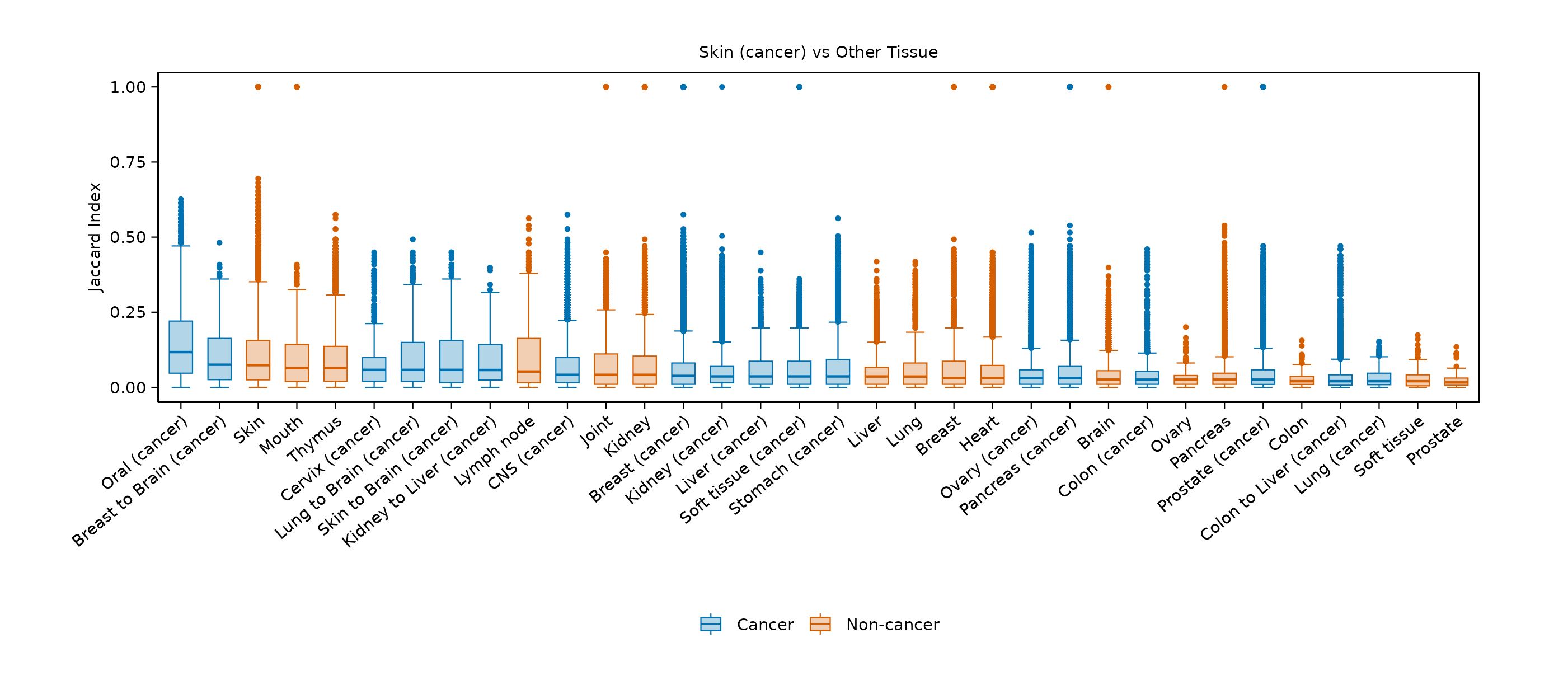}
    \caption{ Jaccard Index similarity between Skin (cancer) vs Other Tissue}
    \label{supp_fig:tissue_sim_Skin__cancer__vsother}
\end{figure}

\begin{figure}[H]
    \centering
    \includegraphics[width=0.8\textwidth]{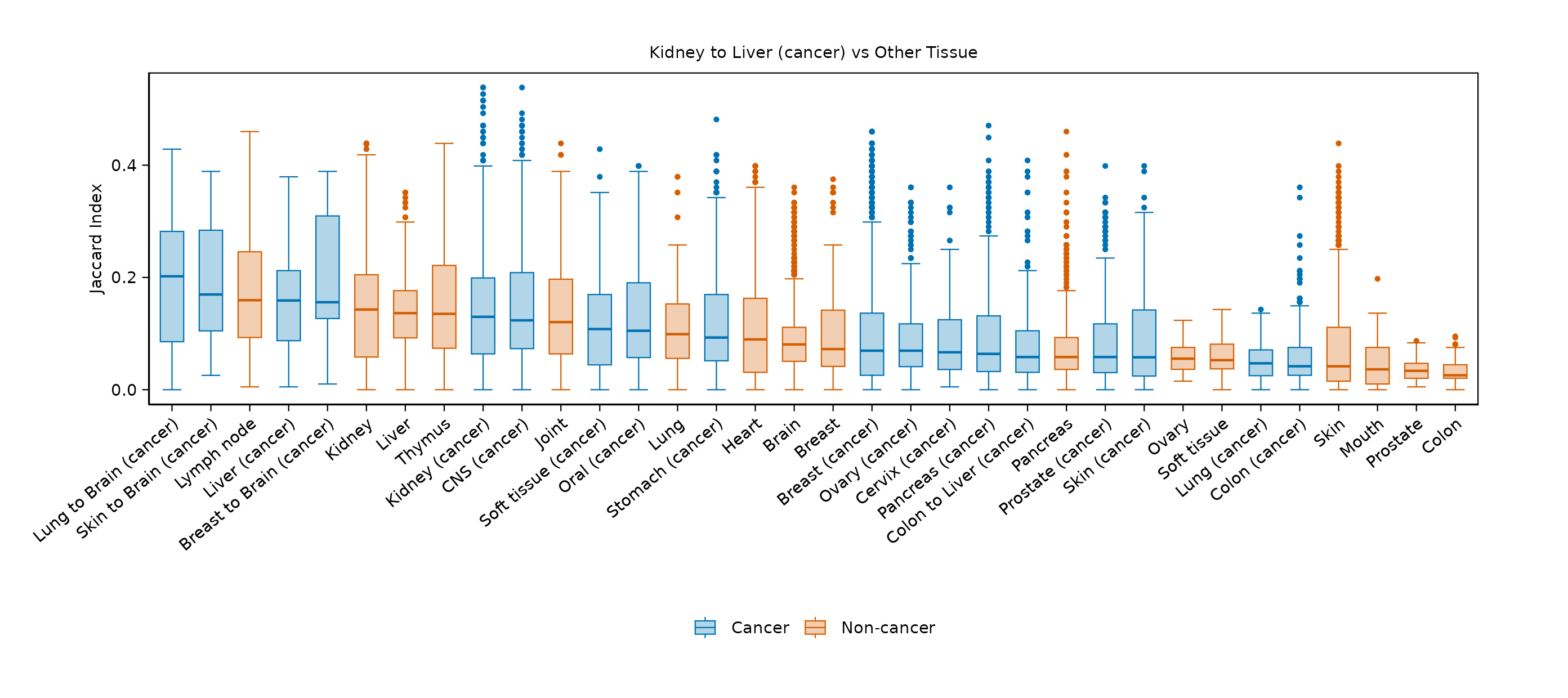}
    \caption{ Jaccard Index similarity between Kidney to Liver (cancer) vs Other Tissue}
    \label{supp_fig:tissue_sim_Kidney_to_Liver__cancer__vsother}
\end{figure}

\begin{figure}[H]
    \centering
    \includegraphics[width=0.8\textwidth]{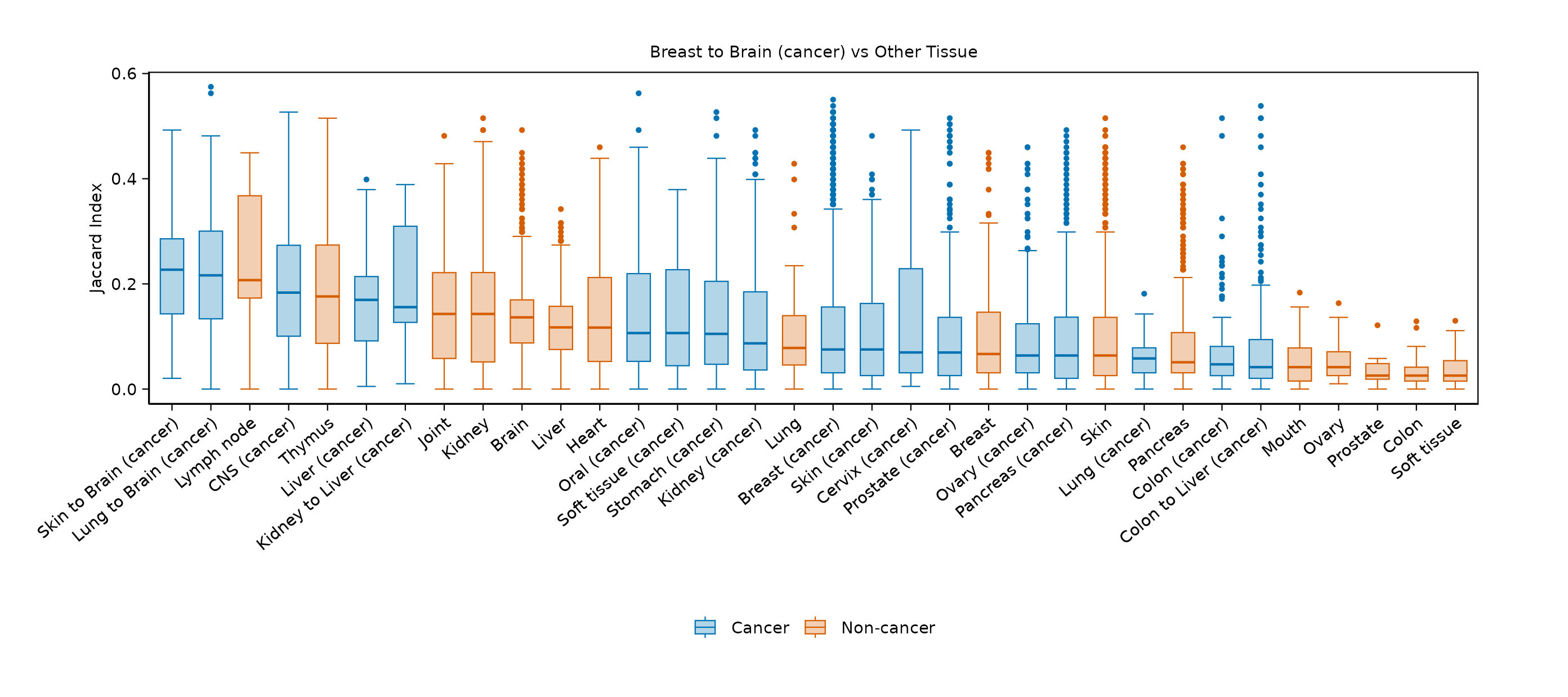}
    \caption{ Jaccard Index similarity between Breast to Brain (cancer) vs Other Tissue}
    \label{supp_fig:tissue_sim_Breast_to_Brain__cancer__vsother}
\end{figure}

\begin{figure}[H]
    \centering
    \includegraphics[width=0.8\textwidth]{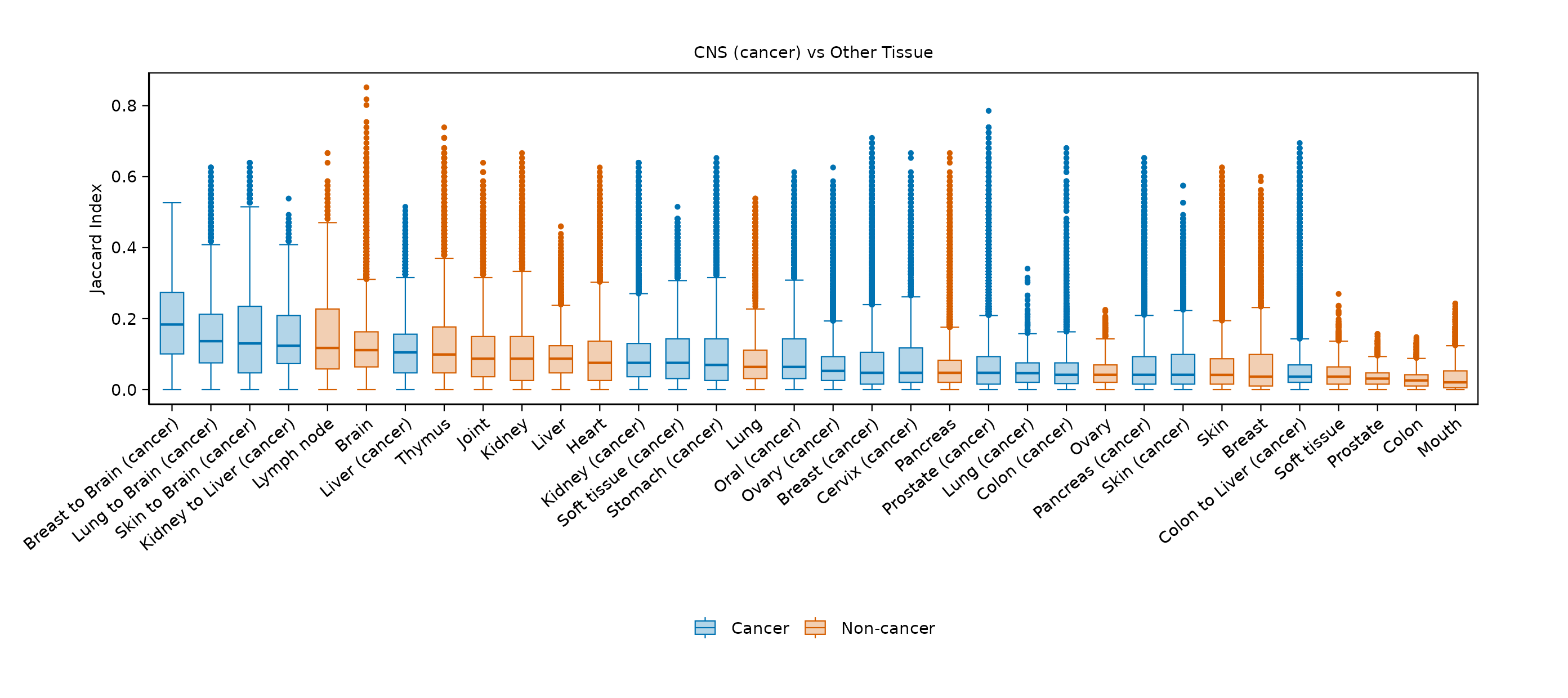}
    \caption{ Jaccard Index similarity between CNS (cancer) vs Other Tissue}
    \label{supp_fig:tissue_sim_CNS__cancer__vsother}
\end{figure}

\begin{figure}[H]
    \centering
    \includegraphics[width=0.8\textwidth]{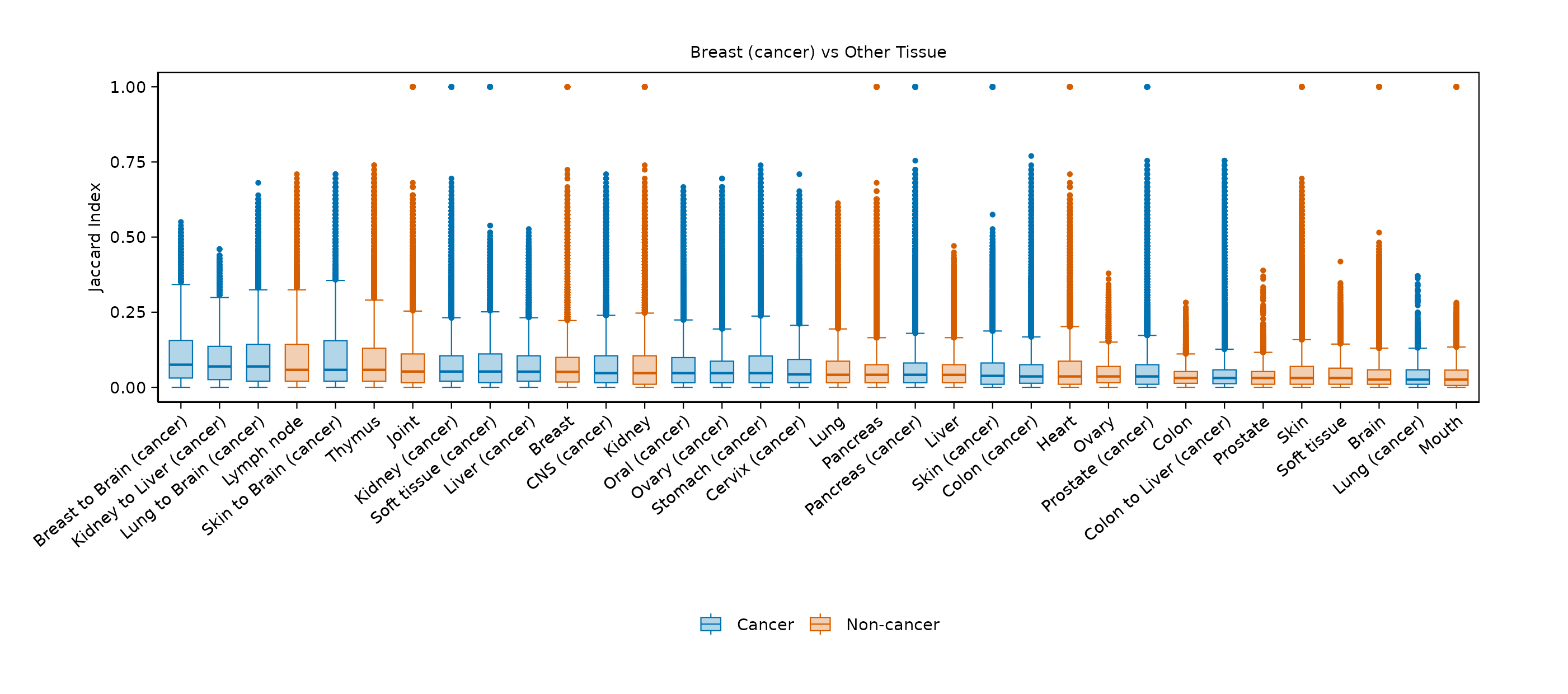}
    \caption{ Jaccard Index similarity between Breast (cancer) vs Other Tissue}
    \label{supp_fig:tissue_sim_Breast__cancer__vsother}
\end{figure}

\begin{figure}[H]
    \centering
    \includegraphics[width=0.8\textwidth]{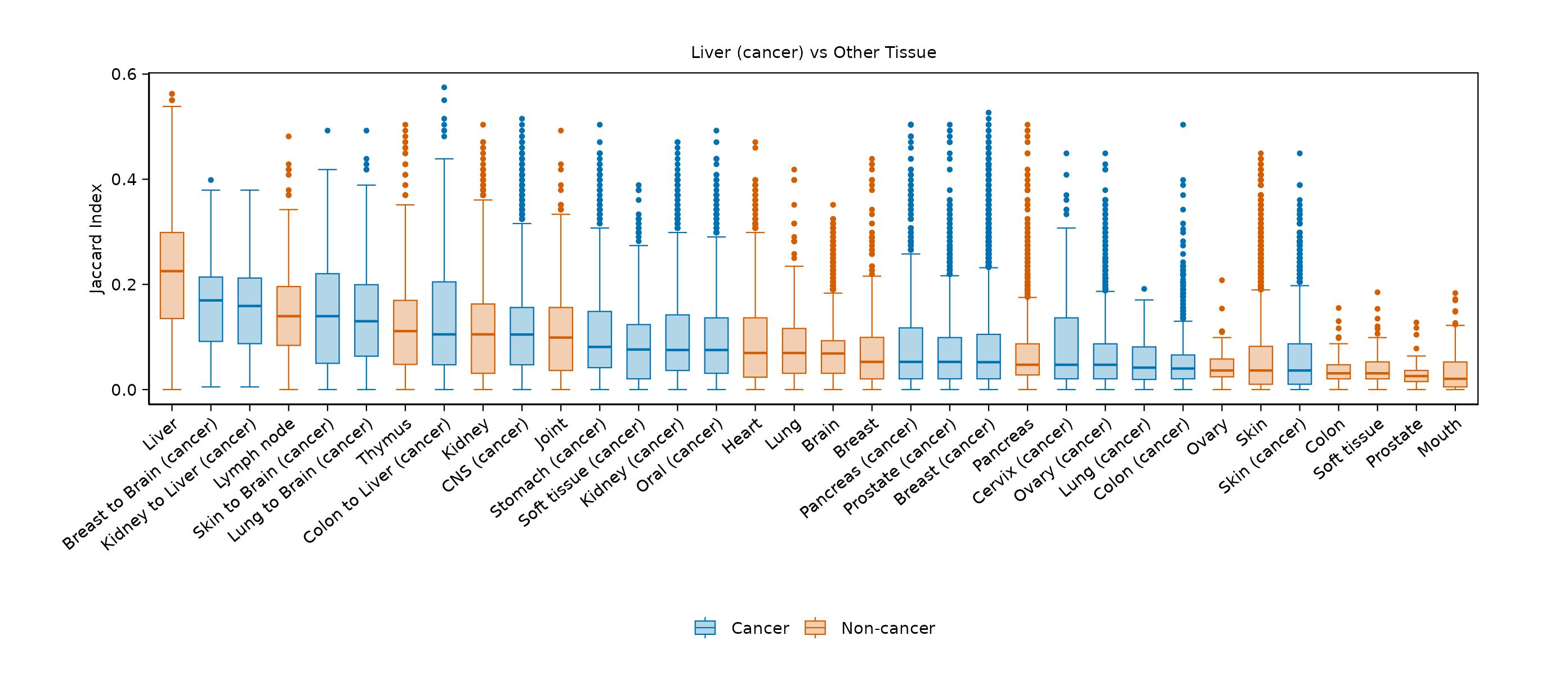}
    \caption{ Jaccard Index similarity between Liver (cancer) vs Other Tissue}
    \label{supp_fig:tissue_sim_Liver__cancer__vsother}
\end{figure}

\begin{figure}[H]
    \centering
    \includegraphics[width=0.8\textwidth]{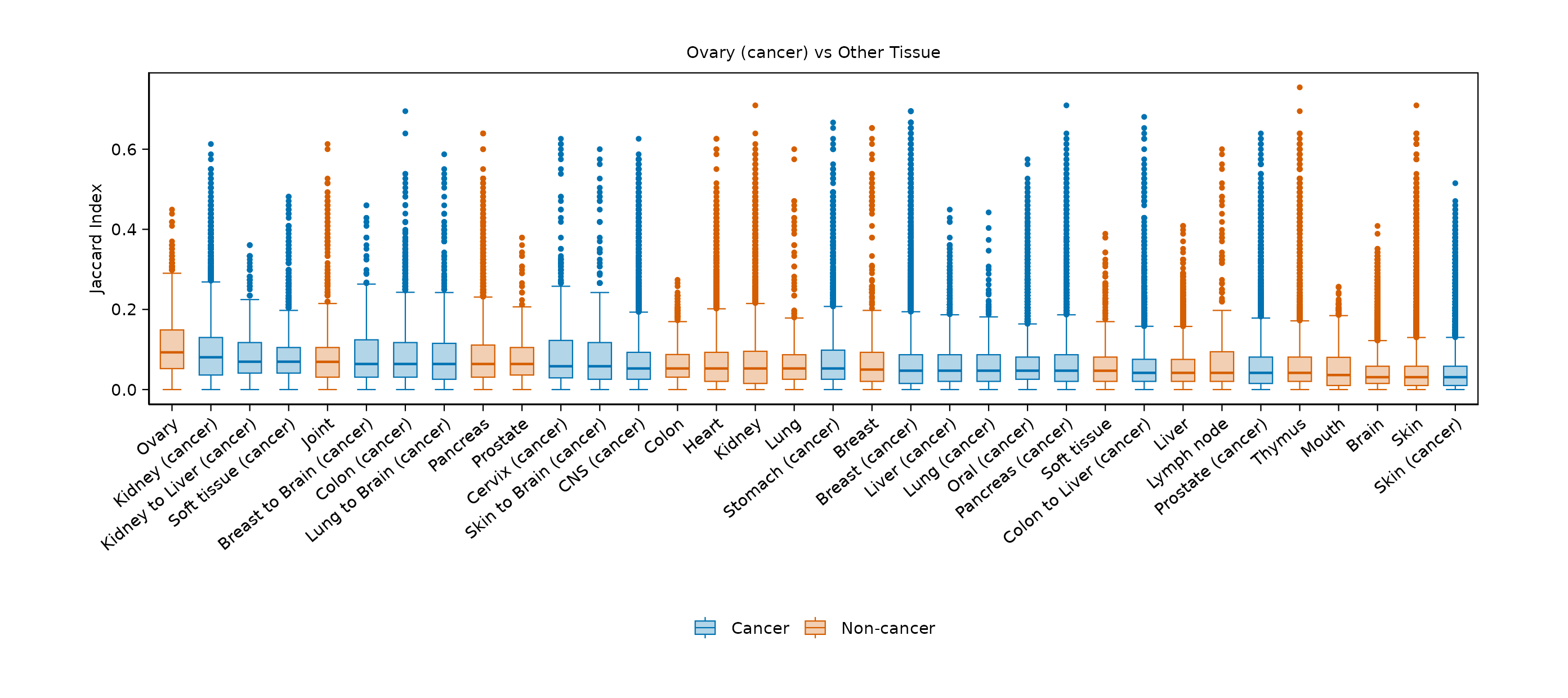}
    \caption{ Jaccard Index similarity between Ovary (cancer) vs Other Tissue}
    \label{supp_fig:tissue_sim_Ovary__cancer__vsother}
\end{figure}

\begin{figure}[H]
    \centering
    \includegraphics[width=0.8\textwidth]{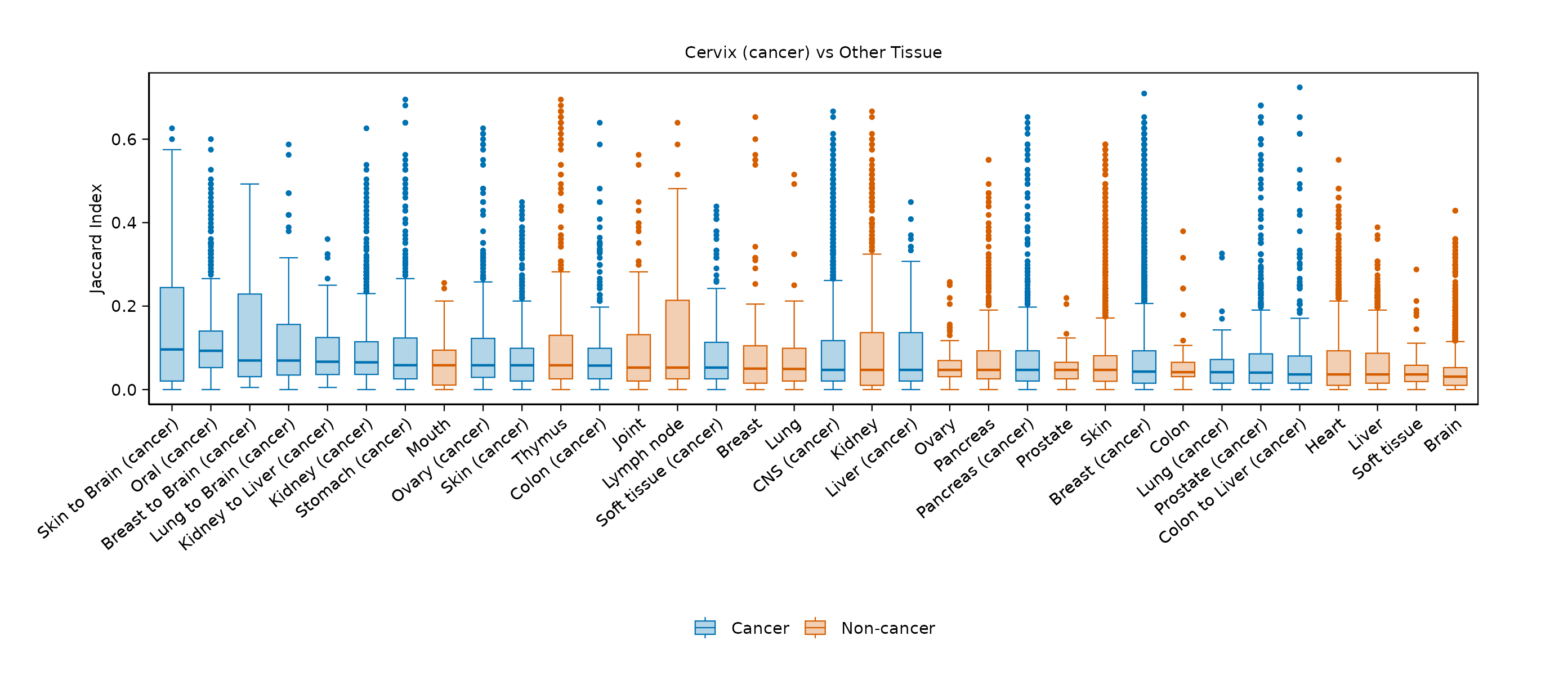}
    \caption{ Jaccard Index similarity between Cervix (cancer) vs Other Tissue}
    \label{supp_fig:tissue_sim_Cervix__cancer__vsother}
\end{figure}

\begin{figure}[H]
    \centering
    \includegraphics[width=0.8\textwidth]{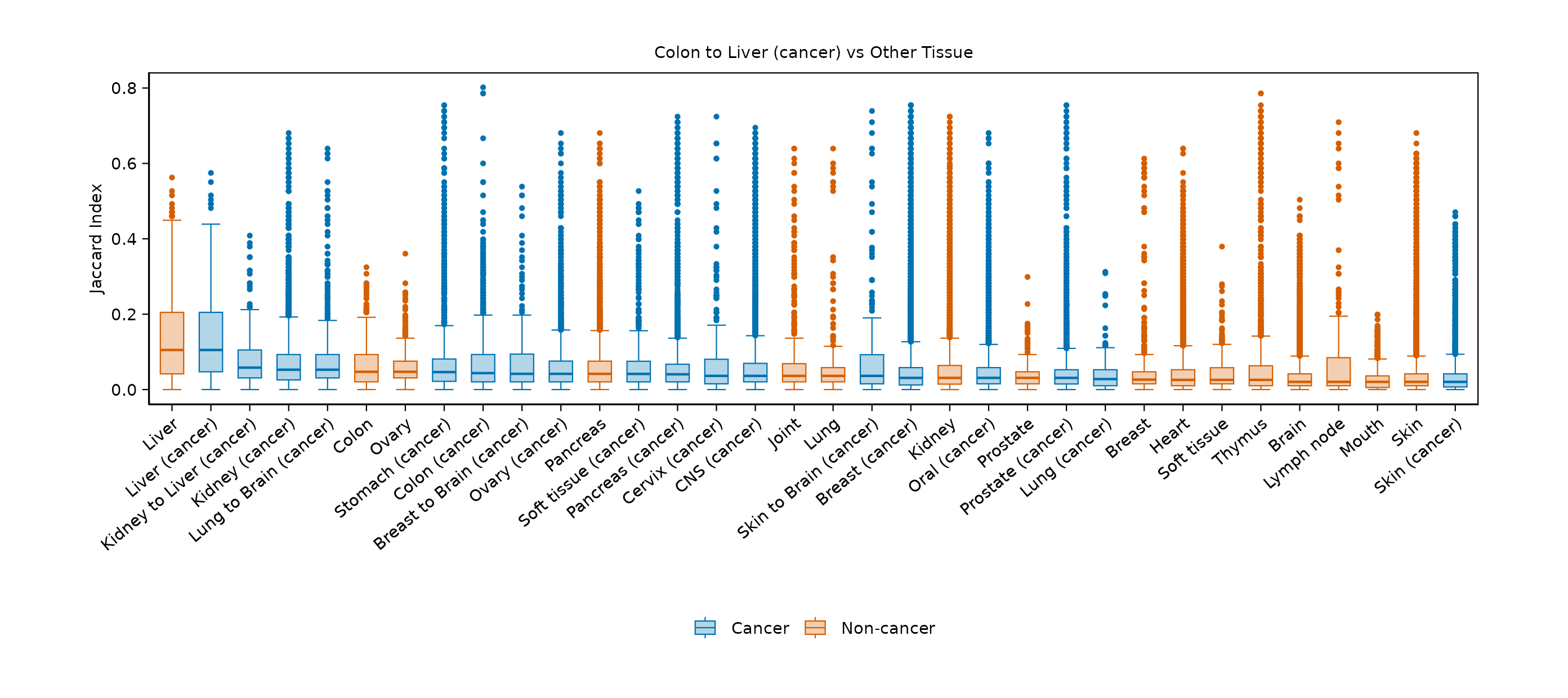}
    \caption{ Jaccard Index similarity between Colon to Liver (cancer) vs Other Tissue}
    \label{supp_fig:tissue_sim_Colon_to_Liver__cancer__vsother}
\end{figure}

\begin{figure}[H]
    \centering
    \includegraphics[width=0.8\textwidth]{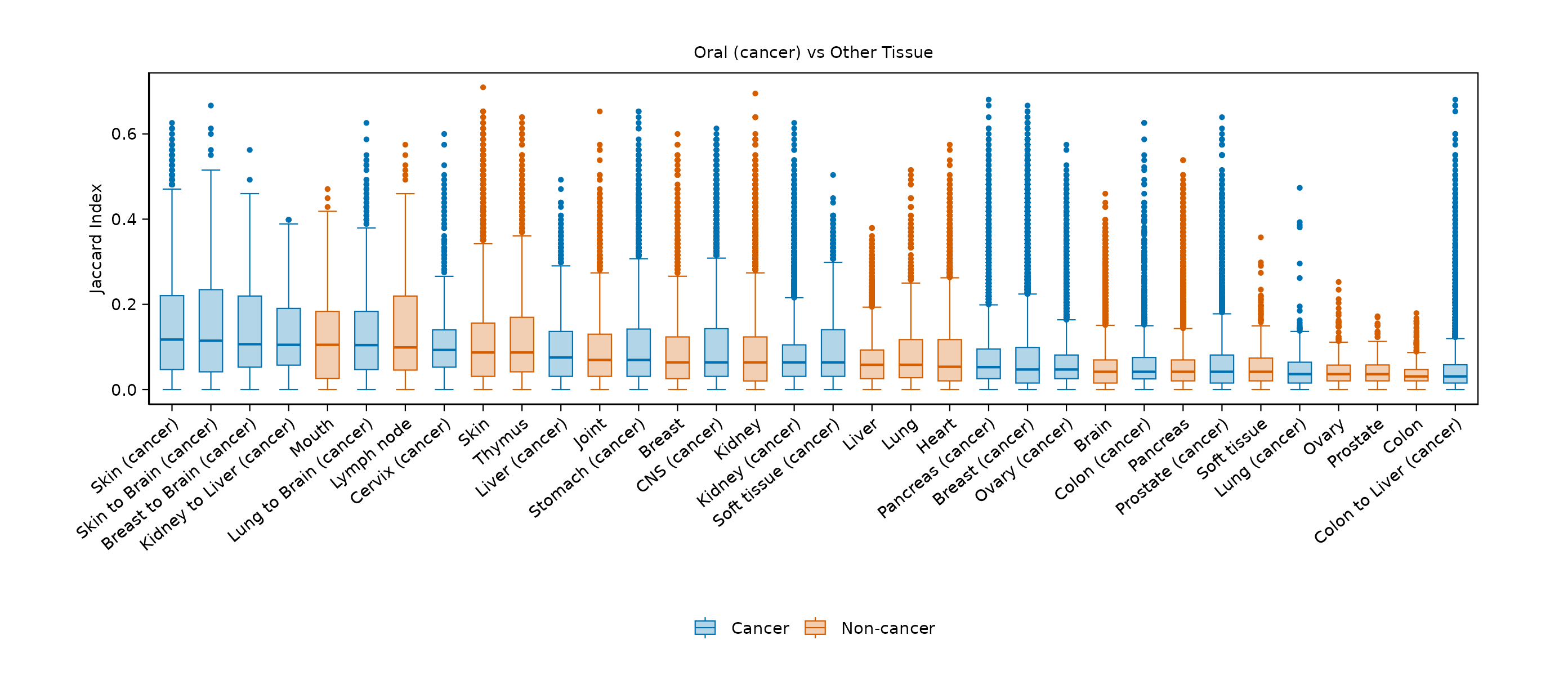}
    \caption{ Jaccard Index similarity between Oral (cancer) vs Other Tissue}
    \label{supp_fig:tissue_sim_Oral__cancer__vsother}
\end{figure}

\begin{figure}[H]
    \centering
    \includegraphics[width=0.8\textwidth]{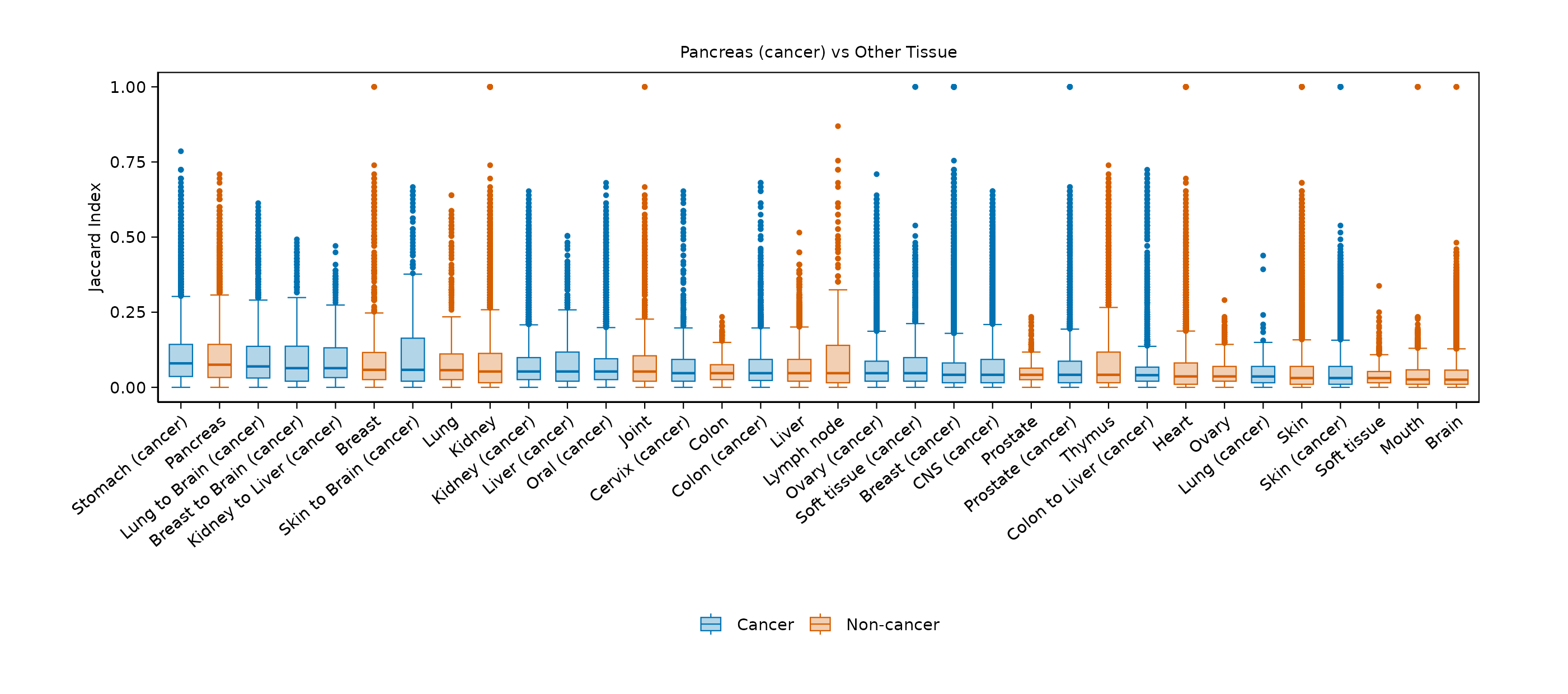}
    \caption{ Jaccard Index similarity between Pancreas (cancer) vs Other Tissue}
    \label{supp_fig:tissue_sim_Pancreas__cancer__vsother}
\end{figure}

\begin{figure}[H]
    \centering
    \includegraphics[width=0.8\textwidth]{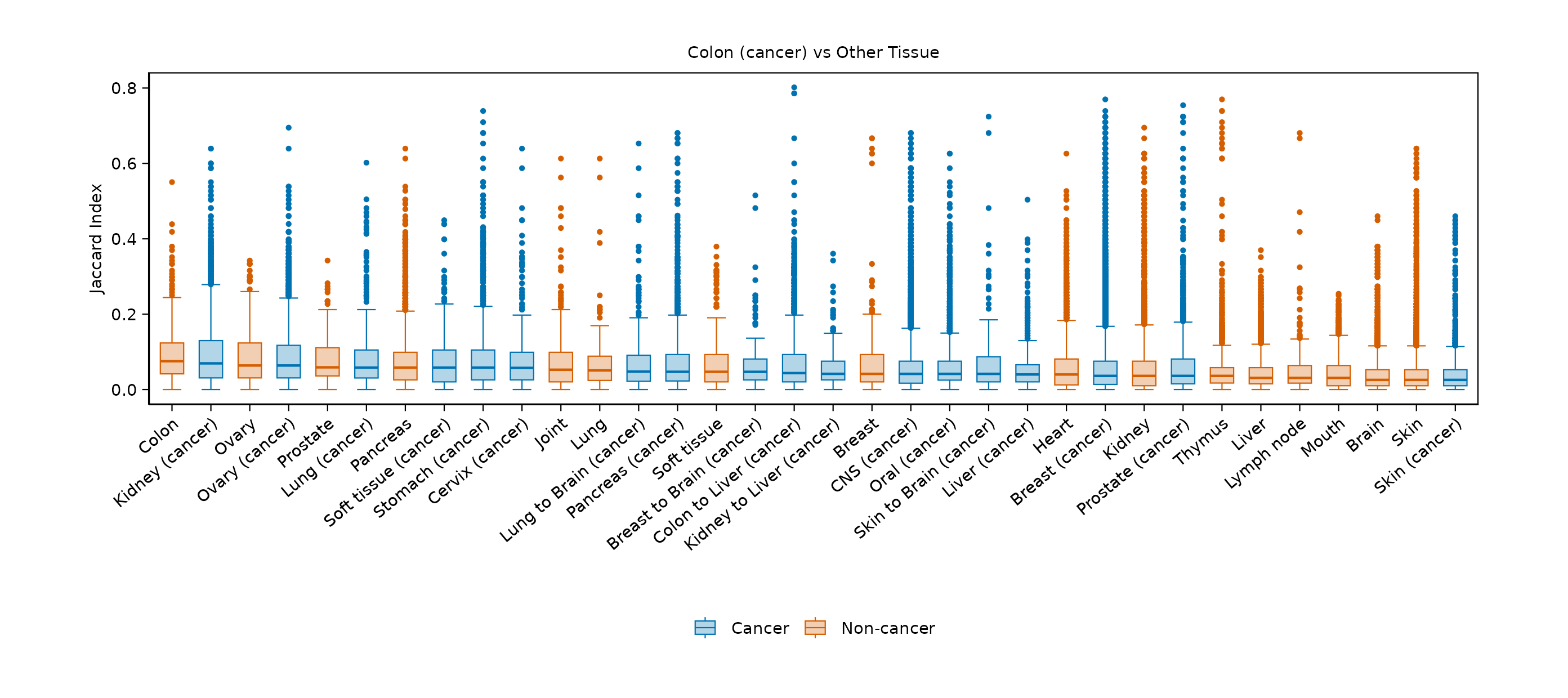}
    \caption{ Jaccard Index similarity between Colon (cancer) vs Other Tissue}
    \label{supp_fig:tissue_sim_Colon__cancer__vsother}
\end{figure}

\begin{figure}[H]
    \centering
    \includegraphics[width=0.8\textwidth]{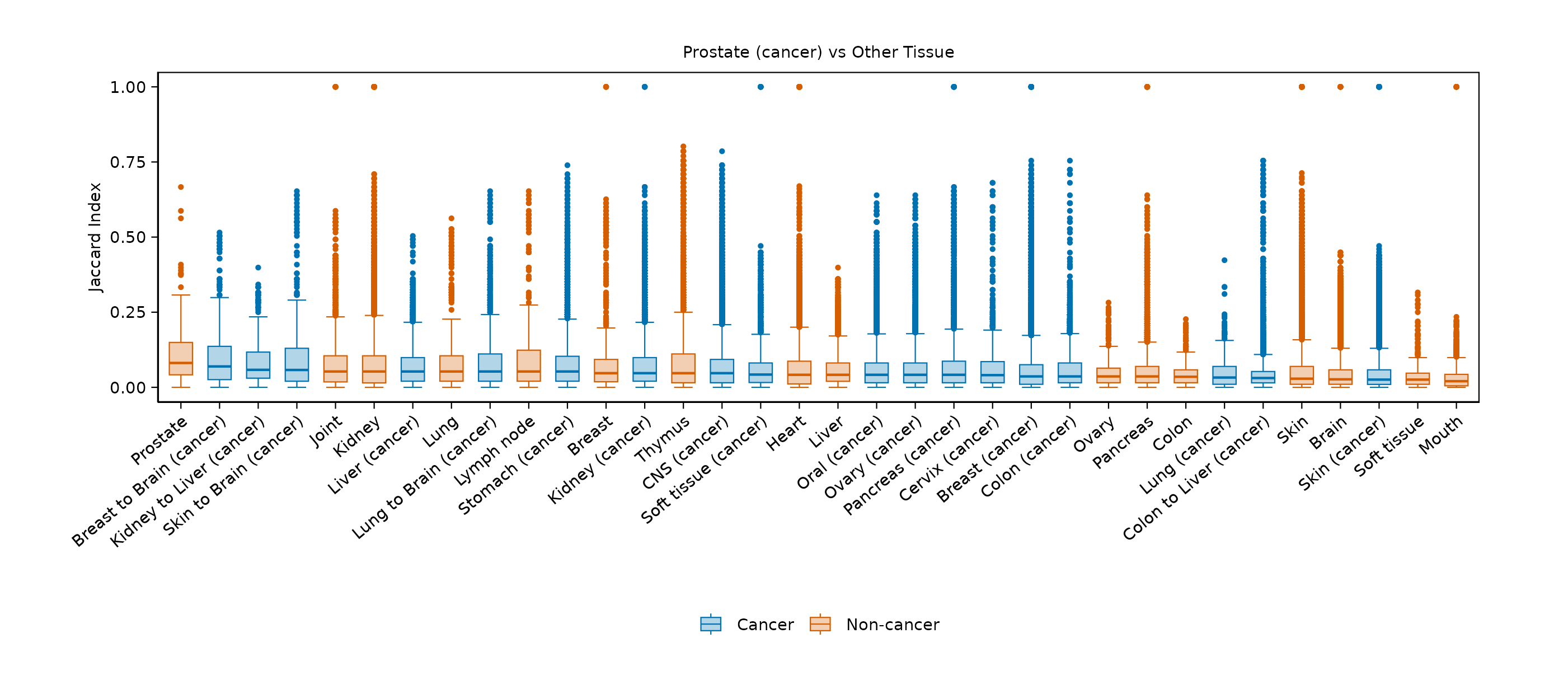}
    \caption{ Jaccard Index similarity between Prostate (cancer) vs Other Tissue}
    \label{supp_fig:tissue_sim_Prostate__cancer__vsother}
\end{figure}

\begin{figure}[H]
    \centering
    \includegraphics[width=0.8\textwidth]{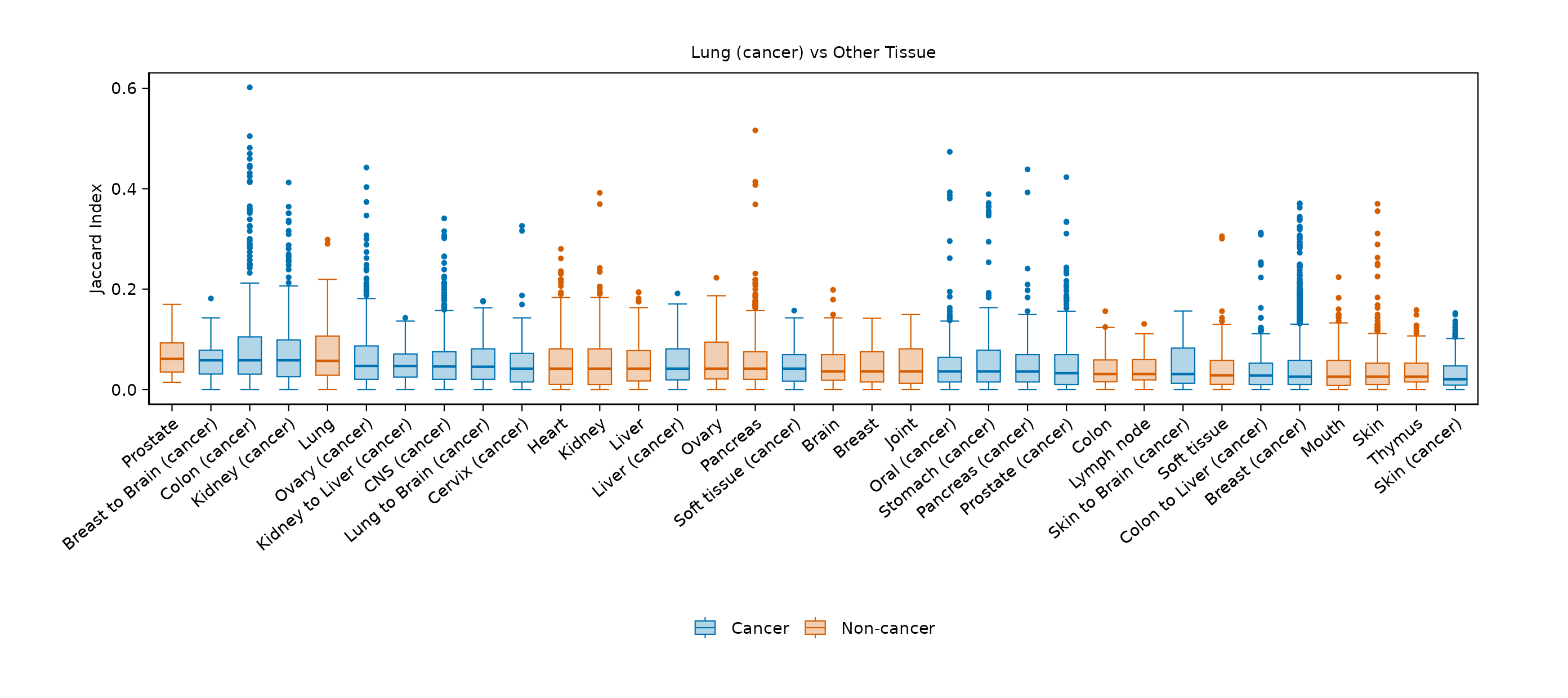}
    \caption{ Jaccard Index similarity between Lung (cancer) vs Other Tissue}
    \label{supp_fig:tissue_sim_Lung__cancer__vsother}
\end{figure}

\begin{figure}[H]
    \centering
    \includegraphics[width=0.8\textwidth]{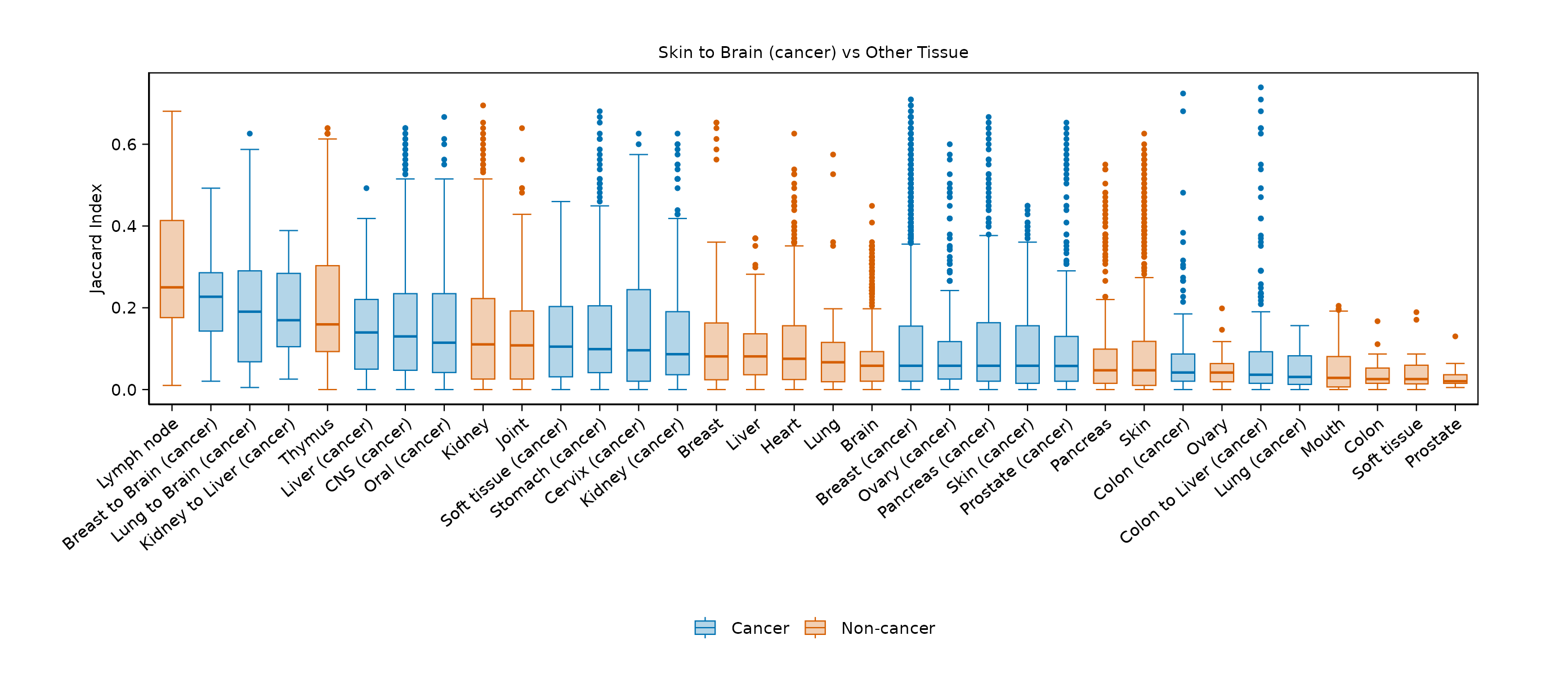}
    \caption{ Jaccard Index similarity between Skin to Brain (cancer) vs Other Tissue}
    \label{supp_fig:tissue_sim_Skin_to_Brain__cancer__vsother}
\end{figure}

\begin{figure}[H]
    \centering
    \includegraphics[width=0.8\textwidth]{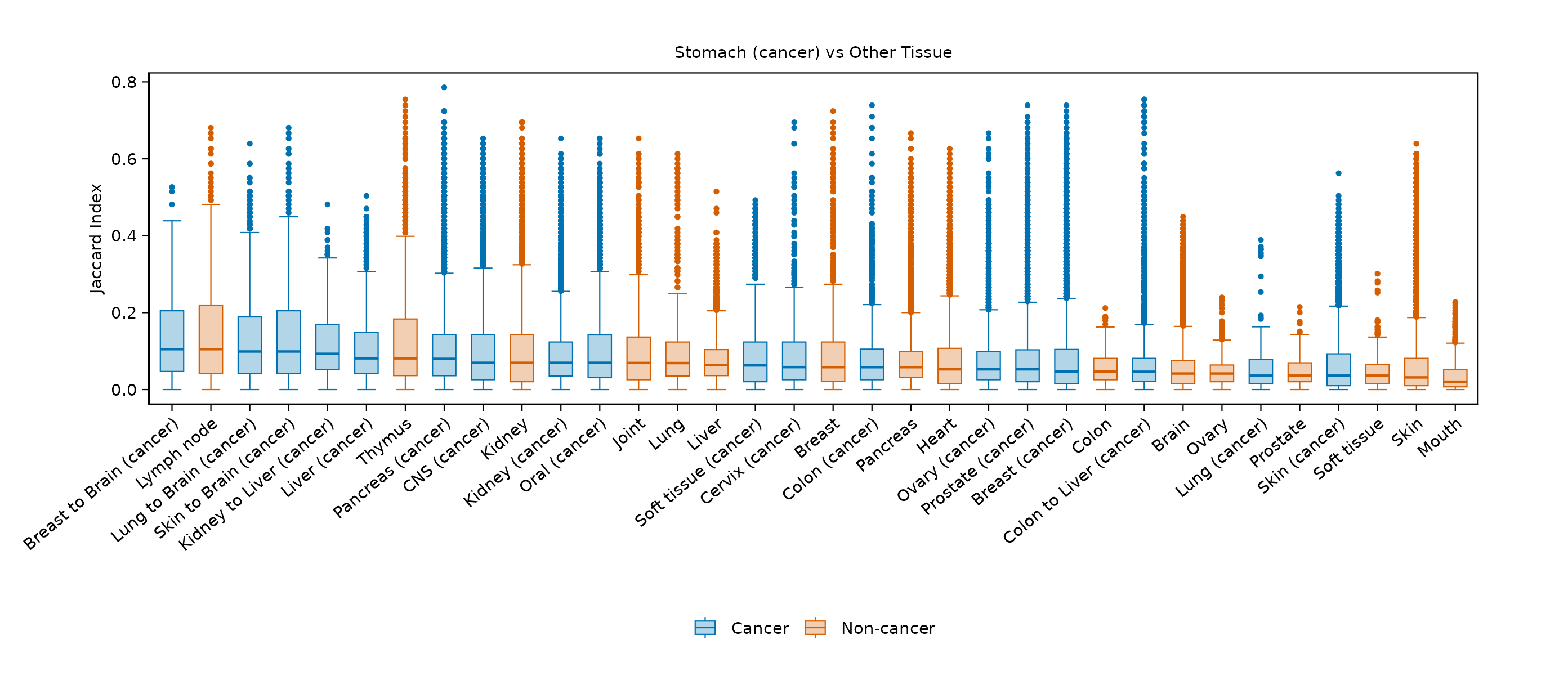}
    \caption{ Jaccard Index similarity between Stomach (cancer) vs Other Tissue}
    \label{supp_fig:tissue_sim_Stomach__cancer__vsother}
\end{figure}

\begin{figure}[H]
    \centering
    \includegraphics[width=0.8\textwidth]{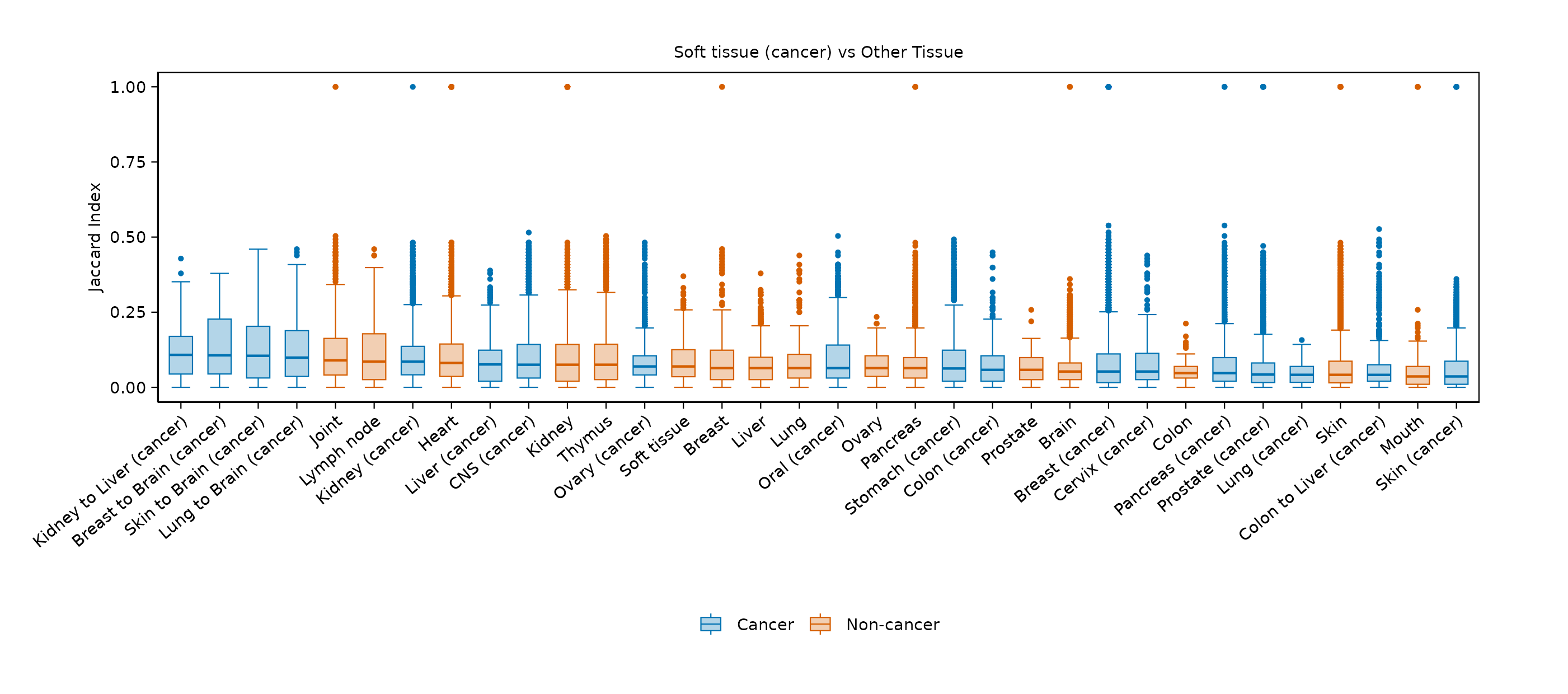}
    \caption{ Jaccard Index similarity between Soft tissue (cancer) vs Other Tissue}
    \label{supp_fig:tissue_sim_Soft_tissue__cancer__vsother}
\end{figure}


\begin{figure}[H]
    \centering
    \includegraphics[]{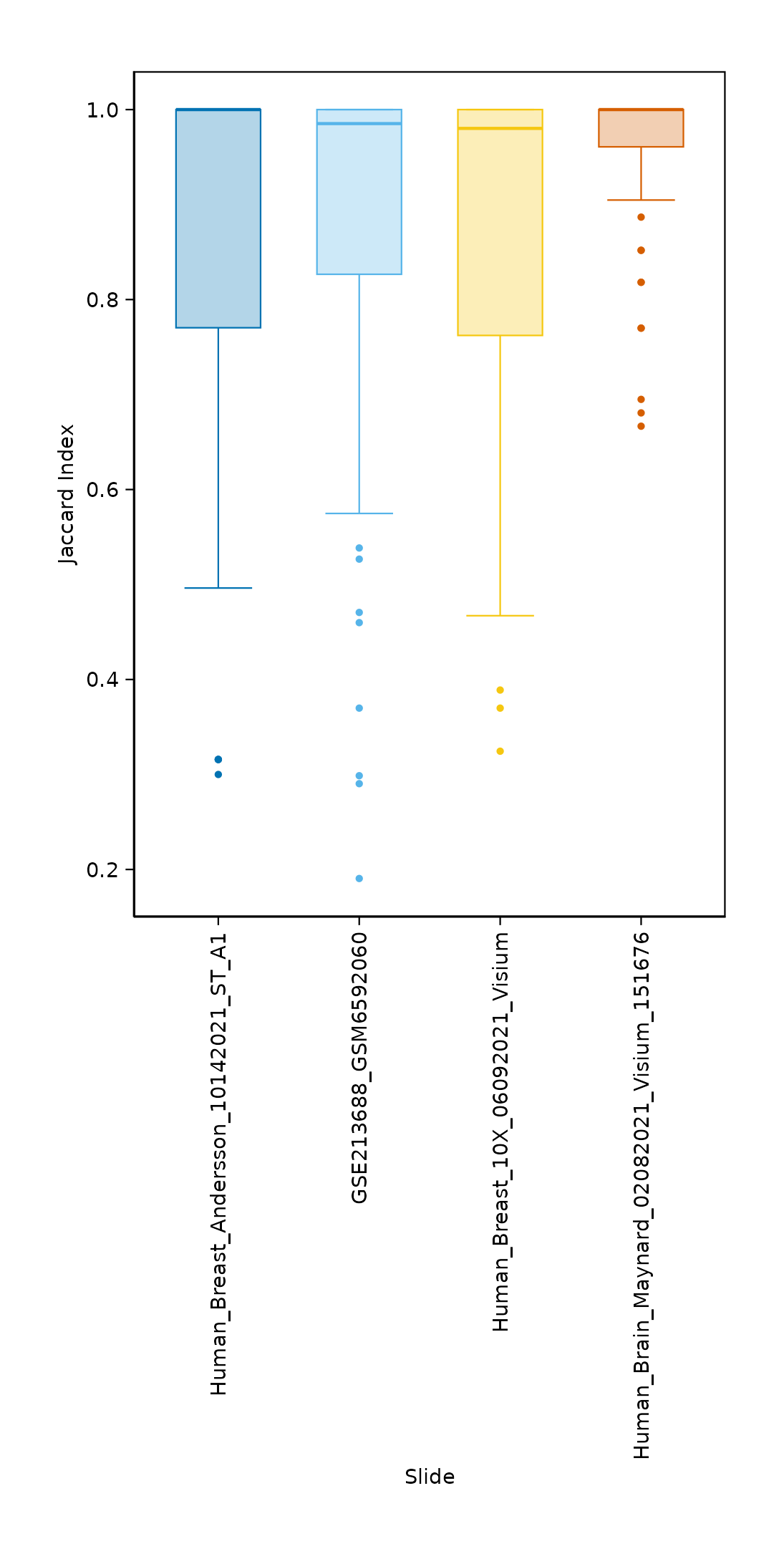}
    \caption{Rotation robustness stratified by slides.}
    \label{supp_fig:rotation_by_tissue}
\end{figure}

\begin{figure}[H]
    \centering
    \includegraphics[width=\textwidth]{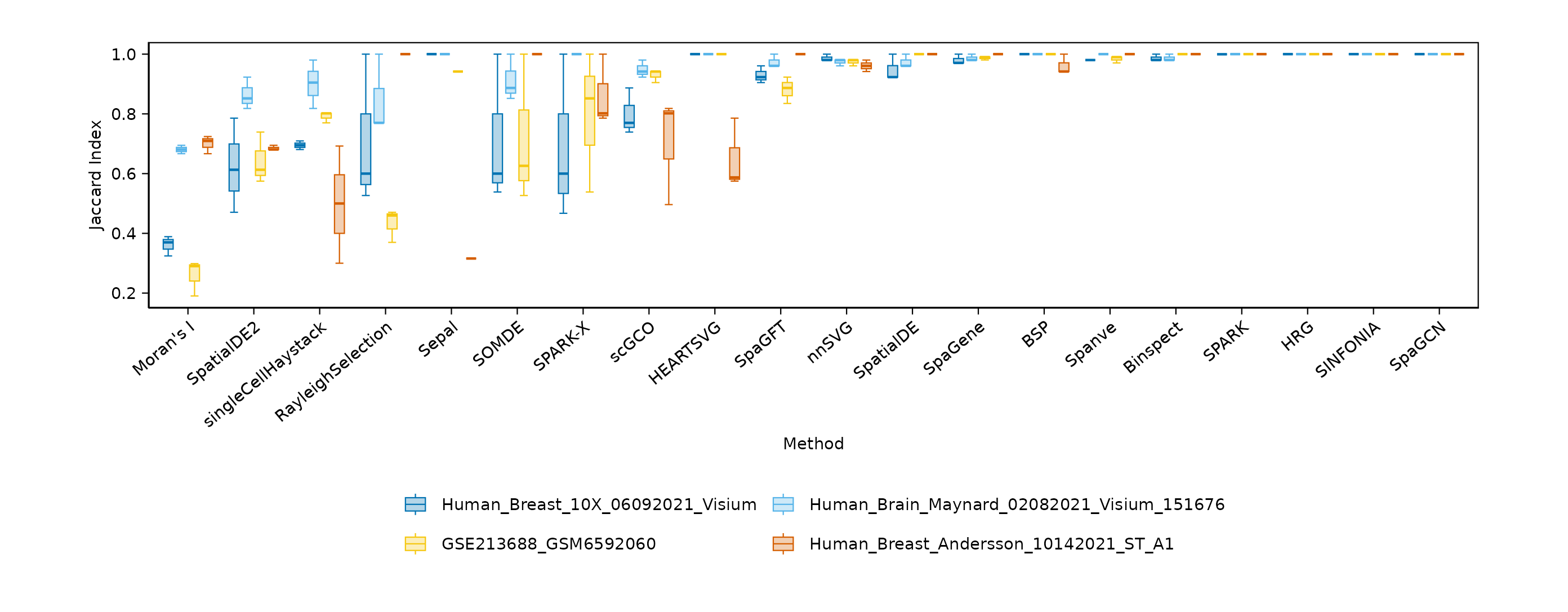}
    \caption{Rotation robustness stratified by slides and methods.}
    \label{supp_fig:rotation_by_method}
\end{figure}

\end{document}